\documentclass[a4paper,fleqn]{cas-sc}

\usepackage[numbers,sort&compress]{natbib}

\usepackage{amsthm,physics}
\usepackage{algorithm, algorithmicx, algpseudocode}
\algdef{S}[FOR]{ParFor}[1]
    {\textbf{parallel for}\ #1\ \algorithmicdo}
\usepackage{chngcntr}
\usepackage{etoolbox}
\pretocmd{\appendix}{%
  \counterwithin{algorithm}{section}%
}{}{}

\usepackage{bm}
\usepackage{bm,pifont}
\usepackage{bm,pifont}
\usepackage{tikz}
\usepackage{tabularx}
\usepackage{subcaption}
\usepackage{float}
\usepackage[final,commandnameprefix=ifneeded]{changes}
\usepackage[scr=boondoxo]{mathalpha}
\usepackage{comment}

\usetikzlibrary{arrows.meta,positioning,calc,patterns,decorations.markings}

\newtheorem{lemma}{Lemma}

\algdef{SE}[SWITCH]{Switch}{EndSwitch}[1]
  {\textbf{switch} (#1)}{\textbf{end switch}}
\algdef{SE}[CASE]{Case}{EndCase}[1]
  {\textbf{case} #1 \textbf{:}}{\textbf{end case}}
\algdef{SE}[DEFAULT]{Default}{EndDefault}
  {\textbf{default:}}{\textbf{end default}}
\algtext*{EndCase}
\algtext*{EndDefault}

\definecolor{Koryn}{rgb}{0.25,0.55,0.2}
\definecolor{r1color}{rgb}{0.8,0.1,0.1}
\definecolor{r2color}{rgb}{0.15,0.2,0.65}
\definechangesauthor[name={Keyu Xu},color=Koryn]{xky}
\definechangesauthor[name={Qian Wang},color=r1color]{qian}
\definechangesauthor[name={Nan Zhao},color=r2color]{nan}

\begin{document}
\let\WriteBookmarks\relax
\def\floatpagepagefraction{1}
\def\textpagefraction{.001}
\shorttitle{High-fidelity simulation and reduced-order modeling of optical pumping}
\shortauthors{Xu et~al.}

\title[mode=title]{High-fidelity simulation and reduced-order modeling of optical pumping in quantum devices}

% \tnotemark[1,2]
% \tnotetext[1]{This document is the results of the research
%    project funded by the National Science Foundation.}
% \tnotetext[2]{The second title footnote which is a longer text matter
%    to fill through the whole text width and overflow into
%    another line in the footnotes area of the first page.}

\author[Mechanics]{Keyu Xu} [orcid=0009-0009-4803-1504]
\credit{Methodology, Software, Validation, Formal analysis, Visualization, Writing – original draft, Writing – review \& editing}

\author[Quantum]{Kezheng Yan} [orcid=0009-0000-3226-6995]
\credit{Methodology, Validation}

\author[Quantum]{Jinbo Hu} [orcid=0000-0001-8555-8166]
\credit{Methodology, Validation}

\author[Mechanics]{Yanqing Shi} [orcid=0009-0007-7117-5528]
\credit{Writing – review \& editing}

\author[Mechanics]{Xiaorui Xu} [orcid=0009-0001-1776-8687]
\credit{Writing – review \& editing}

\author[Quantum]{Nan Zhao} [orcid=0000-0002-4521-709X]
\credit{Conceptualization, Supervision, Writing – review \& editing}

\author[Mechanics]{Qian Wang}[orcid=0000-0001-5409-1663]
\credit{Conceptualization, Methodology, Writing – review \& editing, Supervision, Funding acquisition}
\cormark[1]
\ead{qian.wang@csrc.ac.cn}

\affiliation[Mechanics]{organization={Mechanics Division, Beijing Computational Science Research Center},%Department and Organization
    city={Beijing},
    postcode={100193},
    country={China}}
    % country={People's Republic of China}}

\affiliation[Quantum]{organization={Quantum Physics and Quantum Information Division, Beijing Computational Science Research Center},%Department and Organization
    city={Beijing},
    postcode={100193},
    country={China}}

\cortext[correspond]{Corresponding author.}

\begin{abstract}
Predicting spin polarization and light attenuation in optically pumped atomic devices requires solving the coupled diffusive quantum master equation (DQME)--light system. Its nonlinearity, spatially nonlocal coupling, and physical admissibility constraints make high-fidelity numerical simulation challenging. We develop a structure-preserving high-fidelity solver that combines second-order finite differences for three-dimensional atomic diffusion, explicit pseudo-time marching for atomic populations, and an exponential integrator for light propagation. Under a derived time-step restriction, the solver preserves population normalization and nonnegativity, as well as the positivity and monotonic attenuation of the light intensity. Three complementary test cases based on an analytical reduction, asymptotic limits, and an analytical invariant are constructed to verify the solver. To accelerate repeated parameter queries, we eliminate the light field and construct a reduced-order model (ROM) using proper orthogonal decomposition and Galerkin projection of the resulting integro-differential formulation involving only atomic populations. An exact quadratic expansion of the spin-exchange term and discrete empirical interpolation of the optical-pumping term make the cost of evaluating the reduced operators independent of the full-order dimension. Numerical tests verify the solver and demonstrate ROM accuracy outside the training set. The GPU implementation completes simulations with millions of unknowns in minutes, while single-core ROM evaluations take seconds, achieving respective speedup factors exceeding $150$ and $7000$ relative to the single-core high-fidelity solver.
\end{abstract}


\maketitle
% \linenumbers

% Main text
% !TeX root = main.tex

\section{Introduction}  \label{sec:intro}

Quantum sensors exploit the properties of atomic and other quantum
systems to achieve high measurement sensitivity in applications
ranging from fundamental physics research to navigation, timekeeping,
and biomedical imaging~\cite{aslam2023quantum}.
Optical pumping provides an essential means of preparing and controlling
atomic spin polarization in a broad class of quantum devices, including
spin-exchange-relaxation-free (SERF) atomic
magnetometers~\cite{budker2007optical,allred2002high},
coherent-population-trapping (CPT) atomic clocks~\cite{vanier2005atomic},
and atomic co-magnetometer-based nuclear-spin
gyroscopes~\cite{kornack2005nuclear}.
In optically pumped alkali-metal vapors, atomic spin polarization
evolves under the combined effects of optical pumping, spin-exchange
collisions, spin relaxation, and spatial
diffusion~\cite{happer1972optical,happer1973spin,bhaskar1981light,appelt1998theory,happer2010optically,walker2011fundamentals}.
These processes can be modeled by a nonlinear reaction--diffusion
partial differential equation (PDE) derived from the quantum master
equation, referred to here as the diffusive quantum master equation
(DQME).
The propagation of pump light through the vapor is governed by a
first-order ordinary differential equation (ODE) along the propagation
direction.
The two equations are coupled: the atomic state determines optical
absorption along the propagation path, while the resulting light field
controls the local optical-pumping
rate~\cite{appelt1998theory,seltzer2008developments}.
The resulting coupled DQME--light system combines nonlinear atomic
population dynamics, three-dimensional (3D) diffusion, and spatially
nonlocal coupling through light propagation.
Accurate and efficient solution of this system is therefore important
for predicting the spatial distributions of atomic spin polarization
and pump-light intensity and for assessing their dependence on device
parameters and operating conditions.

However, a high-fidelity numerical solver for the coupled
DQME--light system is still lacking.
Previous analytical and numerical studies have investigated simpler
quantum master equations and optically pumped atomic
systems~\cite{happer1973spin,appelt1998theory,Savukov2005,PhysRevA.111.013103,campaioli2024quantum}.
Spatially resolved models have also been developed to incorporate
atomic diffusion, light propagation, and other transport
processes~\cite{walker1997spin,fink2005production,olsen2011optical,schrank2019novel,schrank2020fluid,kelley2020resolving,yan2026opticalpumpingalkalimetalvapor}.
Developing a solver for the fully coupled system poses several
challenges.
First, the mutual dependence of the atomic state and the light field
creates nonlinear feedback that must be resolved self-consistently.
Second, the numerical discretization must preserve the intrinsic
physical constraints on atomic populations and light attenuation.
Third, weak diffusion can produce thin polarization boundary layers
near depolarizing walls, requiring fine spatial
resolution~\cite{torrey1956bloch,bevington2024optical}.
Finally, the lack of general analytical reference solutions for the
fully coupled nonlinear system complicates systematic numerical
verification.

To address these challenges, we develop an efficient,
structure-preserving finite-difference solver for high-fidelity
simulation of the fully coupled 3D DQME--light system.
The DQME is discretized on a Cartesian mesh using second-order
central differences for atomic diffusion. The remaining terms,
including optical pumping (OP), spin destruction (SD),
F-destruction (FD), and spin exchange (SE), are evaluated locally
at each grid point.
The resulting semi-discrete system is advanced in pseudo-time
until a steady state is reached.
At each pseudo-time step, the light-propagation equation is solved
along each axial ray using an exponential integrator derived from
its integral representation, and the atomic populations are updated
using the explicit forward Euler method.
The population discretization preserves the density-matrix trace,
while the exponential integrator preserves the positivity and
monotonic attenuation of the light intensity.
We also derive a global restriction on the pseudo-time step that
ensures all population components remain within their physically
admissible bounds. Together, these properties establish the
structure-preserving character of the solver.
The scheme requires only local reaction evaluations, a compact
diffusion stencil, and independent axial light-propagation solves,
making it well suited to tensorization and parallel execution on
graphics processing units (GPUs).
Numerical results demonstrate that the tensorized GPU implementation
achieves a speedup of more than two orders of magnitude over a
single-core CPU implementation, enabling high-fidelity simulations
with millions of unknowns to be completed in minutes.

Systematic verification of the developed high-fidelity solver is
challenging because the nonlinear coupling between population
dynamics and light propagation makes general analytical reference
solutions difficult to obtain.
We therefore construct three complementary test cases based on an
analytical reduction, asymptotic limits, and an analytical invariant.
These cases progressively assess local atomic dynamics, atom--light
coupling in limiting regimes, and coupled behavior beyond those limits.
First, with diffusion neglected and the light field prescribed
analytically, we verify that the steady-state populations agree
with the spin-temperature distribution (STD) and exhibit the
expected independence from the alkali--alkali SE
rate~\cite{appelt1998theory}.
Second, we restore the atom--light coupling and test Beer--Lambert
attenuation and weak attenuation in the low- and high-pump-rate
limits, respectively~\cite{seltzer2008developments}.
Third, we reformulate the Lambert-$W$ relation for circularly
polarized light propagation as an analytical invariant to verify
the coupled response in a nonasymptotic
regime~\cite{seltzer2008developments}.
The numerical solutions reproduce the analytical populations,
recover the expected asymptotic behavior, and preserve the
invariant, supporting the correctness of the solver across
the tested regimes.

Although GPU acceleration substantially reduces the cost of individual
high-fidelity simulations, repeated simulations remain computationally
demanding for applications such as experimental design, parameter
exploration, sensitivity analysis, and optimization.
We therefore develop a reduced-order model (ROM) using the
reduced-basis method~\cite{quarteroni2015reduced,hesthaven2016certified}
to enable rapid computation of solutions at new parameter values.
A key challenge in model reduction is coupling atomic evolution in
pseudo-time with spatial light propagation. To address this challenge,
we eliminate the light field, which is uniquely determined by the
atomic populations and the prescribed incident-light condition, to
obtain an integro-differential formulation involving only the populations.
The ROM is then constructed using proper orthogonal decomposition
(POD) and Galerkin
projection~\cite{lucia2004reduced,iollo2000stability,benner2015survey}:
a POD basis is extracted from high-fidelity snapshots at sampled
parameter points, and the semi-discrete model is projected onto the
resulting reduced subspace.
However, direct evaluation of the nonlinear SE and OP terms still
depends on the dimension of the full-order model (FOM).
To remove this dependence, we derive an exact offline--online
decomposition of the SE term using quadratic expansion (QE) and
approximate the nonpolynomial, nonlocal OP term using the discrete
empirical interpolation method
(DEIM)~\cite{barrault2004empirical,chaturantabut2010nonlinear}.
These treatments make the cost of evaluating the reduced operators
independent of the full-order dimension.
Numerical tests demonstrate that the ROM accurately reproduces
high-fidelity solutions at parameter points outside the training set.
On a single CPU core, each online ROM evaluation takes only seconds,
with a speedup factor exceeding $7000$ relative to a single-core CPU
implementation of the FOM.

The remainder of the paper is organized as follows.
Section~\ref{sec:FOM} presents the governing equations and
high-fidelity discretization, analyzes the properties of the
resulting numerical scheme, and verifies the solver.
Section~\ref{sec:ROM} develops the POD--Galerkin reduced-order
model with nonlinear hyperreduction and evaluates its accuracy
and computational efficiency across multiple parameter configurations.
Finally, Section~\ref{sec:conclusion} summarizes the main findings
and outlines future research directions.

% !TeX root = main.tex

\section{High-fidelity simulation} 
\label{sec:FOM}

\subsection{Governing equations}
\label{sec:FOM_govering}
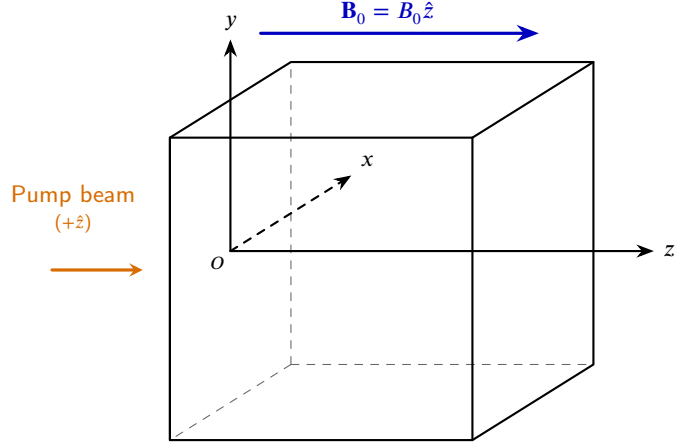
\begin{figure}
\centering
\begin{tikzpicture}[scale=3.5, >=Stealth, line join=round, line cap=round,
    % font=\footnotesize,
    % --- Keeping original perspective ---
    % Physical z (flow) -> TikZ x (horizontal right)
    % Physical y (height) -> TikZ y (vertical up)
    % Physical x (depth) -> TikZ z (diagonal in)
    x={(1cm,0cm)},
    y={(0cm,1cm)},
    z={(0.4cm,0.25cm)}
    ]

    % Left surface (z=0)
    \coordinate (L_Top_Front) at (0,  0.5, -0.5);
    \coordinate (L_Top_Back)  at (0,  0.5,  0.5);
    \coordinate (L_Bot_Back)  at (0, -0.5,  0.5);
    \coordinate (L_Bot_Front) at (0, -0.5, -0.5);
    \coordinate (Origin)      at (0, 0, 0);
    \node[anchor=north east, inner sep=2pt] at (Origin) {\scriptsize $O$};
    
    % Right surface (z=1)
    \coordinate (R_Top_Front) at (1,  0.5, -0.5);
    \coordinate (R_Top_Back)  at (1,  0.5,  0.5);
    \coordinate (R_Bot_Back)  at (1, -0.5,  0.5);
    \coordinate (R_Bot_Front) at (1, -0.5, -0.5);
    
    \draw[dashed, thin, black!60] (L_Bot_Back) -- (L_Top_Back);
    \draw[dashed, thin, black!60] (L_Bot_Back) -- (R_Bot_Back);
    \draw[dashed, thin, black!60] (L_Bot_Back) -- (L_Bot_Front);
    
    % Axes
    \draw[->, thick] (Origin) -- (1.4, 0, 0) node[right] {$z$}; % z axis (flow direction)
    \draw[->, thick] (Origin) -- (0, 0.7, 0) node[above] {$y$}; % y axis (height)
    \draw[->, thick, dashed] (Origin) -- (0, 0, 1.0) node[above right] {$x$}; % x axis (depth)
    
    % Incident Boundary Conditions (arrows for light)
    % \foreach \dy in {-0.5, -0.3, -0.10, 0.1} {
    %     \foreach \dx in {-0.2, 0.1, 0.4, 0.7} {
    %     \draw[->, very thick, orange!80!black] (-0.7, \dy, \dx) -- (-0.5, \dy, \dx);
    %     }
    % }
    % \node[anchor=east, align=right] at (-0.3, 0, 0) {\textbf{Incident B.C.}\\$\xi(z=0,\tau)=1$};

    %-----------------------------
    % Pump beam (longitudinal pumping)
    %-----------------------------
    % \foreach \yy in {-0.32,-0.1,0.12} {
    %     \foreach \zz in {-0.02,0.32} {
    %         \draw[->, very thick, orange!85!black]
    %             (-0.75,\yy,\zz) -- (-0.35,\yy,\zz);
    %     }
    % }
    \draw[->, very thick, orange!85!black, >=stealth] (-0.65, -0.1, 0.15) -- (-0.35, -0.1, 0.15);
    \node[orange!85!black, align=center] at (-0.52,0.14,0)
        {Pump beam\\[-1pt]\scriptsize ($+\hat z$)};

    %-----------------------------
    % Magnetic field B0 || z
    %-----------------------------
    \draw[->, very thick, blue!75!black]
        (0.1,0.72,0) -- (1.02,0.72,0);
    \node[blue!75!black, above] at (0.52,0.72,0)
        {$\mathbf{B}_0 = B_0\hat z$};

    %-----------------------------
    % Spin polarization inside cell
    %-----------------------------
    % \draw[->, very thick, green!50!black]
    %     (0.25,-0.15,0) -- (0.75,-0.15,0);
    % \node[green!50!black, below] at (0.5,-0.02,0)
    %     {$\mathbf{S}=s\hat z$};

    \draw[thick] (R_Top_Front) -- (L_Top_Front);
    \draw[thick] (R_Top_Front) -- (R_Bot_Front);
    \draw[thick] (R_Top_Front) -- (R_Top_Back);
    \draw[thick] (L_Bot_Front) -- (R_Bot_Front) -- (R_Bot_Back) -- (R_Top_Back) -- (L_Top_Back) -- (L_Top_Front) -- cycle;
    \draw[thick] (L_Top_Front) -- (L_Bot_Front);
    
    % \draw[thin, |-|] ($(L_Bot_Front)+(0,-0.05,0)$) -- ($(R_Bot_Front)+(0,-0.05,0)$) node[midway, below] {$L=1$};
    % \draw[thin, |-|] ($(R_Top_Front)+(0,0.05,0)$) -- ($(R_Top_Back)+(0,0.05,0)$) node[midway, below right] {$W=1$};
    % \draw[thin, |-|] ($(R_Bot_Front)+(0.05,0,0)$) -- ($(R_Top_Front)+(0.05,0,0)$) node[midway, right] {$H=1$};
    
    % \node[anchor=south] (BC_Label) at (0.5, 0.56, -0.25) {$\bm{\rho}_h|_{\partial\Omega} = \frac{1}{8}$};
    % \draw[->, thin] (BC_Label) -- (0.5, 0.5, -0.25);
\end{tikzpicture}
\caption{Schematic of SEOP under longitudinal pumping. A circularly polarized pump beam enters the vapor cell at $z=0$ and propagates along the $+z$ direction.}
\label{fig:configuration}
\end{figure}

We consider the spatially resolved spin-polarization dynamics of \(^{87}\mathrm{Rb}\) atoms in a buffer-gas-filled vapor cell under longitudinal optical pumping, as illustrated in Fig.~\ref{fig:configuration}. 
% The static magnetic field $\mathbf{B}_0$ is aligned with the $z$-axis, which is taken as the quantization axis.
The static magnetic field $\mathbf{B}_0$ defines the quantization axis, chosen here along the $z$-direction.
A circularly polarized pump beam propagates along the $z$-direction and transfers angular momentum to atomic electron spins, generating spin polarization along this axis.
The resulting spin dynamics are governed by optical pumping (OP), spin-exchange (SE) collisions that redistribute angular momentum among atoms, and spin-destruction (SD) collisions that relax the spin polarization~\cite{happer1972optical,appelt1998theory,happer2010optically}. In the presence of buffer gas, atomic diffusion further transports the internal spin states between different spatial locations. Under the longitudinal pumping configuration considered here, the evolution is restricted to the population subspace of the ground-state density matrix. For \(^{87}\mathrm{Rb}\), this subspace is represented by an eight-dimensional population vector \(\bm\rho(\mathbf r,t)\in\mathbb R^8\), whose components correspond to the populations of the ground-state hyperfine and Zeeman sublevels. The population basis and its ordering are specified in Appendix~\ref{appendix-sec:evolutionmatrix}.

% We consider the spatially resolved spin dynamics of $^{87}\mathrm{Rb}$ vapor under longitudinal optical pumping in the high-pressure (HP) regime, as illustrated in Fig.~\ref{fig:configuration}.
% The pump beam and the static magnetic field $\mathbf B_0$, when present, are taken to be collinear with the $z$ axis, which defines the quantization axis. 
% No transverse or radio-frequency magnetic driving is considered, so the atomic dynamics are restricted to the population subspace of the ground-state hyperfine manifolds.
% For $^{87}\mathrm{Rb}$, the local atomic state is therefore represented by the eight-dimensional population vector 
% \[
%     \bm{\rho} = [\rho_1,\ldots,\rho_8]^{\top} \in\mathbb R^8 ,
% \]
% corresponding to the eight Zeeman sublevels of the ground-state
% hyperfine manifolds.
% The detailed definition and ordering of these population components are given in Appendix~\ref{appendix-sec:evolutionmatrix}.

For numerical simulations, the governing equations are nondimensionalized using the characteristic length $L$ of the vapor cell and the spin-destruction relaxation time scale $\Gamma_{\rm SD}^{-1}$. 
Hereafter, $\mathbf r=(x,y,z)\in\Omega\subset\mathbb R^3$ denotes the dimensionless spatial coordinate and $\tau=\Gamma_{\rm SD}t$ the dimensionless time.
The resulting dimensionless parameters are defined as
\begin{equation*}
    \mathbb D = \frac{D}{L^2\Gamma_{\rm SD}}, \qquad
    \eta = \frac{\Gamma_{\rm SE}}{\Gamma_{\rm SD}}, \qquad
    f_D = \frac{\Gamma_{\rm FD}}{\Gamma_{\rm SD}}, \qquad
    \mathbb R_0 = \frac{R_0}{\Gamma_{\rm SD}},
\end{equation*}
where $D$, $\Gamma_{\rm SE}$, $\Gamma_{\rm FD}$, and $R_{0}$ denote the diffusion coefficient, spin-exchange rate, F-destruction rate, and incident optical pumping rate, respectively.
% The spatiotemporal evolution of the atomic population in the computational domain $\Omega \subset \mathbb{R}^3$ is described by the quantum master equation with diffusive term (DQME) in the population subspace

The population field $\bm\rho(\mathbf r,\tau)$ is governed by the diffusive quantum master equation (DQME)
\begin{equation}
    \frac{\partial\bm{\rho}}{\partial\tau} = 
        \mathbb{D}\nabla^{2} \bm{\rho}
        - \Bigl[ \mathbb{G}_{0}
        + \mathbb{G}_{\mathrm{OP}}(\xi)
        + \mathbb{G}_{\mathrm{SE}}(\bm{\rho}) \Bigr]
        \bm{\rho}.
\label{eq:DME}
\end{equation}

\added[id=xky]{Let $\mathbf S_z\in\mathbb R^8$ denote the vector formed by the diagonal matrix elements of the electron-spin operator $S_z$ in the population basis.}
The dimensionless electron-spin polarization along the pumping direction is defined by
\begin{equation}
    P(\mathbf r,\tau)\equiv 2\langle S_z\rangle
    =2\mathbf S_z^{\top}\bm\rho(\mathbf r,\tau).
\label{eq:electron_spin_polarization}
\end{equation}
For compactness, the local collision and pumping contributions are grouped as
\begin{subequations}
\begin{align}
    \mathbb G_{0} &= (1+\eta)A_{\mathrm{SD}} + f_DA_{\mathrm{FD}},
    \label{eq:define_G0}\\
    \mathbb G_{\mathrm{OP}}(\xi) &= \mathbb R_0\,\xi(\mathbf{r},\tau)A_{\mathrm{OP}},
    \label{eq:define_GOP}\\
    \mathbb G_{\mathrm{SE}}(\bm{\rho}) &= -\frac{\eta}{2}P A^{(z)}_{\mathrm{SE}}.
    \label{eq:define_GSE}
\end{align}    
\end{subequations}
\replaced[id=xky]{Here, \(\mathbb G_0\) collects the polarization-independent collision contributions, including spin destruction, the linear part of spin exchange, and F-destruction. The optical-pumping operator \(\mathbb G_{\rm OP}\) describes the redistribution of atomic populations induced by the absorption of circularly polarized light, with its rate determined by the local light intensity. The polarization-dependent operator \(\mathbb G_{\rm SE}\) accounts for the nonlinear contribution of spin-exchange collisions to the population dynamics.}{Here, $\mathbb G_0$ collects the polarization-independent collision terms, including spin destruction, the linear part of spin exchange, and F-destruction; $\mathbb G_{\rm OP}$ represents optical pumping; and $\mathbb G_{\rm SE}$ contains the polarization-dependent nonlinear spin-exchange contribution.}
The matrices $A_{\rm SD}$, $A_{\rm FD}$, $A_{\rm OP}$, and $A_{\rm SE}^{(z)}$ are fixed $8\times8$ \replaced[id=xky]{superoperators}{operators} acting on the population vector.
Their explicit forms are determined by the internal atomic structure and are provided in Appendix~\ref{appendix-sec:evolutionmatrix}.

We express the normalized pump intensity as
\[
    \xi(\mathbf{r},\tau)=\xi_z(\mathbf{r},\tau) \xi_{\perp}(x,y),
\]
where \(\xi_z\) describes the longitudinal propagation of the pump light through the atomic medium, and \(\xi_\perp\) is the prescribed transverse Gaussian profile
\[
    \xi_{\perp}(x,y)=\exp\!\left[-\frac{2(x^2+y^2)}{w^2}\right],
\]
with \(w\) denoting the dimensionless beam waist.
Although the light propagates along the $z$ direction, $\xi_z=\xi_z(\mathbf r,\tau)$ remains dependent on all three spatial coordinates, since the local absorption depends on the atomic
polarization $P(\mathbf r,\tau)$.
In the high-pressure (HP) regime, $\xi_z$ satisfies the light-propagation equation
\begin{equation}
    \frac{\partial \xi_{z}}{\partial z}
    = -\mathrm{OD}\xi_{z}(1-P),
    \label{eq:light}
\end{equation}
where $\mathrm{OD}$ denotes the optical depth.
For an unpolarized medium, $P=0$, and \eqref{eq:light} reduces to the standard Beer--Lambert law of exponential attenuation. 
% By their physical definitions, the dimensionless parameters satisfy $\mathbb D,\eta,f_D,\mathbb R_0,\mathrm{OD}\geq 0$, while the dimensionless beam waist satisfies $w>0$.

Equations~\eqref{eq:DME} and~\eqref{eq:light} form the coupled DQME--light system.
The numerical solution of this system presents several challenges.
First, the solution variables satisfy strict physical constraints.
The population vector represents probabilities and therefore must remain bounded $0 \leq \rho_i \leq 1$ and normalized $\sum_{i=1}^{8}\rho_i=1$.
Moreover, the longitudinal propagation factor satisfies $0<\xi_z \leq 1$ and 
\(\partial_z\xi_z\le0\), which indicates that $\xi_z$ remains bounded and monotonically decreases along the propagation direction.
Therefore, numerical discretizations should preserve these intrinsic physical structures.
Second, the spin-diffusion length can be much smaller than the cell dimension, giving rise to thin polarization boundary layers near depolarizing walls.
Adequate spatial resolution is therefore required to resolve the steep gradients associated with these layers.
Third, the nonlinear spin-exchange term and the population-dependent light attenuation make the governing equations strongly coupled and nonlinear.
For explicit time integration, the admissible time step is further constrained by both the diffusion operator and the local reaction rates.
Consequently, these features motivate the structure-preserving finite-difference discretization presented below.

\subsection{Numerical method}
\label{sec:FOM_numericalmethod}

The simulation setup is illustrated in Fig.~\ref{fig:configuration}. 
The coupled DQME--light system is solved on the dimensionless domain $\Omega = \left[-1/2,1/2\right] \times \left[-1/2,1/2\right] \times \left[0,1\right]$. 
At the incident plane ($z = 0$), the longitudinal propagation factor is prescribed as $\xi_z(x,y,0,\tau) = 1$.
For the atomic state, a Dirichlet boundary condition is imposed on the entire boundary of the computational domain. 
The boundary conditions are summarized as follows:
\begin{equation}
\begin{cases}
    \bm{\rho}(\mathbf{r}, \tau) = \tfrac{1}{8}\mathbf{1}, & \forall\mathbf{r} \in \partial\Omega, \\
    \xi_{z}(x,y,0, \tau) = 1, & \forall\tau > 0, 
\end{cases}
\label{eq:boundary-condition}
\end{equation}
where $\mathbf{1} \in \mathbb{R}^8$ denotes the all-ones vector.
\replaced[id=xky]
{This boundary condition models fully depolarizing walls, providing an idealized description of spin relaxation at the uncoated glass surfaces of conventional vapor cells}
{This boundary condition models fully depolarizing cell walls, which is commonly used as an idealized wall-relaxation model in high-pressure SEOP}~\cite{appelt1998theory,happer2010optically}.

Although the governing equations are written in time-dependent form, this work focuses on steady-state solutions. These solutions are obtained using pseudo-time iteration, in which $\tau$ serves as an artificial time variable and $\bm{\rho}$ is advanced until the steady-state condition $\partial\bm{\rho}/\partial\tau=0$ is satisfied. The iteration is initialized from a spatially uniform, unpolarized atomic state,
\begin{equation}
    \bm{\rho}(\mathbf{r},0) = \tfrac{1}{8}\mathbf{1}, \quad \forall\mathbf{r}\in \Omega.
    \label{eq:initial-condition}
\end{equation}
The corresponding initial light field is obtained by solving the light-propagation equation.

% In this work, we solve a steady-state problem, using a pseudo-time iteration in the sense that $\tau$ is a pseudo time and the iterations in the pseudo-time direction is performed until the pseudo-time derivative $\partial \bm{\rho}/\partial \tau$ vanishes. 
% The pseudo-time iteration is initialized from a spatially uniform, unpolarized atomic state, 
% \begin{equation}
%     \bm{\rho}(\mathbf{r},0) = \tfrac{1}{8}\mathbf{1}, \quad \forall\mathbf{r}\in \Omega,
% \label{eq:initial-condition}
% \end{equation}
% and the initial light-field can be obtained by solving the light-propagation equation.

The DQME is discretized on a uniform Cartesian grid using finite differences, while the light-propagation equation is discretized separately along the \(z\) direction.
At the pseudo-time step $n$, the numerical solution at the grid point $(i, j, k)$ is denoted as $\bm{\rho}^n_h(i, j, k) \in \mathbb{R}^{8}$, where $0 \le i \le N_x$, $0 \le j \le N_y$, and $0 \le k \le N_z$ are the indices in the $x$, $y$, and $z$ directions, respectively.
The numerical solution of the light-propagation equation is denoted as $\xi^n_z(i, j, k) \in \mathbb{R}$.
The detailed numerical discretization will be presented in the remainder of this subsection.

\subsubsection{Spatial discretization and pseudo-time integration}
\label{sec:FOM_spatialtime}

The Laplacian term on the right-hand side (RHS) of~\eqref{eq:DME} is approximated on the uniform Cartesian grid using a standard second-order central finite-difference scheme. At the grid point $\left(i,j,k\right)$, this approximation is written as
\begin{align*}
    \nabla^{2} \bm{\rho} \big|_{\mathbf{r}=(x_i,y_j,z_k)} \approx 2\Bigg[
    & \frac{\bm{\rho}^n_h(i+1,j,k)-\bm{\rho}^n_h(i,j,k)}
        {\Delta x_i (\Delta x_i+\Delta x_{i-1})}
      - \frac{\bm{\rho}^n_h(i,j,k)-\bm{\rho}^n_h(i-1,j,k)}
        {\Delta x_{i-1} (\Delta x_i+\Delta x_{i-1})} +\\
    & \frac{\bm{\rho}^n_h(i,j+1,k)-\bm{\rho}^n_h(i,j,k)}
        {\Delta y_j (\Delta y_j+\Delta y_{j-1})}
      - \frac{\bm{\rho}^n_h(i,j,k)-\bm{\rho}^n_h(i,j-1,k)}
        {\Delta y_{j-1} (\Delta y_j+\Delta y_{j-1})} +\\
    & \frac{\bm{\rho}^n_h(i,j,k+1)-\bm{\rho}^n_h(i,j,k)}
        {\Delta z_k (\Delta z_k+\Delta z_{k-1})}
      - \frac{\bm{\rho}^n_h(i,j,k)-\bm{\rho}^n_h(i,j,k-1)}
        {\Delta z_{k-1} (\Delta z_k+\Delta z_{k-1})}
    \Bigg],
\end{align*}
where $\Delta x_i=x_{i+1}-x_i$, $\Delta y_j=y_{j+1}-y_j$, and $\Delta z_k=z_{k+1}-z_k$ denote the grid spacings in the $x$-, $y$-, and $z$-directions, respectively. 
% In all computations and in the analysis below, these spacings are uniform and equal to $h$, so the expression reduces to the standard second-order central-difference stencil.
The remaining local source terms on the RHS of~\eqref{eq:DME} are evaluated pointwise as
\begin{align*}
    \left. \mathbb G_{0} \bm{\rho} \right|_{\mathbf{r}=(x_i,y_j,z_k)}
    &\approx
    \left(\left(1+\eta\right)A_{\mathrm{SD}} + f_DA_{\mathrm{FD}}\right)
    \bm{\rho}^n_h(i,j,k),
    \\[0.5em]
    \left. \mathbb G_{\mathrm{OP}}(\xi) \bm{\rho} \right|_{\mathbf{r}=(x_i,y_j,z_k)}
    &\approx
    \mathbb R_0\,\xi_{\perp}(i,j)\xi^n_z(i,j,k) A_{\mathrm{OP}}
    \bm{\rho}^n_h(i,j,k),
    \\[0.5em]
    \left. \mathbb G_{\mathrm{SE}}(\bm{\rho})\bm{\rho} \right|_{\mathbf{r}=(x_i,y_j,z_k)}
    &\approx
    -\eta \mathbf S_z^\top \bm{\rho}^n_h(i,j,k)
    A^{(z)}_{\mathrm{SE}} \bm{\rho}^n_h(i,j,k).
\end{align*}
Accordingly, the spatially discretized RHS at the grid point $(i,j,k)$ is defined as
\begin{equation}
    \mathcal{R}^n_{h}(i,j,k)
    =
    \left.
    \left(
    \mathbb{D} \nabla^{2} \bm{\rho}
    -\mathbb G_{0} \bm{\rho}
    -\mathbb G_{\mathrm{OP}}(\xi) \bm{\rho}
    -\mathbb G_{\mathrm{SE}}(\bm{\rho})\bm{\rho}
    \right)
    \right|_{\mathbf{r}=(x_i,y_j,z_k)} .
    \label{eq:rhs}
\end{equation}

The semi-discrete system is advanced in pseudo-time using an explicit forward Euler scheme,
\begin{equation*}
    \bm{\rho}^{n+1}_h(i,j,k) = \bm{\rho}^n_h(i,j,k) + \Delta\tau \mathcal{R}^n_{h}(i,j,k),
\end{equation*}
where $\Delta \tau$ is the uniform admissible time-step size determined from the analysis in Section~\ref{sec:FOM_stability}. This time-integration scheme yields a simple, fully local update at each grid point, which can be readily vectorized and efficiently implemented.
Before evaluating \eqref{eq:rhs}, the light field $\xi_z$ is updated by solving \eqref{eq:light} using the current atomic population field, as described in Section \ref{sec:FOM_light}.
The changes between successive iterates are measured by the Euclidean norms over all grid points and population components:
\begin{align*}
    \Delta\rho^{n+1} &:=\left\| \bm{\rho}_h^{n+1}-\bm{\rho}_h^n \right\|_2=
    \left[\sum_{i=0}^{N_x} \sum_{j=0}^{N_y} \sum_{k=0}^{N_z}\sum_{\ell=1}^{8}
        \left(\rho_{h,\ell}^{n+1}(i,j,k)-\rho_{h,\ell}^{n}(i,j,k)\right)^2
    \right]^{1/2},
    \\
    \Delta\xi_z^{n+1} &:=\left\| \bm{\xi}_z^{n+1} - \bm{\xi}_z^{n} \right\|_2=
    \left[\sum_{i=0}^{N_x} \sum_{j=0}^{N_y} \sum_{k=0}^{N_z}
        \left(\xi_z^{n+1}(i,j,k)-\xi_z^{n}(i,j,k)\right)^2
    \right]^{1/2},
\end{align*}
where $\rho_{h,\ell}^{n}(i,j,k)$ denotes the $\ell$th component of $\bm\rho_h^{n}(i,j,k)$.
The iteration is terminated when
\[
    \Delta\rho^{n+1}<\varepsilon
    \qquad\text{and}\qquad
    \Delta\xi_z^{n+1}<\varepsilon,
\]
with $\varepsilon=10^{-8}$.
To reduce computational cost, convergence is checked every $M=200$ pseudo-time steps.

\subsubsection{Axial discretization of the light-propagation equation}
\label{sec:FOM_light}

For each fixed transverse grid point $(x_i,y_j)$, \eqref{eq:light} reduces to a one-dimensional light-propagation equation along the \(z\) direction.
The light field and the atomic state are nonlinearly coupled in the full system.
For a fixed iterate of $\bm\rho_h$, however, the light-propagation equation becomes a scalar linear ODE with a known spatially varying coefficient.
It can therefore be advanced independently along each axial ray.

For this one-way propagation problem, only the incident condition \(\xi_z=1\) at \(z=0\) is prescribed.
The value of \(\xi_z\) at \(z=1\) is determined by axial marching and is not imposed as an additional boundary condition.
To simplify the notation, we omit the pseudo-time superscript $n$ and the fixed transverse indices $(i,j)$ in the remainder of this subsection, and use the shorthand notation
\[
    \xi_z(k) := \xi_z(i,j,k),
    \qquad
    \bm{\rho}_h(k) := \bm{\rho}_h(i,j,k).
\]

For a fixed population field, ~\eqref{eq:light} is a scalar linear ODE in \(z\). 
We therefore construct an exponential-integrator (EI) discretization from its integral representation.
Integrating its logarithmic form from the incident plane to an arbitrary axial location $z$ gives
\begin{equation}
    \xi_z(z) = \xi_z(0)
    \exp\left(
        -\mathrm{OD}\,z
        +2\mathrm{OD}
        \int_0^z
        \mathbf S_z^\top\bm\rho_h(s)\,\mathrm ds
    \right).
    \label{eq:light_integral}
\end{equation}
Together with $\xi_z(0)=1$, this integral representation uniquely determines $\xi_z$ along each axial ray.
Taking the ratio of the representations at two consecutive nodes, $z_k$ and $z_{k+1}$, yields an exact one-step propagation relation.
The unpolarized Beer--Lambert factor \(e^{-\mathrm{OD}\Delta z_k}\) is represented exactly, and the only approximation arises from the quadrature of the polarization-dependent integral.
Applying the trapezoidal rule gives

% Taking the ratio of Eq.~\eqref{eq:light_integral} at two consecutive
% nodes $z_k$ and $z_{k+1}$ gives the exact one-step relation
% \begin{equation}
%     \xi_z(k+1)
%     =
%     \xi_z(k)
%     \exp\left[
%         -\mathrm{OD}\Delta z_k
%         +2\mathrm{OD}
%         \int_{z_k}^{z_{k+1}}
%         \mathbf S_z^{\top}\bm\rho_h(s)\,\mathrm ds
%     \right],
%     \label{eq:light_exact_step}
% \end{equation}
% where $\Delta z_k=z_{k+1}-z_k$.
% Thus, the unpolarized Beer--Lambert factor
% $e^{-\mathrm{OD}\Delta z_k}$ is represented exactly, and the only
% approximation arises from the quadrature of the polarization-dependent
% integral.

\begin{equation}
    \int_{z_k}^{z_{k+1}}
    \mathbf S_z^\top\bm\rho_h(s)\,\mathrm ds
    =
    \frac{\Delta z_k}{2}
    \mathbf S_z^\top
    \left[
        \bm\rho_h(k)+\bm\rho_h(k+1)
    \right]
    +\mathcal O(\Delta z_k^3).
\label{eq:light_integral_trap}
\end{equation}
Substituting \eqref{eq:light_integral_trap} into the one-step propagation relation gives
\begin{equation}
    \xi_z(k+1)
    =
    \xi_z(k)e^{-\mathrm{OD}\Delta z_k}
    \exp\left[
        \mathrm{OD}\Delta z_k\,
        \mathbf S_z^\top
        \left(
            \bm\rho_h(k)+\bm\rho_h(k+1)
        \right)
    \right].
\label{eq:light_integration_trap}
\end{equation}
Since $\bm\rho_h(k)$ and $\bm\rho_h(k+1)$ are available from the current iterate, \eqref{eq:light_integration_trap} provides an explicit step-by-step update from $z_k$ to $z_{k+1}$.

The resulting scheme has three properties that are particularly relevant to the present problem. 
First, provided that $\mathbf S_z^\top\bm\rho_h$ is sufficiently smooth along each axial ray, the trapezoidal approximation has a local error of $\mathcal O(\Delta z_k^3)$, resulting in a globally second-order-accurate axial discretization. 
Second, the update preserves both positivity and monotone attenuation.
Indeed, ~\eqref{eq:light_integration_trap} gives
\[
    \frac{\xi_z(k+1)}{\xi_z(k)}
    =
    \exp\left[
        -\mathrm{OD}\Delta z_k
        \left(
            1-
            \mathbf S_z^{\top}\bm\rho_h(k)
            -
            \mathbf S_z^{\top}\bm\rho_h(k+1)
        \right)
    \right].
\]
For physically admissible population states, $|\mathbf S_z^{\top}\bm\rho_h|\leq 1/2$, so the exponent is non-positive and therefore
\[
    0<\xi_z(k+1)\leq\xi_z(k)
\]
Third, in the Beer--Lambert limit \(\mathbf S_z^\top\bm\rho_h=0\), the scheme reduces to \(\xi_z(k+1) = \xi_z(k)e^{-\mathrm{OD}\Delta z_k}\), which is the exact one-step solution of the limiting Beer--Lambert equation.
Thus, the numerical approximation enters only through the quadrature of the polarization-dependent contribution, whereas the Beer--Lambert factor \(e^{-\mathrm{OD}\Delta z_k}\) is represented exactly.

The axial light solve can be carried out efficiently as, for a fixed population field, the one-dimensional propagation problems associated with different transverse grid points are independent.
The axial rays can therefore be processed in parallel, while each ray requires only a single sequential sweep from $z=0$ to $z=1$.
Moreover, \eqref{eq:light_integration_trap} avoids the solution of a nonlinear algebraic system.
For a mesh with $N_x N_y$ axial rays and $N_z$ grid points per ray, the total computational complexity is $\mathcal O(N_x N_y N_z)$.
% For a mesh with $N_xN_y$ axial rays and $N_z$ along each ray, the resulting computational complexity is $\mathcal O(N_xN_yN_z)$.

The EI scheme in \eqref{eq:light_integration_trap} is used in all high-fidelity simulations that require numerical solution of the light-propagation equation, including verification Cases~2 and~3 in Section~\ref{sec:FOM_verification} and snapshot generation in Section~\ref{sec:ROM}.
Case~1 instead prescribes the light field analytically.
The complete high-fidelity solver is summarized in Algorithms~\ref{alg:LightPropagationSolver} and ~\ref{alg:FOM}.

\subsection{Analysis}
\label{sec:FOM_analysis}

In this subsection, we analyze the structural properties of the finite-difference scheme presented in Section~\ref{sec:FOM_numericalmethod}. We first establish the sign and conservation properties of the local reaction operators. Using these properties, we show that the fully discrete scheme exactly preserves population normalization, or equivalently, the density-matrix trace in the present population representation. We then derive a sufficient pseudo-time-step restriction under which the forward-Euler update preserves population nonnegativity. Together with trace preservation, this ensures that all population components remain within their physically admissible bounds. 

% In this subsection, we analyze the discrete structural properties of the finite-difference scheme presented in Section \ref{sec:FOM_numericalmethod}. 
% We first establish the sign and conservation properties of the local reaction operators that are required in the subsequent analysis.
% Using these properties, we show that the fully discrete scheme preserves population normalization, equivalently the density-matrix trace in the present population representation, exactly.
% We then derive a sufficient pseudo-time-step restriction under which the forward-Euler update preserves nonnegative populations.
% Combined with the trace preservation, this yields preservation of the physical population bounds.
% Finally, under the same time-step restriction, a frozen-coefficient analysis establishes stability of the homogeneous perturbation equation in the induced $1$-norm.

The analysis is presented for a uniform Cartesian mesh with $\Delta x_i=\Delta y_j=\Delta z_k=h$.
Extending the analysis to nonuniform Cartesian meshes is straightforward but requires a corresponding diffusion stencil and a mesh-dependent time-step restriction. The details are omitted for brevity.
% Extension to nonuniform Cartesian meshes would require the corresponding nonuniform diffusion stencil, a mesh-dependent
% time-step restriction, and a modified stability argument, and is not considered here.

% Preservation of the density-matrix trace is a well-established requirement in numerical discretizations of quantum master equations~\cite{alund2021trace}.
% In the present population representation, this requirement is expressed as preservation of the normalization $\mathbf 1^\top\bm\rho=1$.
% % In the present population-space DQME, the relevant discrete structure also includes nonnegativity of the population components and the associated physical bounds.
% We therefore analyze trace preservation, bound preservation, and the pseudo-time-step restriction required for the explicit solver.

\subsubsection{Structural properties of the superoperators}
\label{sec:superoperators_structure}

Before analyzing trace preservation and the time-step restriction, we establish the structural properties of the local reaction matrix
\begin{equation}
    B(P,\xi) = (1+\eta)A_{\mathrm{SD}} + f_D A_{\mathrm{FD}}
     + \mathbb R_0 \xi A_{\mathrm{OP}} - \frac{\eta P}{2} A_{\mathrm{SE}}^{(z)}.
\label{eq:frozen_reaction_matrix}
\end{equation}
At pseudo-time step $n$ and grid point $(i,j,k)$, we define
\begin{align*}
    P^n(i,j,k) &:=2\mathbf S_z^{\top}\bm\rho_h^n(i,j,k),\\
    \xi^n(i,j,k) &:=\xi_\perp(i,j)\xi_z^n(i,j,k),\\
    B^n(i,j,k) &:=B\bigl(P^n(i,j,k),\xi^n(i,j,k)\bigr)
    \in\mathbb R^{8\times8}.
\end{align*}
The local reaction coefficients are held fixed during each forward Euler update.
Throughout this analysis, $i,j,k$ denote spatial indices, while $\mathscr{q},\ell\in\{1,\ldots,8\}$ denote population-component indices.
The zero-column-sum identities given in \eqref{eq:appendix_column_sums} of Appendix~\ref{appendix-sec:evolutionmatrix} imply
\begin{equation}
    \mathbf1^\top B(P,\xi)=\mathbf0^\top,
\label{eq:B_column_conservation}
\end{equation}
showing that the local reaction matrix satisfies column conservation independently of the values of $P$ and $\xi$.

A population vector $\bm\rho$ is physically admissible if it satisfies
\begin{equation}
    0 \le \rho_\ell \le 1, \qquad \mathbf 1^\top\bm\rho= \sum^8_{\ell=1} \rho_\ell=1.
\label{eq:admissible_population}
\end{equation}
Under this condition, $P/2=\sum_{\ell=1}^{8}s_\ell\rho_\ell$ is a convex combination of the entries $s_\ell$ of $\mathbf S_z$.
As shown in Appendix~\ref{appendix-sec:evolutionmatrix}, these entries lie in the interval $[-1/2,1/2]$, hence
\begin{equation}
    -1\leq P\leq1.
\label{eq:P_physical_bound}
\end{equation}
The initial and boundary data used in this work satisfy \eqref{eq:admissible_population}. 
The analyses in Sections \ref{sec:HF_trace} and \ref{sec:FOM_stability} show that the fully discrete update preserves these physical conditions, provided that the derived pseudo-time-step restriction is satisfied.

For a physically admissible population field, the EI
update~\eqref{eq:light_integration_trap}, together with the
incident condition $\xi_z^n(i,j,0)=1$, yields
\begin{equation*}
    0<\xi_z^n(i,j,k+1)\leq\xi_z^n(i,j,k)\leq1,
    \qquad 0\leq k\leq N_z-1,
\end{equation*}
for every axial ray $(i,j)$.
Thus, the discrete longitudinal intensity is nonincreasing along the $z$-direction, which is consistent with the light-propagation dynamics.
% Between successive light recomputations, Algorithm~\ref{alg:FOM} retains the previously stored light field, so the same bound is inherited during every frozen-light pseudo-time step. 
% Since $\bm\xi_z^0=\mathbf1$, the stored numerical field satisfies $0<\xi_z^n\leq1$ whenever population admissibility is preserved. 
The transverse Gaussian profile satisfies $0<\xi_\perp(x,y)\leq1$; therefore, the total normalized intensity $\xi=\xi_\perp\xi_z$ also obeys $0<\xi\leq1$.
To include the limiting case of complete attenuation, the estimates below are taken over the closed parameter set
\begin{equation}
    (P,\xi) \in [-1,1]\times[0,1].
\label{eq:physical_parameter_domain}
\end{equation}

% Independently of the admissibility bounds on $P$ and $\xi$, the zero-column-sum identities in \eqref{eq:appendix_column_sums} imply
% \begin{equation}
%     \mathbf1^\top B(P,\xi)=\mathbf0^\top
%     \qquad\text{for all }P,\xi.
% \label{eq:B_column_conservation}
% \end{equation}
% This conservation identity is structural and does not require \eqref{eq:physical_parameter_domain}.

\begin{lemma}
\label{lem:frozen_reaction_structure}
Assume that $\eta,f_D,\mathbb R_0\geq0$.
Then, for any $(P,\xi)$ satisfying \eqref{eq:physical_parameter_domain}, the frozen local reaction matrix $B(P,\xi)$ satisfies
\begin{equation}
    B(P,\xi)_{\mathscr{q}\ell}\leq 0,
    \quad \mathscr{q}\neq\ell,
    \qquad B(P,\xi)_{\ell\ell}\geq0.
\label{eq:B_sign_structure}
\end{equation}
\end{lemma}

\begin{proof}
Equations~\eqref{eq:appendix_offdiagonal_signs} and \eqref{eq:appendix_SD_SE_bound} in Appendix~\ref{appendix-sec:evolutionmatrix} give the required off-diagonal sign and magnitude relations.
For any off-diagonal entry of $B(P,\xi)$, the nonnegativity of $f_D$, $\mathbb R_0$, and $\xi$, together with the preceding sign properties, gives
\begin{align}
    B(P,\xi)_{\mathscr{q}\ell} &= (1+\eta)(A_{\mathrm{SD}})_{\mathscr{q}\ell} + f_D(A_{\mathrm{FD}})_{\mathscr{q}\ell}
    +\mathbb R_0\xi(A_{\mathrm{OP}})_{\mathscr{q}\ell} - \frac{\eta P}{2}(A_{\mathrm{SE}}^{(z)})_{\mathscr{q}\ell}
    \notag\\
    &\leq (1+\eta)(A_{\mathrm{SD}})_{\mathscr{q}\ell} + \frac{\eta|P|}{2} \left|(A_{\mathrm{SE}}^{(z)})_{\mathscr{q}\ell}\right|
    \notag\\
    &\leq \left(1+\eta-\eta|P|\right)(A_{\mathrm{SD}})_{\mathscr{q}\ell}.
\label{eq:B_offdiag_bound}
\end{align}

Using $|P|\leq1$, we have $1+\eta-\eta|P|\geq1$.
Since $(A_{\mathrm{SD}})_{\mathscr{q}\ell}\leq0$, the right-hand side of \eqref{eq:B_offdiag_bound} is nonpositive, and therefore $B_{\mathscr{q}\ell}(P,\xi)\leq0$ for all $\mathscr{q}\neq \ell$.

Using the unconditional column-conservation identity \eqref{eq:B_column_conservation}, every column of $B(P,\xi)$ sums to zero. 
Combining this identity with the nonpositivity of the off-diagonal entries gives
\[
    B(P,\xi)_{\ell\ell} = -\sum_{\substack{\mathscr{q}=1\\\mathscr{q}\neq\ell}}^{8}B(P,\xi)_{\mathscr{q}\ell} \geq 0,
\]
which completes the proof.
\end{proof}

\subsubsection{Trace-preserving property}
\label{sec:HF_trace}

A fundamental physical constraint of the quantum master equation is the conservation of total probability.
In the present population representation, the density-matrix trace is represented by the population sum, and the conservation condition is expressed as
\begin{equation*}
    T(\mathbf{r},\tau):=\mathbf 1^\top\bm\rho(\mathbf{r},\tau),
    \qquad
    T_h^n(i,j,k):=\mathbf 1^\top\bm\rho_h^n(i,j,k).
\end{equation*}
The physical normalization condition is $T=1$.
The following analysis shows that the high-fidelity solver preserves this conservation exactly.

At grid point $(i,j,k)$, the forward Euler update can be written as
\begin{equation}
    \bm{\rho}^{n+1}_h(i,j,k) = \bm{\rho}^n_h(i,j,k) + \Delta\tau \left[\mathbb D \left(\Delta_h \bm \rho_h^n(i,j,k)\right) - B^n(i,j,k)\bm\rho_h^n(i,j,k) \right],
\label{eq:forward-euler-update-B-matrix}
\end{equation}
where $\Delta_h$ denotes the discrete Laplacian based on central differences.
The column-conservation identity in \eqref{eq:B_column_conservation}, $\mathbf1^\top B(P,\xi)=\mathbf0^\top$, holds for all $P$ and $\xi$, ensuring that the reaction terms do not contribute to changes in the population sum.
% The column-conservation identity $\mathbf1^\top B(P,\xi)=\mathbf0^\top$ in \eqref{eq:B_column_conservation} ensures that the reaction terms do not contribute to changes in the population sum.
Left-multiplying \eqref{eq:forward-euler-update-B-matrix} by $\mathbf1^\top$ therefore yields
\begin{equation}
    T_h^{n+1}(i,j,k) = T_h^n(i,j,k) + \Delta\tau\,\mathbb D(\Delta_hT_h^n)_{i,j,k}.
\label{eq:discrete_trace_equation}
\end{equation}

\begin{lemma}
\label{lem:trace_preservation}
Suppose that the initial and Dirichlet boundary conditions satisfy
\(\bm\rho_h^0(i,j,k) = \frac18\mathbf1\), and 
\(\left.\bm\rho_h^n\right|_{\partial\Omega} = \frac18\mathbf1\).
Then the fully discrete scheme satisfies \(T_h^n(i,j,k)=1\) for every pseudo-time step $n$ and every spatial grid point.
\end{lemma}

\begin{proof}
At $n=0$, \(T_h^0(i,j,k) = \mathbf1^\top\left(\frac18\mathbf1\right) = 1\).
Similarly, the trace on the Dirichlet boundary satisfies \(\left.T_h^n\right|_{\partial\Omega}=1\).
Assume that \(T_h^n(i,j,k)=1\) at every grid point. 
Then $T_h^n$ is constant throughout the computational domain, and hence \(\Delta_hT_h^n=0\).
It follows from \eqref{eq:discrete_trace_equation} that \(T_h^{n+1}(i,j,k) = T_h^n(i,j,k) = 1\).
The result therefore follows by induction.
\end{proof}

% Trace preservation itself is independent of the pseudo-time-step size. In exact arithmetic, it holds for any $\Delta\tau$, provided that the initial and boundary traces are equal to one.
% For completeness, the corresponding continuous problem has the same trace-preservation property. Applying $\mathbf 1^\top$ to \eqref{eq:DME} yields a scalar diffusion equation for $T$. Letting $u=T-1$, we obtain a homogeneous diffusion equation with zero initial and boundary data. By the parabolic maximum principle, $u\equiv 0$, and hence
% $T(\mathbf r,\tau)\equiv 1$, $\forall \ \tau\ge 0$.
% Preservation of nonnegative populations and frozen-coefficient stability in the $1$-norm additionally require the time-step restriction derived next.

Trace preservation is independent of the pseudo-time-step size and holds in exact arithmetic for any $\Delta\tau$ under the prescribed initial and Dirichlet boundary conditions. The continuous model has the same property. Applying $\mathbf1^\top$ to \eqref{eq:DME} and using \eqref{eq:B_column_conservation} yields
\[
\frac{\partial T}{\partial\tau}=\mathbb D\nabla^2 T.
\]
Since $T\equiv1$ satisfies this equation and the prescribed initial and boundary conditions, uniqueness of the diffusion problem implies $T(\mathbf r,\tau)\equiv1$ for all $\tau\geq0$.

\subsubsection{Bound preservation}
\label{sec:FOM_stability}

We derive a sufficient time-step restriction under which all coefficients in the pointwise forward Euler update are nonnegative.
Combined with trace preservation established in Section~\ref{sec:HF_trace} and the EI light update, this ensures physical admissibility of the numerical solution.

% We then use the zero-column-sum structure of the reaction operators and the conservative diffusion stencil to show that the amplification matrix is \mathscr{q}、el l, yielding $\|G^n\|_1\leq1$.
% Finally, boundedness of the corresponding inhomogeneous update over finite pseudo-time intervals is established in Appendix~\ref{app:inhomogeneous_stability}.

Using the local reaction matrix defined in Section~\ref{sec:superoperators_structure}, the update is
% \begin{equation}
%     \frac{\bm\rho_h^{n+1}-\bm\rho_h^n}{\Delta\tau}
%     = \mathbb D \Delta_h \bm\rho_h^n - B^n\bm\rho_h^n.
% \label{eq:FD_amplification_matrix}
% \end{equation}
% \begin{equation}
%     \frac{\bm\rho_h^{n+1}(i,j,k)-\bm\rho_h^n(i,j,k)}{\Delta\tau}
%     =\mathbb D(\Delta_h\bm\rho_h^n)(i,j,k)
%      -B^n(i,j,k)\bm\rho_h^n(i,j,k).
%     \label{eq:FD_amplification_matrix}
% \end{equation}
% As the physical reaction operator \eqref{eq:frozen_reaction_matrix} depends on the current population vector \(\rho\) and light intensity \(\xi\), its coefficients are frozen at each pseudo-time step. The resulting forward Euler update can be written as
\begin{equation}
    \frac{\bm\rho_h^{n+1}-\bm\rho_h^n}{\Delta\tau}
    = \mathbb D \Delta_h \bm\rho_h^n - B^n\bm\rho_h^n + \bm b.
\label{eq:FD_amplification_matrix}
\end{equation}
Here, $\bm\rho_h^n$ denotes the vector containing the unknowns at all interior grid points, while $\bm b$ contains the contributions from the prescribed Dirichlet boundary values. 
At boundary nodes, the solution is set directly to the values specified in \eqref{eq:boundary-condition}. 
Since these boundary values are nonnegative, $\bm b\geq0$ componentwise.
Define $M_h^n=\mathbb D\Delta_h-B^n$. 
The difference scheme can then be written as
\begin{equation}
    \bm\rho_h^{n+1}
    =G^n\bm\rho_h^n+\Delta\tau\bm b,
    \label{eq:difference_scheme}
\end{equation}
where $G^n:=I+\Delta\tau M_h^n$ is the amplification matrix.
On a three-dimensional uniform mesh, the action of $M_h^n$ at an interior grid point $(i,j,k)$ is
\begin{align}
    \left(M_h^n\bm\rho_h^n\right)_{i,j,k}
    ={}&\frac{\mathbb D}{h^2}\bigl[
        \bm\rho_h^n(i+1,j,k)+\bm\rho_h^n(i-1,j,k)
        +\bm\rho_h^n(i,j+1,k)+\bm\rho_h^n(i,j-1,k)
        \notag\\
    &\qquad
        +\bm\rho_h^n(i,j,k+1)+\bm\rho_h^n(i,j,k-1)
        -6\bm\rho_h^n(i,j,k)\bigr]
        -B^n(i,j,k)\bm\rho_h^n(i,j,k).
\label{eq:frozen_operator_pointwise}
\end{align}
For a stencil that intersects the Dirichlet boundary, the corresponding boundary-neighbor values in~\eqref{eq:frozen_operator_pointwise} are moved into the source vector \(\bm b\) and therefore do not appear as off-diagonal couplings in the interior matrix \(M_h^n\). 
This operation leaves the diagonal diffusion coefficient unchanged and introduces no negative off-diagonal entries.
% ; its effect on the column sums is examined later in this subsection.

The diffusion operator produces the nonnegative coupling coefficient
$\mathbb D/h^2$ between neighboring grid points. In addition, Lemma~\ref{lem:frozen_reaction_structure} gives
\[
    B^n(i,j,k)_{\mathscr{q}\ell}\leq0,\qquad \mathscr{q}\neq \ell.
\]
% implying that the reaction contribution $-B^n(i,j,k)$ has nonnegative off-diagonal entries. 
Consequently, every off-diagonal entry of the global frozen matrix satisfies
\begin{equation}
    (M_h^n)_{\mathscr{q}\ell}\geq0,\qquad \mathscr{q}\neq \ell.
\label{eq:M_offdiagonal}
\end{equation}
The diagonal entry associated with the grid point $(i,j,k)$ is
\begin{equation}
    M_h^n(i,j,k)_{\ell\ell}
    =-\frac{6\mathbb D}{h^2}-B^n(i,j,k)_{\ell\ell}\leq0,
    \qquad \ell=1,\ldots,8.
\label{eq:M_diagonal}
\end{equation}

For the frozen homogeneous update we require the physically motivated condition \(G^n\geq0\). Together with the nonnegative boundary source $\bm b$, this condition ensures that the full update maps a nonnegative interior state to a nonnegative interior state.
From~\eqref{eq:M_offdiagonal} it follows that all off-diagonal entries of \(G^n\) are automatically nonnegative:
\[
    G_{\mathscr{q}\ell}^n=\Delta\tau\,(M_h^n)_{\mathscr{q}\ell}\geq0,\qquad \mathscr{q}\neq \ell.
\]
It therefore remains only to ensure nonnegativity of the diagonal entries.
Using~\eqref{eq:M_diagonal}, the diagonal entry at \((i,j,k)\) is
\begin{equation}
    G^n(i,j,k)_{\ell\ell}
    =1-\Delta\tau\left[\frac{6\mathbb D}{h^2}+B^n(i,j,k)_{\ell\ell}\right].
\label{eq:G_diagonal}
\end{equation}
% Since all off-diagonal entries of $G^n$ are already nonnegative, entrywise nonnegativity of the frozen amplification matrix is equivalent to nonnegativity of all its diagonal entries. 
Hence, for a fixed admissible frozen coefficient field,
\begin{equation}
    G^n\geq0
    \quad\Longleftrightarrow\quad
    \Delta\tau\leq
    \left[
        \frac{6\mathbb D}{h^2}
        +\max_{\substack{i,j,k\\1\leq \ell\leq8}}B^n(i,j,k)_{\ell\ell}
    \right]^{-1}.
\label{eq:local_timestep_condition}
\end{equation}

To obtain a conservative global pseudo-time step that is independent of the spatial location and pseudo-time iteration, we construct a uniform upper bound for the diagonal entries of the frozen reaction matrix.
From the explicit matrices given in ~Appendix~\ref{appendix-sec:evolutionmatrix}, the eight diagonal entries of $B(P,\xi)$ are
\begin{equation}
    \begin{aligned}
    B_{11}(P,\xi) &= \frac{1+\eta}{2}+2f_D-\frac{\eta P}{2},
    \\
    B_{22}(P,\xi) &= \frac{11(1+\eta)}{16}+5f_D+\frac{7\mathbb R_0\xi}{16}-\frac{\eta P}{4},
    \\
    B_{33}(P,\xi) &= \frac{3(1+\eta)}{4}+6f_D+\frac{3\mathbb R_0\xi}{4},
    \\
    B_{44}(P,\xi) &= \frac{11(1+\eta)}{16}+5f_D+\frac{15\mathbb R_0\xi}{16}+\frac{\eta P}{4},
    \\
    B_{55}(P,\xi) &= \frac{1+\eta}{2}+2f_D+\mathbb R_0\xi+\frac{\eta P}{2},
    \\
    B_{66}(P,\xi) &= \frac{11(1+\eta)}{16}+f_D+\frac{7\mathbb R_0\xi}{16}-\frac{\eta P}{4},
    \\
    B_{77}(P,\xi) &= \frac{3(1+\eta)}{4}+2f_D+\frac{3\mathbb R_0\xi}{4},
    \\
    B_{88}(P,\xi) &= \frac{11(1+\eta)}{16}+f_D+\frac{15\mathbb R_0\xi}{16}+\frac{\eta P}{4}.
\end{aligned}
\label{eq:B_diagonal_entries}
\end{equation}
Using $0\leq\xi\leq1$ and $|P|\le1$, we obtain
\begin{equation}
    B^n(i,j,k)_{\ell\ell}\leq\beta_{\max}
    :=\frac{3(1+\eta)}{4}+6f_D
    +\mathbb R_0
    +\frac{\eta}{2}.
\label{eq:beta_max}
\end{equation}
Therefore, a global pseudo-time step that guarantees \(G^n\geq0\) is
\begin{equation}
    \Delta\tau = c
    \left(\frac{6\mathbb D}{h^2}+\beta_{\max}\right)^{-1},
    \qquad 0<c\leq1.
\label{eq:global_timestep}
\end{equation}
The choice $c=1$ gives the largest step size permitted by the sufficient positivity condition.
For all production computations, we use $c=0.8$, which satisfies this condition while providing an additional safety margin for floating-point errors and implementation effects.

We complete the admissibility argument by joint induction on the population and light fields.
Initially, $\bm\rho_h^0=\frac18\mathbf1$ and $0<\xi_z^0\leq1$ at every grid point.
Suppose that, at step $n$, the populations are nonnegative and normalized, and $0<\xi_z^n\leq1$.
Then \eqref{eq:P_physical_bound} and $0<\xi_\perp\leq1$ imply
\[
    |P^n(i,j,k)|\leq1,
    \qquad
    0<\xi^n(i,j,k)\leq1,
\]
so the structural properties of $B^n$ established in Lemma~\ref{lem:frozen_reaction_structure} apply.
Under the global time-step restriction in \eqref{eq:global_timestep}, $G^n\geq0$.
Together with $\bm b\geq0$, this ensures that the interior update satisfies $\bm\rho_h^{n+1}\geq0$.
The prescribed Dirichlet boundary values are imposed directly and remain nonnegative.
Combining nonnegativity with trace preservation from Lemma~\ref{lem:trace_preservation} gives, at every grid point,
\begin{equation}
    0\leq\rho_{h,\ell}^{n+1}(i,j,k)\leq1,
    \qquad
    \sum_{\ell=1}^8\rho_{h,\ell}^{n+1}(i,j,k)=1,
    \qquad \ell=1,\ldots,8.
    \label{eq:population_bound_preservation}
\end{equation}
Applying the EI scheme to the admissible population field $\bm\rho_h^{n+1}$ yields a light propagation factor $\xi_z^{n+1}$ that satisfies $0<\xi_z^{n+1}\leq1$ and $\partial \xi_z^{n+1}/\partial z \le 0 $. This completes the induction, establishing the physical admissibility of both fields at every pseudo-time step. Thus, the finite-difference solver is structure-preserving.

\subsection{GPU implementation}
\label{sec:HF_gpu}

The high-fidelity simulation of optical pumping is computationally expensive, as a large number of degrees of freedom (DoFs) are required to resolve three-dimensional spatial inhomogeneity and, where required, increasingly thin diffusion layers at small diffusion coefficients. 
Therefore, the high-fidelity solver presented in Section \ref{sec:FOM_numericalmethod} must be carefully implemented for computational efficiency.

In the high-fidelity solver, the spatial discretization employs a central finite-difference approximation of the diffusion operator on a local stencil, while all remaining source terms are evaluated pointwise. Time integration is performed using an explicit forward Euler scheme. The light propagation equation is solved via a stepwise update along the axial direction. Overall, the resulting algorithm is compact in stencil width and readily amenable to vectorization, making it well-suited for implementation on graphics processing units (GPUs).

The high-fidelity solver is implemented on GPUs using PyTorch \cite{paszke2019pytorch}. The explicit forward Euler update is inherently data-parallel, as it involves only nearest-neighbor stencil operations and local matrix--vector products. This structure is leveraged by reformulating the update as a sequence of tensor contractions. The $8 \times 8$ \replaced[id=xky]{superoperaters}{operators} are applied at all grid points via batched Einstein summation (\texttt{torch.einsum}), eliminating explicit spatial loops. The finite-difference Laplacian is assembled from second derivatives in each spatial direction, each approximated using a three-point stencil implemented through analogous strided tensor operations, without Python-level loops.

This fully tensorized formulation maps efficiently onto the single-instruction, multiple-thread (SIMT) architecture of modern GPUs. In this regime, performance is primarily limited by memory bandwidth rather than arithmetic intensity. 
The numerical results in Section~\ref{sec:ROM_efficiency} show that the GPU-based solver achieves more than a 100-fold speedup over the CPU implementation.
% Numerical results in Section \ref{sec:ROM_efficiency} demonstrate that, compared with the CPU implementation, the GPU-based solver achieves more than a 100-fold speedup, highlighting the substantial computational efficiency advantage of the GPU implementation.

\subsection{Verification of the high-fidelity solver}
\label{sec:FOM_verification}

Owing to the strong nonlinear coupling between spatial transport and internal spin dynamics, a closed-form solution is not available for the full system.
We therefore consider three complementary verification cases based on an analytical reduction, asymptotic limits, and an analytical invariant, respectively.
Here we give only the formulas needed for verification.
For longitudinal pumping sufficiently far from depolarizing walls, Appelt et al. show that neglecting spatial diffusion reduces the steady-state problem to a local equation and yields the spin-temperature distribution [Sec.~IX, Eqs.~(111)--(114)] \cite{appelt1998theory}. They further show that the alkali--alkali spin-exchange contribution vanishes for the longitudinal spin-temperature steady state in the absence of diffusion, radio-frequency fields, and slow processes [Eqs.~(121)--(123)]. Under the additional HP-limit assumptions used in the present verification, including fully circularly polarized pumping and no additional polarization-transfer contribution, their steady-state result reduces in the present nondimensional variables to
\begin{equation}
    P(\mathbf{r})=\frac{\mathbb R_0\xi(\mathbf{r})}{\mathbb R_0\xi(\mathbf{r})+1}.
\label{eq:P}
\end{equation}
The analytical population vector follows directly from the general spin-temperature distribution of Appelt et al.~\cite{appelt1998theory} by setting $I=3/2$ for ${}^{87}\mathrm{Rb}$ and adopting the population ordering specified in Appendix~\ref{appendix-sec:evolutionmatrix}. Thus, when $\mathbb D\to0$ and $f_D=0$, the diffusionless bulk population state is the following deterministic function of $P$:
\begin{equation} \label{eq:rho_exact}
    \bm{\rho}_{\mathrm{exact}} = \frac{1}{8(P^2 + 1)}
    \begin{bmatrix}
        (P+1)^4 \\
        -(P-1)(P+1)^3 \\
        (P^2 - 1)^2 \\
        -(P-1)^3 (P+1) \\
        (P-1)^4 \\
        -(P-1)^3 (P+1) \\
        (P^2 - 1)^2 \\
        -(P-1)(P+1)^3
    \end{bmatrix}.
\end{equation}

Equations~\eqref{eq:P} and \eqref{eq:rho_exact} describe the diffusionless bulk equilibrium, or outer solution, obtained by neglecting the wall-adjacent diffusion layers.
The comparisons presented below assess the agreement of the numerical solutions with this analytical reduction and the associated propagation limits or invariant.
The prescribed depolarizing-wall boundary conditions remain enforced in the numerical solver.

A local reaction--diffusion balance provides an estimate of the small-diffusion wall-layer scale.
Using the local reaction matrix $B(P,\xi)$ defined in \eqref{eq:frozen_reaction_matrix}, let $\bm\rho_*$ denote a diffusionless equilibrium at a locally fixed light intensity $\xi_*$:
\begin{equation*}
    B(P(\bm\rho_*),\xi_*)\bm\rho_*=0,
    \qquad
    J_*:= \left. \frac{\partial \left[B(P(\bm\rho),\xi_*)\bm\rho\right]}
    {\partial\bm\rho}\right|_{\bm\rho=\bm\rho_*}.
\end{equation*}
The perturbation $\bm u=\bm\rho-\bm\rho_*$ satisfies $\mathbf1^\top\bm u=0$, since both states have unit total population.
Neglecting tangential variation, linearization of the steady DQME in the inward wall-normal coordinate $s$ gives
\begin{equation*}
    \mathbb D\,\frac{\mathrm d^2\bm u}{\mathrm ds^2}-J_*\bm u=0.
\end{equation*}
The Jacobian includes the state dependence of $B$ through $P(\bm\rho)$ and is generally not the frozen matrix $B$.
For an eigenmode $J_*\bm v_j=\lambda_j\bm v_j$ in the normalization-preserving subspace with $\lambda_j>0$ real, substitution of $\bm u(s)=a_j(s)\bm v_j$ gives
\begin{equation*}
    \mathbb D\,a_j''-\lambda_j a_j=0,
    \qquad
    a_j(s)\propto e^{-s\sqrt{\lambda_j/\mathbb D}},
    \qquad
    \delta_j=\sqrt{\mathbb D/\lambda_j},
\end{equation*}
where the negative exponential branch is selected to decay toward the bulk.
Appelt et al. derive diffusion-layer eigenmodes with the same exponential spatial dependence [Sec.~IX, ``Diffusion layer,'' pp.~1424--1425, Eqs.~(128)--(130)] \cite{appelt1998theory}; here this modal balance is applied to the local DQME reaction Jacobian.
This conditional estimate neither establishes the spectrum of the full coupled problem nor describes its nonlinear wall-layer profile.
For fixed reaction parameters and decay eigenvalues of order one, $\delta=O(\sqrt{\mathbb D})$ gives a scale of $O(10^{-3})$ at $\mathbb D=10^{-6}$, below the baseline axial spacing $\Delta z=10^{-2}$.
The actual modal thickness depends on $\lambda_j$, therefore, this estimate does not establish adequate resolution of the wall-adjacent diffusion layers.
Their spatial resolution is examined separately in Appendix~\ref{appendix-sec:gridrefinement}.

Cases~2 and~3 fix $\mathbb D=10^{-6}$, whereas Case~1 uses the values specified for its individual tests and prescribes the longitudinal propagation factor analytically without a numerical light solve.
Unless otherwise stated, the default parameter values in this subsection are $\eta=1.0$, $f_D=0.0$, and $w=0.5$.
The case-dependent values of $\mathrm{OD}$ and $\mathbb R_0$ are given in the respective case descriptions and, for Case~3, Table~\ref{tab:FOM_Lambert_errors}.
Based on the mesh refinement study in Appendix~\ref{appendix-sec:gridrefinement}, we adopt a uniform Cartesian mesh with spacing $\Delta x=\Delta y=\Delta z=1/100$.

\subsubsection{Case 1: Decoupled-light, diffusionless benchmark}
\label{sec:FOM_verify_case1}

Case~1 isolates the DQME solver by prescribing the axial light field analytically as
\begin{equation}
    \xi_z=e^{-\mathrm{OD}  \cdot z}.
\end{equation}
This prescription decouples the axial light-propagation equation from the DQME. 
It is an idealized verification benchmark with a known reference solution, rather than a claim that the prescribed field is an exact limit of the coupled DQME--light system; such analytical benchmark problems are appropriate for testing a numerical implementation \cite{roache2002code}. 
Substitution into Eqs.~\eqref{eq:P} and~\eqref{eq:rho_exact} gives the deterministic reference solution derived from the spin-temperature reduction.
This idealized setting supports two checks.
First, we verify convergence of the bulk solution toward the diffusionless local reduction as $\mathbb{D} \to 0$.
Second, the solution depends only on $P(\mathbf{r})$ in this limit; the steady state is therefore analytically independent of the nonlinear spin-exchange coefficient $\eta$.

For the first test, we decrease the diffusion coefficient over the range $\mathbb D\in[10^{-2},10^{-7}]$ while fixing $\mathbb R_0=3.0$ and $\mathrm{OD}=0.5$.
We then examine the expected independence with respect to $\eta$ by varying $\eta\in[0,1000]$ with $\mathbb{D}=10^{-6}$, $\mathbb R_0=1.5$ and $\mathrm{OD}=0.5$.
The interior-domain error is measured by the $L_\infty$ error, the root-mean-square error, and the relative $L_2$ error:
\begin{subequations} \label{eq:errors}
    \begin{align}
        E_{\infty} &= \left\| \bm{\rho}_{\mathrm{exact}}(\mathbf{r}) - \bm{\rho}_h  (\mathbf{r}) \right\|_\infty, \\
        E_{\mathrm{rms}} &= \frac{1}{\sqrt{8N}}  \left\| \bm{\rho}_{\mathrm{exact}}(\mathbf{r}) - \bm{\rho}_h(\mathbf{r}) \right\|_2, \\
        E_{2,\mathrm{rel}} &= \frac{\left\| \bm{\rho}_{\mathrm{exact}}(\mathbf{r}) - \bm{\rho}_h(\mathbf{r}) \right\|_2} {\left\| \bm{\rho}_{\mathrm{exact}}(\mathbf{r}) \right\|_2},
    \end{align}
\end{subequations}
where $N$ is the number of interior grid points and all norms are evaluated over the stacked vectors containing the eight population components at each interior point.

\begin{figure}
\centering
    \begin{subfigure}[b]{0.45\textwidth}
        \includegraphics[width=1.0\linewidth]{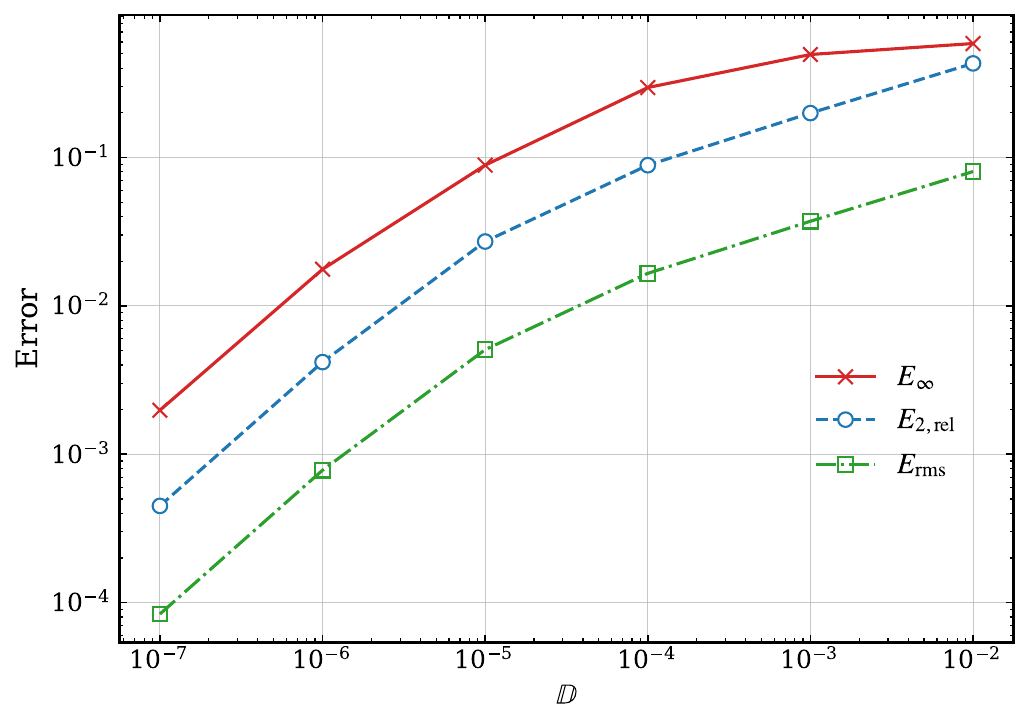}
        \subcaption{Asymptotic behavior as diffusion vanishes.}
        \label{fig:FOM_D}
    \end{subfigure}
% \hfill
    \begin{subfigure}[b]{0.45\textwidth}
        \includegraphics[width=1.0\linewidth]{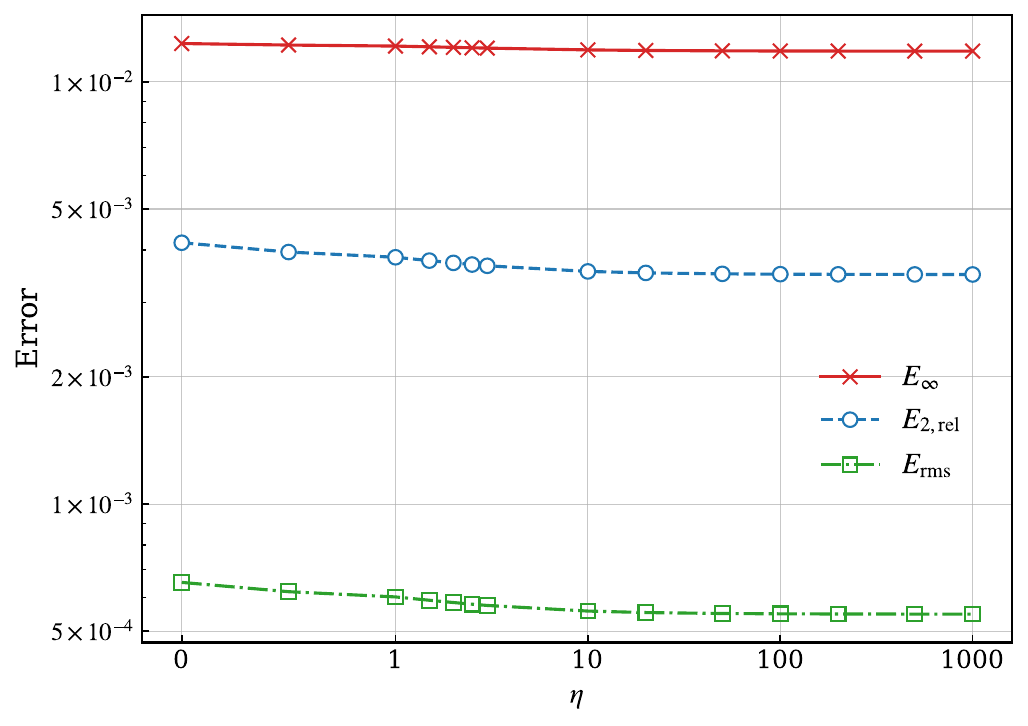}
        \subcaption{Error invariance with respect to $\eta$.}
        \label{fig:FOM_eta}
    \end{subfigure}
    \caption{Verification of the full-order solver using the decoupled-light, diffusionless benchmark.}
\label{fig:FOM_error}
\end{figure}

Figure~\ref{fig:FOM_error} summarizes the two tests within the decoupled-light benchmark.
Figure~\ref{fig:FOM_D} shows that all three error measures decrease monotonically as $\mathbb D\to 0$.
This behavior is consistent with the asymptotic reduction from the full partial differential system to the local algebraic relation.
In particular, the relative $L_2$ error decreases from $43.06\%$ at $\mathbb D=10^{-2}$ to $0.04\%$ at $\mathbb D=10^{-7}$.
These results indicate that, away from the thin wall-adjacent diffusion layers at small diffusion coefficients, the numerical solution approaches the diffusionless local equilibrium.
Figure~\ref{fig:FOM_eta} shows that the error is nearly unchanged over three orders of magnitude in $\eta$.
This is consistent with the analytical decoupling in this limit: since the asymptotic population vector $\bm\rho$ is independent of $\eta$, so is the electron-spin polarization $P(\mathbf r)$.
The flat error profile therefore indicates that the solver does not introduce an artificial dependence on the spin-exchange coefficient.

\subsubsection{Case 2: Low- and high-pump-rate limits of light propagation}
\label{sec:FOM_verify_case2}

Case 1 used a prescribed longitudinal propagation factor and therefore tested only the atomic-state solver.
We now restore the full coupling between $\xi_z$ and $\bm\rho$ and examine two asymptotic pumping regimes of the exact Lambert-$W$ solution. 
Seltzer gives the polarization-dependent propagation equation and its Beer--Lambert reduction in Sec.~2.3.3, Eqs.~(2.54)--(2.55), and derives the transcendental and Lambert-$W$ solutions for circularly polarized light in Eqs.~(2.57)--(2.58) \cite{seltzer2008developments}. 
The same section and Fig.~2.11 identify the associated physical behavior: weak pumping produces low polarization and strong absorption, whereas strong pumping produces nearly complete polarization and weak attenuation.
% These tests assess whether the fully coupled solver recovers the theoretically predicted propagation rates in the two asymptotic pumping regimes.
We use these two asymptotic regimes to assess whether the fully coupled solver recovers the theoretically predicted propagation rates.

Using the local polarization relation \eqref{eq:P}, the light-propagation equation~\eqref{eq:light} reduces to
\begin{equation}\label{eq:light_z}
    \frac{\mathrm d\xi_z}{\mathrm dz}
    =-\mathrm{OD}\,\xi_z\left(1-\frac{\mathbb R_0\xi}{\mathbb R_0\xi+1}\right).
    % =-\frac{\mathrm{OD}\,\xi_z}{\mathbb R_0\xi_z+1}.
\end{equation}
Consistently with Seltzer's Eq.~(2.58), this equation yields the dimensionless Lambert-$W$ solution
\begin{equation}
    \xi_z = \frac{1}{\mathbb{R}_0 \xi_\perp}
            W\!\left[\mathbb{R}_0 \xi_\perp \cdot \exp(\mathbb{R}_0 \xi_\perp -\mathrm{OD}\cdot z)\right].
\end{equation}
To obtain the asymptotic limits in the present nondimensional variables, define
% For the following asymptotic analysis, define the argument of the Lambert-$W$ function as
\begin{equation*}
    X=\mathbb R_0\xi_\perp e^{\mathbb R_0\xi_\perp-\mathrm{OD}\,z}.
\end{equation*}
For fixed $\mathrm{OD}$, $z$, and $\xi_\perp>0$, this argument tends to zero as $\mathbb R_0\to0$ and to infinity as $\mathbb R_0\to\infty$.

In the low-pump-rate limit $\mathbb R_0\to0$, the expansion $W(X)=X+O(X^2)$ gives
% At low pumping, substitution of $W(X)=X+O(X^2)$ recovers Beer--Lambert attenuation:
\begin{equation*}
    \xi_z(z)=\frac{X+O(X^2)}{\mathbb R_0\xi_\perp}
    =e^{\mathbb R_0\xi_\perp}e^{-\mathrm{OD}\,z}+O(\mathbb R_0)
    =e^{-\mathrm{OD}\,z}+O(\mathbb R_0).
\end{equation*}
% Thus, the light propagation approaches the Beer--Lambert law, with the leading-order attenuation rate determined by $\mathrm{OD}$.

In the high-pump-rate limit $\mathbb R_0\to\infty$, consider the central ray $(x,y)=(0,0)$, where $\xi_\perp=1$.
On this ray, $X=\mathbb R_0 e^{\mathbb R_0-\mathrm{OD}\,z}$ and $W(X)=\mathbb R_0\xi_z$.
Using $W(X)e^{W(X)}=X$ then gives
% Using $W(X)e^{W(X)}=X$ and $W(X)=\mathbb R_0\xi_z$ gives
\begin{equation*}
    \mathbb R_0\xi_z e^{\mathbb R_0\xi_z}
    =\mathbb R_0 e^{\mathbb R_0-\mathrm{OD}\,z}.
\end{equation*}
Taking logarithms and rearranging yields
\begin{equation*}
    \mathbb R_0(\xi_z-1)+\ln\xi_z=-\mathrm{OD}\,z.
\end{equation*}
Substitution of $\xi_z=1+c_1(z)/\mathbb R_0+O(\mathbb R_0^{-2})$ gives $c_1(z)=-\mathrm{OD}\,z$, and hence, for fixed $\mathrm{OD}$ and bounded $z$,
\begin{equation*}
    \xi_z(z)=1-\frac{\mathrm{OD}}{\mathbb R_0}z+O(\mathbb R_0^{-2}).
\end{equation*}

% Substituting
% $\xi_z=1+c_1(z)/\mathbb R_0+O(\mathbb R_0^{-2})$
% into the preceding equation, the logarithmic term is
% $O(\mathbb R_0^{-1})$.
% The leading-order balance therefore gives
% $c_1(z)=-\mathrm{OD}\,z$, yielding
% \begin{equation*}
%     \xi_z(z)=1-\frac{\mathrm{OD}}{\mathbb R_0}z
%     +O(\mathbb R_0^{-2}),
% \end{equation*}
% for fixed $\mathrm{OD}$ and bounded $z$.

For the present verification at $\mathbb D=10^{-6}$, we treat the residual influence of the unresolved wall layers on the sampled interior solution as negligible for these comparisons.
This approximation is motivated by the decay of wall corrections in Appelt et al.'s local theory [Sec.~IX, pp.~1424--1425, Eqs.~(128)--(130)] \cite{appelt1998theory}, the population comparisons in Section~\ref{sec:FOM_verify_case1}, and the complementary mesh-refinement results for $\xi_z$ and $P$ over all interior nodes in Appendix~\ref{appendix-sec:gridrefinement}.
We therefore retain all interior axial nodes in $[z_a,z_b]$, where $z_a$ and $z_b$ are the first and last interior nodes, respectively.
Let $\xi_{z,h}(z_k)$ denote the numerical longitudinal propagation factor on the central ray $(0,0,z_k)$ and define $Z_k=z_k-z_a$.
Taking the ratio of \eqref{eq:light_integral} at $z$ and $z_a$ gives
\begin{equation*}
    \frac{\xi_z(z)}{\xi_z(z_a)}
    =
    \exp\left[
        -\mathrm{OD}\int_{z_a}^{z}
        \bigl(1-P(s)\bigr)\,\mathrm ds
    \right].
\end{equation*}
This exact relation motivates anchoring the comparison at $z_a$: the ratio cancels the explicit attenuation factor accumulated over $[0,z_a]$ while leaving the computed solution unchanged.
The polarization $P(s)$ within the remaining integral still reflects the coupled atomic and light fields.

The two asymptotic limits suggest different quantities for measuring the propagation rate.
For low pumping, we use the logarithmic intensity ratio, whereas for high pumping, we use the intensity difference.
Both quantities vanish at $z_a$, and their leading-order relations are
% We use the logarithmic ratio for low pumping and the intensity difference for high pumping, so both transformed quantities vanish at the reference point.
% The corresponding asymptotic relations are
\begin{align*}
    \ln\frac{\xi_z(z)}{\xi_z(z_a)}&\approx-\mathrm{OD}(z-z_a)
    &&\text{(low pumping)},\\
    \xi_z(z)-\xi_z(z_a)&\approx-\frac{\mathrm{OD}}{\mathbb R_0}(z-z_a)
    &&\text{(high pumping)}.
\end{align*}
For each regime we perform a least-squares fit constrained to pass through the origin. The fitted positive slope magnitudes are
\begin{equation}
    \alpha_h=-\frac{\sum_k Z_k\ln[\xi_{z,h}(z_k)/\xi_{z,h}(z_a)]}{\sum_k Z_k^2},
    \qquad
    \beta_h=-\frac{\sum_k Z_k[\xi_{z,h}(z_k)-\xi_{z,h}(z_a)]}{\sum_k Z_k^2},
\label{eq:verification_fitted_slopes}
\end{equation}
with asymptotic predictions $\mathrm{OD}$ and $\mathrm{OD}/\mathbb R_0$, respectively. The sums include all interior axial nodes, including the first and last interior nodes. For low pumping, we fix $\mathbb R_0=10^{-3}$ and take $\mathrm{OD}\in\{0.1,0.5,1,3,5,10,20,40\}$. For high pumping, we fix $\mathrm{OD}=0.5$ and take $\mathbb R_0\in\{20,30,40,50,70,100,150,200\}$.

\begin{figure}
\centering
    \includegraphics[width=0.45\linewidth]{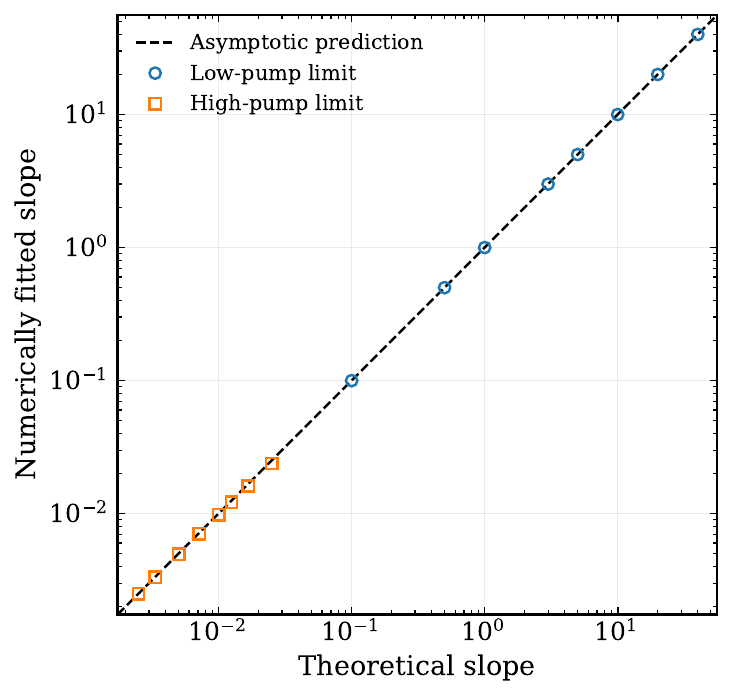}
    \caption{Central-ray slope magnitudes from zero-intercept least-squares fits \eqref{eq:verification_fitted_slopes} versus the theoretical values $\mathrm{OD}$ (low pumping) and $\mathrm{OD}/\mathbb R_0$ (high pumping).} %Both axes are logarithmic; the dashed diagonal denotes perfect agreement.
\label{fig:asymptotic_behavior}
\end{figure}

Figure~\ref{fig:asymptotic_behavior} compares the fitted slope magnitudes $\alpha_h$ and $\beta_h$ with their respective asymptotic predictions $\mathrm{OD}$ and $\mathrm{OD}/\mathbb R_0$ in a single parity plot.
For low pumping, the maximum absolute relative deviation $|\alpha_h/\mathrm{OD}-1|$ is $0.096\%$, so the fitted attenuation coefficients agree with $\mathrm{OD}$ within approximately $0.10\%$ over $\mathrm{OD}\in[0.1,40]$. 
For high pumping, $\beta_h$ approaches $\mathrm{OD}/\mathbb R_0$ from below as $\mathbb R_0$ increases: the absolute relative deviation $|\beta_h/(\mathrm{OD}/\mathbb R_0)-1|$ decreases from $4.79\%$ at $\mathbb R_0=20$ to $0.47\%$ at $\mathbb R_0=200$. 
% These deviations measure departure from leading-order asymptotic predictions and include finite-pumping, finite-diffusion, and numerical effects; they are not estimates of pure discretization error.

The high-pump trend is consistent with the higher-order asymptotic correction obtained from \eqref{eq:light_z} on the central ray:
\begin{equation*}
    -\frac{\mathrm d\xi_z}{\mathrm dz}
    =\frac{\mathrm{OD}}{\mathbb R_0}
    \frac{\mathbb R_0\xi_z}{1+\mathbb R_0\xi_z}
    =\frac{\mathrm{OD}}{\mathbb R_0}
    \left[1-\frac{1}{\mathbb R_0\xi_z}+O(\mathbb R_0^{-2})\right].
\end{equation*}
The negative correction accounts for the lower attenuation rate.
For fixed $\mathrm{OD}$ on a bounded interval with $\xi_z$ bounded away from zero, the relative correction is $O(\mathbb R_0^{-1})$, consistent with the observed approach as pumping increases.
% The finite-pumping slope magnitude is below its leading-order prediction.

The results support recovery of the predicted propagation rates by the coupled solver in both asymptotic pumping regimes.

\subsubsection{Case 3: General regime and Lambert invariant}
\label{sec:FOM_verify_case3}

Case~2 verifies the coupled solver in asymptotic pumping regimes. 
Case~3 tests the coupled FOM away from those limits using an exact invariant of the diffusionless Lambert-$W$ light-propagation relation.
For this verification, we subtract Seltzer's logarithmic propagation relation at two axial locations $z$ and $z_a$ on the same ray [Sec.~2.3.3, Eqs.~(2.57)--(2.58) \cite{seltzer2008developments}] to obtain, in the present nondimensional variables,
\begin{equation}
    \Delta B(z;z_a) = 
    \mathbb R_0\xi_\perp\left[\xi_z(z)-\xi_z(z_a)\right]
    +\ln\frac{\xi_z(z)}{\xi_z(z_a)}
    =-\mathrm{OD}(z-z_a).
\label{eq:transcendental_light}
\end{equation}
We use the same interval $[z_a,z_b]$ and reference point as in Section~\ref{sec:FOM_verify_case2}, retaining all interior axial nodes.
% Equation~\eqref{eq:transcendental_light} is exact for the diffusionless propagation relation; it need not hold identically for the finite-diffusion case.
To assess this relation, we evaluate the numerical and analytical expressions at each axial node $z_k$:
% Using the numerical longitudinal propagation factor, we construct
\begin{equation*}
    \Delta B_h(z_k;z_a)
    =\mathbb R_0\xi_\perp
    \left[\xi_{z,h}(z_k)-\xi_{z,h}(z_a)\right]
    +\ln\frac{\xi_{z,h}(z_k)}{\xi_{z,h}(z_a)},
    \qquad
    \Delta B_{\mathrm{analytical}}(z_k;z_a)=-\mathrm{OD}(z_k-z_a).
\end{equation*}
The full invariant curves are compared over the interior axial interval. 
We quantify the agreement between the numerical and analytical profiles over the interior axial interval using the relative $L_\infty$ and $L_2$ errors
\begin{equation}
    E_{B,\infty}
    =\frac{\|\Delta B_h-\Delta B_{\mathrm{analytical}}\|_\infty}
    {\|\Delta B_{\mathrm{analytical}}\|_\infty},
    \qquad
    E_{B,2}
    =\frac{\|\Delta B_h-\Delta B_{\mathrm{analytical}}\|_2}
    {\|\Delta B_{\mathrm{analytical}}\|_2}.
\label{eq:Lambert_invariant_errors}
\end{equation}
The norms are evaluated over the sampled axial nodes.
% The norms are evaluated over the sampled axial nodes; the normalized $L_2$ error is equivalently the relative RMSE (RMSE of the difference divided by RMS of the analytical invariant). These errors include finite-diffusion and numerical effects, rather than pure discretization error.

\begin{table}
\centering
\caption{Relative errors of the anchored Lambert invariant for all nine parameter combinations, with $\mathbb D=10^{-6}$, evaluated using \eqref{eq:Lambert_invariant_errors}.}
\label{tab:FOM_Lambert_errors}
\begin{tabular}{cccc}
\hline
$\mathbb R_0$ & $\mathrm{OD}$ & $E_{B,\infty}$ (\%) & $E_{B,2}$ (\%) \\
\hline
0.5 & 5 & 0.0103 & 0.0173 \\
1.0 & 5 & 0.0161 & 0.0257 \\
2.0 & 1 & 0.0338 & 0.0305 \\
2.0 & 3 & 0.0302 & 0.0318 \\
2.0 & 5 & 0.0240 & 0.0328 \\
2.0 & 7 & 0.0205 & 0.0327 \\
2.0 & 9 & 0.0195 & 0.0321 \\
3.0 & 5 & 0.0305 & 0.0348 \\
5.0 & 5 & 0.0373 & 0.0350 \\
\hline
\end{tabular}
\end{table}

\begin{figure}
\centering
    \includegraphics[width=0.45\linewidth]{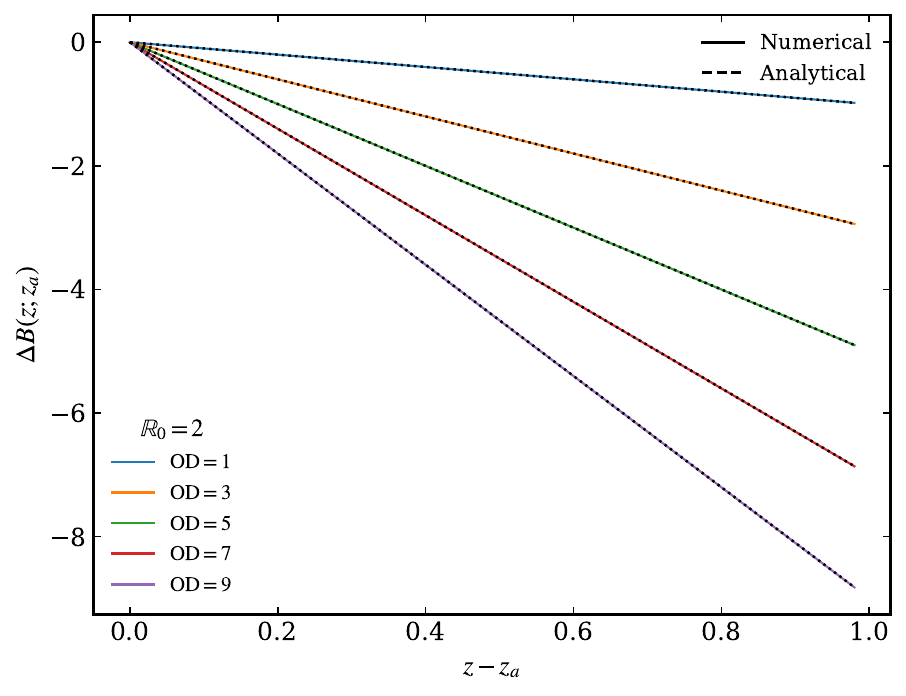}
        \caption{Comparison of the numerical and analytical Lambert-invariant profiles for $\mathbb R_0=2$ and $\mathrm{OD}=1,3,5,7,9$. Solid and dashed lines represent $\Delta B_h$ and $\Delta B=-\mathrm{OD}(z-z_a)$, respectively.}
    % \caption{Verification of the anchored Lambert invariant in the general non-asymptotic regime for $\mathbb R_0=2$ and $\mathrm{OD}=1,3,5,7,9$. Solid and dashed lines denote the numerical invariant and the analytical relation $\Delta B=-\mathrm{OD}(z-z_a)$, respectively; the curves nearly overlap. Quantitative error statistics cover all nine cross-sampled parameter combinations.}
\label{fig:FOM_Bz}
\end{figure}

We use two intersecting parameter sweeps: $\mathbb R_0=2$ with $\mathrm{OD}\in\{1,3,5,7,9\}$, and $\mathrm{OD}=5$ with $\mathbb R_0\in\{0.5,1,2,3,5\}$. The shared point $(\mathbb R_0,\mathrm{OD})=(2,5)$ gives nine distinct combinations. The two sweeps separately assess changes in attenuation strength and pumping strength in the general regime, without taking a low- or high-pump asymptotic limit.

Figure~\ref{fig:FOM_Bz} compares the numerical and analytical profiles for the five OD-sweep cases at fixed $\mathbb R_0=2$.
At fixed $\mathrm{OD}=5$, the analytical relation is independent of $\mathbb R_0$, so the corresponding analytical profiles coincide.
The results for both parameter sweeps are quantified in Table~\ref{tab:FOM_Lambert_errors}, where both errors remain below $0.04\%$ for all nine combinations.
The largest errors occur at $(\mathbb R_0,\mathrm{OD})=(5,5)$, with $E_{B,\infty}\approx0.037\%$ and $E_{B,2}\approx0.035\%$.
These results demonstrate close agreement between the coupled FOM and the analytical Lambert-invariant relation over the interior verification interval in the general non-asymptotic regime.

\section{Reduced-order modeling}
\label{sec:ROM}
The high-fidelity formulation presented in Section~\ref{sec:FOM} involves an $8N_h$-dimensional atomic population field, where $N_h$ denotes the number of grid points in the computational mesh. For the baseline discretization, this corresponds to approximately $8\times10^6$ degrees of freedom. The large state dimension makes repeated high-fidelity simulations costly for parameter studies and sensitivity analysis. This computational cost motivates reduced-order modeling~\cite{lucia2004reduced}, which seeks to construct low-dimensional models that are inexpensive to evaluate while accurately approximating the underlying high-fidelity solutions.

% A key challenge in constructing a reduced-order model (ROM) for the coupled DQME–light system is accounting for the coupling between the atomic population and the pump-light field. Evaluating the governing equations requires both the atomic population field $\bm{\rho}$ and the longitudinal propagation factor $\xi_z$, but these fields are not independent dynamical variables. For a given atomic population field and prescribed incident-light condition, the light-propagation equation uniquely determines $\xi_z$ along each axial ray, thereby providing an auxiliary closure relation for the atomic-state equation. Eliminating $\xi_z$ through this relation yields an equivalent formulation in terms of $\bm{\rho}$ alone, at the expense of introducing a nonlocal nonlinear optical-pumping source term into the atomic-state equation.

A key challenge in constructing a reduced-order model (ROM) for the
coupled DQME--light system is accounting for the coupling between
atomic evolution in pseudo-time and spatial light propagation.
For a prescribed incident-light condition, the atomic population field
uniquely determines the light field. Eliminating the latter yields an
integro-differential formulation involving only the atomic populations.
The differential and integro-differential formulations can be
numerically equivalent when the same axial integration scheme is used,
but they offer distinct computational advantages.
The FOM retains the differential light-propagation equation to evaluate
the light field efficiently during pseudo-time marching.
For model reduction, the integro-differential formulation requires only
the atomic population field to be approximated in a reduced space,
simplifying the treatment of the coupled nonlinear system.

In this work, we develop the ROM using a POD basis and Galerkin projection. During the offline stage, a POD basis is extracted from high-fidelity snapshots generated at sampled points in parameter space. The high-fidelity model is then projected onto the resulting reduced subspace to obtain the ROM. During the online stage, the reduced solution for a new parameter instance is computed by solving the ROM. The construction of the POD-Galerkin ROM is described in detail in the remainder of this section.

\subsection{POD--Galerkin reduced-order model}
\label{sec:ROM_PODGP}

A key challenge in constructing a ROM for the coupled DQME--light
system arises from the different roles of the two governing equations.
The DQME describes the evolution of atomic populations in pseudo-time
$\tau$, whereas the light-propagation equation determines the pump-light
field along the spatial coordinate $z$ at each pseudo-time.
To address this challenge, we exploit the fact that, for a prescribed
incident-light condition, the atomic population field uniquely
determines the pump-light field.
Eliminating the light variable yields an integro-differential
formulation for the atomic populations alone, allowing the reduced
approximation to be constructed solely for the atomic state.

Imposing the prescribed incident condition $\xi_z(x,y,0,\tau)=1$ in the integral representation \eqref{eq:light_integral} and substituting the resulting expression $\xi_z[\bm\rho]$ into \eqref{eq:DME} yields an equivalent integro-differential equation in terms of $\bm\rho$ alone:
\begin{equation}
\begin{aligned}
    \frac{\partial\bm\rho}{\partial\tau}
    = \mathbb D\nabla^2\bm\rho
    &-\left[(1+\eta)A_{\mathrm{SD}}+f_DA_{\mathrm{FD}}\right]\bm\rho
    + \eta\left(\mathbf S_z^{\top}\bm\rho\right)A_{\mathrm{SE}}^{(z)}\bm\rho
    \\
    &-\mathbb R_0\xi_\perp(x,y)
    \exp\!\left[-\mathrm{OD}\,z+2\mathrm{OD}\int_0^z
    \mathbf S_z^{\top}\bm\rho(x,y,s,\tau)\,\mathrm ds\right]
    A_{\mathrm{OP}}\bm\rho.
\end{aligned}
\label{eq:DME_rho_only}
\end{equation}
The first two terms on the right-hand side represent linear diffusion and relaxation, respectively, while the third describes local quadratic spin exchange. The final term describes nonlinear optical pumping and depends nonlocally on $\bm\rho$ through the upstream integral.

Combining the exponential integration in Section~\ref{sec:FOM_light} for the optical-pumping term with the spatial discretization in Section~\ref{sec:FOM_spatialtime} for the remaining terms yields the semi-discrete finite-difference system
\begin{equation}
    \frac{\mathrm d\bm\rho_h}{\mathrm d\tau}
    =\mathbb L_h(\bm\mu)\bm\rho_h
    -\mathbb G_{\mathrm{SE}}(\bm\rho_h;\bm\mu)\bm\rho_h
    -\mathbb G_{\mathrm{OP}}(\bm\rho_h;\bm\mu)\bm\rho_h
    =:\mathcal R_h(\bm\rho_h;\bm\mu),
\label{eq:FOM_compact}
\end{equation}
where $\bm\rho_h \in\mathbb R^{8N_h}$ is the high-dimensional solution vector, $\bm\mu$ denotes the parameter vector, and
\begin{equation*}
    \mathbb L_h(\bm\mu):=\mathbb D(\Delta_h\otimes\mathbf I_8)
    -(1+\eta)(\mathbf I_{N_h}\otimes A_{\mathrm{SD}})
    -f_D(\mathbf I_{N_h}\otimes A_{\mathrm{FD}}).
\end{equation*}
Applying the forward Euler method in pseudo-time to \eqref{eq:FOM_compact} recovers the high-fidelity numerical scheme described in Section~\ref{sec:FOM}. This $\bm\rho$-only formulation provides a convenient starting point for reduced-order modeling.

In the reduced basis method, the high-fidelity solution $\bm\rho_h(\bm\mu) \in\mathbb R^{8N_h}$, defined as the converged steady state of \eqref{eq:FOM_compact} at parameter value $\bm\mu$, is approximated by the affine ansatz
\begin{equation}
    \bm\rho_h(\bm\mu)\approx\bm\rho_r(\bm\mu)
    =\mathbf V\bm\alpha(\bm\mu)+\bar{\bm\rho},
\label{eq:affine_ansatz}
\end{equation}
where $\bm\rho_r(\bm\mu) \in\mathbb R^{8N_h}$ is the reduced basis solution, $\mathbf V=[\mathbf v_1,\ldots,\mathbf v_r]\in\mathbb R^{8N_h\times r}$ is a matrix with the reduced basis vectors as columns, $\bm\alpha(\bm\mu)\in\mathbb R^r$ is the vector of reduced coefficients, and $\bar{\bm\rho} \in\mathbb R^{8N_h}$ is a reference state. The reduced coefficients constitute the degrees of freedom of the ROM.
For the parameter domain $\mathcal P$, define the high-fidelity solution manifold
\begin{equation*}
    \mathcal M_h
    :=\left\{\bm\rho_h(\bm\mu):\bm\mu\in\mathcal P\right\}
    \subset\mathbb R^{8N_h}.
\end{equation*}
The reduced-basis approximation assumes that $\mathcal M_h$ can be accurately approximated by the parameter-independent affine space $\bar{\bm\rho}+\mathcal V_r$, where
$\mathcal V_r=\operatorname{span}\{\mathbf v_1,\ldots,\mathbf v_r\}$
and $r\ll8N_h$.

The reduced basis is constructed using proper orthogonal decomposition (POD), which provides a least-squares-optimal approximation of snapshot data for a prescribed reduced dimension. Let
$\mathcal P_h=\{\bm\mu^{(1)},\ldots,\bm\mu^{(N_\mu)}\}\subset\mathcal P$
be the parameter samples generated by Latin hypercube sampling. We choose the snapshot mean as the reference state:
\begin{equation*}
    \bar{\bm\rho}
    =\frac{1}{N_\mu}\sum_{k=1}^{N_\mu}
    \bm\rho_h(\bm\mu^{(k)}).
\end{equation*}
The centered snapshot matrix is then
\begin{equation*}
    \mathbb S
    =\left[
    \bm\rho_h(\bm\mu^{(1)})-\bar{\bm\rho}
    \mid\cdots\mid
    \bm\rho_h(\bm\mu^{(N_\mu)})-\bar{\bm\rho}
    \right]
    \in\mathbb R^{8N_h\times N_\mu}.
\end{equation*}
Only converged steady-state high-fidelity solutions are included in the snapshot matrix; intermediate pseudo-time iterates and the auxiliary light field are excluded.
A thin singular value decomposition (SVD) of $\mathbb S$ gives
\begin{equation*}
    \mathbb S=\mathbf U\bm\Sigma\mathbf W^\top,
    \qquad
    \bm\Sigma=\operatorname{diag}(\sigma_1,\ldots,\sigma_q),
    \qquad
    \sigma_1\geq\cdots\geq\sigma_q\geq0,
\end{equation*}
where $q=\min(8N_h,N_\mu)$. The reduced dimension is chosen using the standard energy criterion associated with the best rank-$r$ approximation~\cite{Schmidt1907,eckart1936approximation}. Specifically, we select the smallest $r$ satisfying
\begin{equation*}
    \frac{\sum_{i=r+1}^{q}\sigma_i^2}
    {\sum_{i=1}^{q}\sigma_i^2}
    \leq\varepsilon_{\mathrm{energy}}.
\end{equation*}
The leading $r$ left singular vectors form the POD basis,
$\mathbf V=[\mathbf u_1,\ldots,\mathbf u_r]$,
which satisfies $\mathbf V^\top\mathbf V=\mathbf I_r$.

The ROM is obtained by Galerkin projection of the high-fidelity
model onto the reduced space $\mathcal V_r$. We first extend
the affine ansatz \eqref{eq:affine_ansatz} to pseudo-time:
\[
    \bm\rho_r(\tau;\bm\mu)
    =\mathbf V\bm\alpha(\tau;\bm\mu)+\bar{\bm\rho}.
\]
Substituting this approximation into \eqref{eq:FOM_compact}
and requiring the residual to be orthogonal to $\mathcal V_r$
gives
\begin{equation}
    \mathbf V^\top\left[
        \mathbf V\frac{\mathrm d\bm\alpha}{\mathrm d\tau}
        -\mathcal R_h(
            \mathbf V\bm\alpha+\bar{\bm\rho};\bm\mu
        )
    \right]=\bm0.
\label{eq:Galerkin_residual}
\end{equation}
Using the orthonormality condition
$\mathbf V^\top\mathbf V=\mathbf I_r$, we obtain the reduced
evolution equation
\begin{equation}
     \frac{\mathrm{d}\bm{\alpha}}{\mathrm{d}\tau}
        = \underbrace{\mathbf{V}^\top\mathbb L_h(\bm\mu)\mathbf{V}}_{\mathbb L_r(\bm\mu)}\bm{\alpha}
            + \underbrace{\mathbf{V}^\top\mathbb L_h(\bm\mu)\bar{\bm{\rho}}}_{\mathbb C_r(\bm\mu)}
        \quad - \mathbf{V}^\top\mathbb{G}_{\mathrm{SE}}(\bm\rho_r;\bm\mu)\bm\rho_r
            - \mathbf{V}^\top\mathbb{G}_{\mathrm{OP}}(\bm\rho_r;\bm\mu)\bm\rho_r
        =:\mathcal{R}_{r}(\bm\alpha;\bm\mu).
\label{eq:ROM_general}   
\end{equation}
The complete offline stage is described in
Algorithm~\ref{alg:offline}.

% The ROM governing the reduced coefficients is obtained by Galerkin projection of the high-fidelity model onto $\mathcal V_r$. To derive the reduced pseudo-time evolution, we extend \eqref{eq:affine_ansatz} to
% $\bm\rho_r(\tau;\bm\mu)=\mathbf V\bm\alpha(\tau;\bm\mu)+\bar{\bm\rho}$.
% Substituting this approximation into \eqref{eq:FOM_compact} and imposing residual orthogonality to $\mathcal V_r$ gives
% \begin{equation}
%     \mathbf V^\top\left[
%     \mathbf V\frac{\mathrm d\bm\alpha}{\mathrm d\tau}
%     -\mathcal R_h(\mathbf V\bm\alpha+\bar{\bm\rho};\bm\mu)
%     \right]=\bm0.
% \label{eq:Galerkin_residual}
% \end{equation}
% Using $\mathbf V^\top\mathbf V=\mathbf I_r$, we obtain the ROM as
% \begin{equation}
%      \frac{\mathrm{d}\bm{\alpha}}{\mathrm{d}\tau}
%         = \underbrace{\mathbf{V}^\top\mathbb L_h(\bm\mu)\mathbf{V}}_{\mathbb L_r(\bm\mu)}\bm{\alpha}
%             + \underbrace{\mathbf{V}^\top\mathbb L_h(\bm\mu)\bar{\bm{\rho}}}_{\mathbb C_r(\bm\mu)}
%         \quad - \mathbf{V}^\top\mathbb{G}_{\mathrm{SE}}(\bm\rho_r;\bm\mu)\bm\rho_r
%             - \mathbf{V}^\top\mathbb{G}_{\mathrm{OP}}(\bm\rho_r;\bm\mu)\bm\rho_r
%         =:\mathcal{R}_{r}(\bm\alpha;\bm\mu).
% \label{eq:ROM_general}   
% \end{equation}
% The complete offline stage is described in Algorithm~\ref{alg:offline}.

For a new parameter value $\bm\mu^\ast$, the online stage
assembles $\mathbb L_r(\bm\mu^\ast)$ and
$\mathbb C_r(\bm\mu^\ast)$ from the projected affine components
precomputed offline. The reduced coefficients are initialized as
$\bm\alpha^0=\mathbf V^\top(\bm\rho_h^0-\bar{\bm\rho})$
and advanced in pseudo-time using the forward Euler method
with the same time step $\Delta\tau$ as the FOM, specified
in \eqref{eq:global_timestep}:
\begin{equation}
    \bm\alpha^{n+1}
    =\bm\alpha^n+\Delta\tau\,
    \mathcal R_r(\bm\alpha^n;\bm\mu^\ast).
\label{eq:ROM_forward_Euler}
\end{equation}
Pseudo-time marching continues until the ROM convergence
criterion is satisfied. The complete online stage is
described in Algorithm~\ref{alg:online}.

% In the online stage, for a new parameter value $\bm\mu^\ast$, the online stage assembles $\mathbb L_r(\bm\mu^\ast)$ and $\mathbb C_r(\bm\mu^\ast)$ from the projected affine components precomputed offline. The reduced coefficients are advanced in pseudo-time using the forward Euler method with the same time step $\Delta\tau$ as the FOM, specified in \eqref{eq:global_timestep}:
% \begin{equation}
%     \bm\alpha^{n+1}
%     =\bm\alpha^n+\Delta\tau\,
%     \mathcal R_r(\bm\alpha^n;\bm\mu).
% \label{eq:ROM_forward_Euler}
% \end{equation}
%  The reduced coefficients are initialized as
% $\bm\alpha^0=\mathbf V^\top(\bm\rho_h^0-\bar{\bm\rho})$
% and advanced using \eqref{eq:ROM_forward_Euler} until the ROM convergence criterion is satisfied. The complete online stage is described in in Algorithm~\ref{alg:online}.

The affine parameter dependence of $\mathbb L_h(\bm\mu)$ allows its parameter-independent components to be projected offline, making online assembly of the reduced matrices and vectors independent of the full-order dimension. The two nonlinear contributions require a different treatment. The spin-exchange term is quadratic in $\bm\rho_h$ and linear in $\eta$, whereas the optical-pumping term is nonlinear in $\bm\rho_h$, nonlocal through the upstream attenuation integral, and non-affine in parameters such as $\mathrm{OD}$ and $w$. Direct Galerkin evaluation at each pseudo-time step would require reconstructing $\bm\rho_r\in\mathbb R^{8N_h}$, evaluating the nonlinear terms on the full grid, and projecting the resulting vectors with $\mathbf V^\top$. The online cost would therefore still scale with the full-order dimension $8N_h$, limiting the computational savings of model reduction. We therefore employ hyperreduction techniques to evaluate the nonlinear contributions at a cost independent of the full-order dimension, enabling efficient online ROM evaluation.

% The affine parameter dependence of $\mathbb L_h(\bm\mu)$ permits its projected matrices and vectors to be precomputed offline, so their online assembly depends only on the reduced dimension.
% The two nonlinear contributions do not admit the same straightforward precomputation.
% The spin-exchange contribution is quadratic in $\bm\rho_h$ and linear in $\eta$, whereas the optical-pumping contribution is nonlinear in $\bm\rho_h$, nonlocal through the upstream attenuation integral, and non-affine in parameters such as $\mathrm{OD}$ and $w$.
% Direct Galerkin evaluation at every pseudo-time step would require reconstructing $\bm\rho_r\in\mathbb R^{8N_h}$, evaluating the nonlinear vectors on the full grid, and projecting them with $\mathbf V^\top$.
% The online cost would therefore continue to scale with the full-order dimension $8N_h$, defeating the intended offline--online decomposition.
% To address this issue, hyperreduction techniques are employed to approximate the nonlinear terms at a cost independent of the full-order dimension, enabling efficient online evaluation of the ROM.

\subsection{Hyperreduction for nonlinear terms}
\label{sec:ROM_hyperreduction}

In this subsection, we introduce hyperreduction strategies~\cite{touze2021model} to efficiently evaluate or approximate the nonlinear terms in the ROM.
% General reviews of nonlinear model-order reduction survey projection-based and hyperreduction strategies for nonlinear systems~\cite{touze2021model}.
In the present problem, the quadratic spin-exchange term admits an exact offline--online decomposition through quadratic expansion (QE), whereas the nonpolynomial and nonlocal optical-pumping term is approximated using the discrete empirical interpolation method (DEIM).

\subsubsection{Quadratic Expansion (QE) for the spin-exchange term}
\label{sec:ROM_NQE}

From \eqref{eq:define_GSE}, the spin-exchange contribution is a node-local quadratic term. For global vectors
\[
    \bm u=[\bm u_1^\top,\ldots,\bm u_{N_h}^\top]^\top,\qquad
    \bm v=[\bm v_1^\top,\ldots,\bm v_{N_h}^\top]^\top,\qquad
    \bm u_q,\bm v_q\in\mathbb R^8,
\]
define the node-local bilinear mapping
\begin{equation}
    \bigl[\mathcal Q(\bm u,\bm v)\bigr]_q :=
    \left(\mathbf S_z^\top\bm u_q\right) (A_{\mathrm{SE}}^{(z)}\bm v_q),
    \qquad q=1,\ldots,N_h,
\label{eq:NQE_bilinear_map}
\end{equation}
where $\mathcal Q(\bm u,\bm v)\in\mathbb R^{8N_h}$ and $\mathbf S_z^\top\bm u_q$ is a scalar multiplying the eight-component vector $A_{\mathrm{SE}}^{(z)}\bm v_q$.
The spin-exchange contribution can then be written as
\begin{equation}
    \mathbb G_{\mathrm{SE}}(\bm\rho)\bm\rho = -\eta\,\mathcal Q(\bm\rho,\bm\rho).
\label{eq:NQE_global_form}
\end{equation}
This definition makes explicit that $\mathbf S_z^\top$ and $A_{\mathrm{SE}}^{(z)}$
act independently on the eight-dimensional state at each spatial node.

% Let $\mathbf V=[\mathbf v_1,\ldots,\mathbf v_r]\in\mathbb R^{8N_h\times r}$. 
Substituting $\bm\rho_r=\mathbf V\bm\alpha+\bar{\bm\rho}$ into \eqref{eq:NQE_global_form} and using the bilinearity of $\mathcal Q$ yields the exact reduced representation
\begin{equation}
    \mathbf V^\top\mathbb G_{\mathrm{SE}}(\bm\rho_r)\bm\rho_r
    =-\eta\left[
    \mathcal A(\bm\alpha\otimes\bm\alpha)
    +\mathcal B\bm\alpha+\mathcal C
    \right],
\label{eq:NQE_expansion}
\end{equation}
where $\bm\alpha\otimes\bm\alpha\in\mathbb R^{r^2}$ and
\begin{align}
    \mathcal A_{\ell,(i-1)r+j}
    &:=\mathbf v_\ell^\top\mathcal Q(\mathbf v_i,\mathbf v_j),
    &&\ell,i,j=1,\ldots,r,
\label{eq:NQE_A}\\
    \mathcal B_{\ell i}
    &:=\mathbf v_\ell^\top\left[
    \mathcal Q(\mathbf v_i,\bar{\bm\rho})
    +\mathcal Q(\bar{\bm\rho},\mathbf v_i)
    \right],
    &&\ell,i=1,\ldots,r,
\label{eq:NQE_B}\\
    \mathcal C_\ell
    &:=\mathbf v_\ell^\top\mathcal Q(\bar{\bm\rho},\bar{\bm\rho}),
    &&\ell=1,\ldots,r.
\label{eq:NQE_C}
\end{align}
The coefficient matrices and vector have dimensions
\[
    \mathcal A\in\mathbb R^{r\times r^2},
    \qquad
    \mathcal B\in\mathbb R^{r\times r},
    \qquad
    \mathcal C\in\mathbb R^r,
\]
and are precomputed offline, as described in Algorithm \ref{alg:offline-nqe}. For dense coefficient matrices, evaluating the quadratic and linear contributions requires $\mathcal O(r^3)$ and $\mathcal O(r^2)$ operations, respectively, while adding the constant contribution costs $\mathcal O(r)$. The overall online cost is therefore $\mathcal O(r^3)$, independent of the full-order dimension $8N_h$.

% Let $\mathbf V=[\mathbf v_1,\ldots,\mathbf v_r]\in\mathbb R^{8N_h\times r}$ and substitute $\bm\rho_r=\mathbf V\bm\alpha+\bar{\bm\rho}$ into \eqref{eq:NQE_global_form}. 
% By the bilinearity of $\mathcal Q$, the quadratic spin-exchange term can be evaluated exactly and efficiently as
% \begin{equation}
%     \mathbf V^\top\mathbb G_{\mathrm{SE}}(\bm\rho_r)\bm\rho_r
%     =-\eta\left[\mathcal A(\bm\alpha\otimes\bm\alpha)+\mathcal B\bm\alpha+\mathcal C\right],
% \label{eq:NQE_expansion}
% \end{equation}
% where $\bm\alpha\otimes\bm\alpha\in\mathbb R^{r^2}$ and
% \begin{align}
%     \mathcal A_{\ell,(i-1)r+j}&:=\mathbf v_\ell^\top\mathcal Q(\mathbf v_i,\mathbf v_j), &&\ell,i,j=1,\ldots,r,
% \label{eq:NQE_A}\\
%     \mathcal B_{\ell i}&:=\mathbf v_\ell^\top\!\left[\mathcal Q(\mathbf v_i,\bar{\bm\rho})+\mathcal Q(\bar{\bm\rho},\mathbf v_i)\right], &&\ell,i=1,\ldots,r,
% \label{eq:NQE_B}\\
%     \mathcal C_\ell&:=\mathbf v_\ell^\top\mathcal Q(\bar{\bm\rho},\bar{\bm\rho}), &&\ell=1,\ldots,r.
% \label{eq:NQE_C}
% \end{align}
% Thus,
% \(
%     \mathcal A\in\mathbb R^{r\times r^2},
%     \mathcal B\in\mathbb R^{r\times r},
%     \mathcal C\in\mathbb R^r.
% \)
% All three operators are precomputed offline. For a general dense reduced tensor, evaluating $\mathcal A(\bm\alpha\otimes\bm\alpha)$, $\mathcal B\bm\alpha$, and $\mathcal C$ requires $\mathcal O(r^3)$, $\mathcal O(r^2)$, and $\mathcal O(r)$ operations, respectively, all independent of the full-order dimension $8N_h$.

\subsubsection{Discrete Empirical Interpolation Method (DEIM) for the optical--pumping term}
\label{sec:ROM_DEIM}

Define the complete optical-pumping nonlinear vector as
\begin{equation*}
    \mathbf f_{\mathrm{OP}}(\bm\rho_r;\bm\mu)
    :=\mathbb G_{\mathrm{OP}}(\bm\rho_r;\bm\mu)\bm\rho_r
    \in\mathbb R^{8N_h}.
\end{equation*}
Its exponential dependence on the upstream axial integral precludes an exact finite polynomial expansion. We therefore use DEIM~\cite{chaturantabut2010nonlinear} to approximate the projected contribution $\mathbf V^\top\mathbf f_{\mathrm{OP}}$.
DEIM approximates $\mathbf f_{\mathrm{OP}}$ in a POD space, with the expansion coefficients determined by interpolation at entries selected through a greedy procedure. Specifically, snapshots of $\mathbf f_{\mathrm{OP}}$ are assembled into the matrix $\mathbf S_{\mathrm{OP}}$, whose leading $m$ left singular vectors form the POD basis $\mathbf U_f\in\mathbb R^{8N_h\times m}$. The DEIM greedy procedure selects $m$ interpolation indices, encoded by the selection matrix $\mathbf P\in\{0,1\}^{8N_h\times m}$. 
The resulting DEIM approximation of the projected optical-pumping term is
\begin{equation}
    \mathbf V^\top\mathbf f_{\mathrm{OP}}(\bm\rho_r;\bm\mu)
    \approx
    \underbrace{\mathbf V^\top\mathbf U_f
    (\mathbf P^\top\mathbf U_f)^{-1}}
    _{\mathcal D\in\mathbb R^{r\times m}}
    \mathbf P^\top\mathbf f_{\mathrm{OP}}(\bm\rho_r;\bm\mu),
\label{eq:DEIM_projected}
\end{equation}
where $\mathbf U_f$, $\mathbf P$, and $\mathcal D$ are constructed offline as described in Algorithm~\ref{alg:offline-deim}.
With $\mathcal D$ precomputed, online evaluation of \eqref{eq:DEIM_projected} requires only the $m$ selected entries $\mathbf P^\top\mathbf f_{\mathrm{OP}}$, avoiding assembly of the full nonlinear vector in $\mathbb R^{8N_h}$.

Let the $s$th DEIM-selected entry correspond to spatial node
$(i_s,j_s,k_s)$ and internal component $q_s\in\{1,\ldots,8\}$,
for $s=1,\ldots,m$. Let $\mathbf p_{i_s,j_s,k}$ collect the
eight global indices associated with node $(i_s,j_s,k)$.
Then $\bm{\rho}_r(\mathbf p_{i_s,j_s,k})\in\mathbb R^8$
denotes the corresponding local population vector, and
$\mathbf V(\mathbf p_{i_s,j_s,k},:)\in\mathbb R^{8\times r}$
denotes the associated block of basis rows.
Using the integral representation of the light field, the selected optical-pumping
entry is
\begin{equation}
    \bigl[
        \mathbf P^\top
        \mathbf f_{\mathrm{OP}}(\bm{\rho}_r;\bm{\mu})
    \bigr]_s
    =
    \mathbb R_0
    \exp\!\left[
        -\frac{2}{w^2}
        \left(x_{i_s}^2+y_{j_s}^2\right)
    \right]
    \exp\!\left[
        -\mathrm{OD}\,z_{k_s}
        +2\mathrm{OD}\,I_s(\bm{\rho}_r)
    \right]
    \mathbf e_{q_s}^\top A_{\mathrm{OP}}
    \bm{\rho}_r(\mathbf p_{i_s,j_s,k_s}),
\label{eq:DEIM_selected_entry}
\end{equation}
where $\mathbf e_{q_s}\in\mathbb R^8$ is the $q_s$th standard
basis vector and $I_s(\bm{\rho}_r)$ is the discrete upstream
integral. Efficient evaluation therefore requires reduced
representations of both the integral and the local optical-pumping action.

For each selected entry, the upstream integral extends from the
incident plane to $z_{k_s}$. Using the trapezoidal rule in
\eqref{eq:light_integral_trap}, we define its discrete
approximation as
\begin{equation}
    I_s(\bm{\rho}_r)
    :=
    \sum_{k=0}^{k_s-1}
    \frac{\Delta z_k}{2}
    \mathbf S_z^\top
    \left[
        \bm{\rho}_r(\mathbf p_{i_s,j_s,k})
        +
        \bm{\rho}_r(\mathbf p_{i_s,j_s,k+1})
    \right].
\label{eq:DEIM_selected_integral}
\end{equation}
Substituting
$\bm{\rho}_r=\mathbf V\bm{\alpha}+\bar{\bm{\rho}}$
yields the affine reduced representation
\begin{equation}
    \mathbf I(\bm{\rho}_r)
    =
    \mathcal E\bm{\alpha}+\bm e,
\label{eq:DEIM_integral_reduced}
\end{equation}
where
$\mathbf I(\bm{\rho}_r)
=[I_1(\bm{\rho}_r),\ldots,I_m(\bm{\rho}_r)]^\top$
and
\begin{align}
    \mathcal E_{s,:}
    &:=
    \sum_{k=0}^{k_s-1}
    \frac{\Delta z_k}{2}
    \mathbf S_z^\top
    \left[
        \mathbf V(\mathbf p_{i_s,j_s,k},:)
        +
        \mathbf V(\mathbf p_{i_s,j_s,k+1},:)
    \right],
\label{eq:DEIM_integral_basis}
    \\
    e_s
    &:=
    \sum_{k=0}^{k_s-1}
    \frac{\Delta z_k}{2}
    \mathbf S_z^\top
    \left[
        \bar{\bm{\rho}}(\mathbf p_{i_s,j_s,k})
        +
        \bar{\bm{\rho}}(\mathbf p_{i_s,j_s,k+1})
    \right].
\label{eq:DEIM_integral_mean}
\end{align}
Here, $\mathcal E\in\mathbb R^{m\times r}$ and
$\bm e\in\mathbb R^m$.

Similarly, the selected component of the local optical-pumping
action admits the affine representation
\begin{equation}
    \left[
        A_{\mathrm{OP}}
        \bm{\rho}_r(\mathbf p_{i_s,j_s,k_s})
    \right]_{q_s}
    =
    (\mathcal F\bm{\alpha})_s+f_s,
\label{eq:DEIM_local_affine}
\end{equation}
where
\begin{equation}
\begin{aligned}
    \mathcal F_{sa}
    &:=
    \mathbf e_{q_s}^\top A_{\mathrm{OP}}
    \mathbf V(\mathbf p_{i_s,j_s,k_s},a),
    \qquad a=1,\ldots,r,
    \\
    f_s
    &:=
    \mathbf e_{q_s}^\top A_{\mathrm{OP}}
    \bar{\bm{\rho}}(\mathbf p_{i_s,j_s,k_s}).
\end{aligned}
\label{eq:DEIM_local_operators}
\end{equation}
Here, $\mathcal F\in\mathbb R^{m\times r}$ and
$\bm f\in\mathbb R^m$.

Collect the selected coordinates in the column vectors
\[
    \bm x_m=[x_{i_1},\ldots,x_{i_m}]^\top,
    \qquad
    \bm y_m=[y_{j_1},\ldots,y_{j_m}]^\top,
    \qquad
    \bm z_m=[z_{k_1},\ldots,z_{k_m}]^\top.
\]
The parameter-dependent baseline light factors are
\begin{equation}
    \bm{\xi}_{\mathcal P}
    :=
    \exp\!\left[
        -\mathrm{OD}\,\bm z_m
        -\frac{2}{w^2}
        \left(
            \bm x_m\circ\bm x_m
            +
            \bm y_m\circ\bm y_m
        \right)
    \right]
    \in\mathbb R^m,
\label{eq:DEIM_selected_light}
\end{equation}
where the exponential is applied componentwise and $\circ$
denotes the Hadamard product. Combining the factors in
\eqref{eq:DEIM_selected_entry} gives the selected nonlinear
vector directly in terms of the reduced coefficients:
\begin{equation}
    \mathbf P^\top
    \mathbf f_{\mathrm{OP}}(\bm{\rho}_r;\bm{\mu})
    =
    \mathbb R_0\,\bm{\xi}_{\mathcal P}
    \circ
    \exp\!\left[
        2\mathrm{OD}
        \left(\mathcal E\bm{\alpha}+\bm e\right)
    \right]
    \circ
    \left(\mathcal F\bm{\alpha}+\bm f\right).
\label{eq:DEIM_selected_online}
\end{equation}
The reduced optical-pumping contribution is therefore
\begin{equation}
\begin{aligned}
    \mathbf V^\top
    \mathbf f_{\mathrm{OP}}(\bm{\rho}_r;\bm{\mu})
    &\approx
    \mathcal D\,\mathbf P^\top
    \mathbf f_{\mathrm{OP}}(\bm{\rho}_r;\bm{\mu})
    \\
    &=
    \mathbb R_0\,\mathcal D
    \left[
        \bm{\xi}_{\mathcal P}
        \circ
        \exp\!\left[
            2\mathrm{OD}
            \left(\mathcal E\bm{\alpha}+\bm e\right)
        \right]
        \circ
        \left(\mathcal F\bm{\alpha}+\bm f\right)
    \right].
\end{aligned}
\label{eq:DEIM_online_reduced}
\end{equation}

The matrices $\mathcal D\in\mathbb R^{r\times m}$ and
$\mathcal E,\mathcal F\in\mathbb R^{m\times r}$, together with
the vectors $\bm e,\bm f\in\mathbb R^m$, are precomputed offline.
For each new parameter value, assembling
$\bm{\xi}_{\mathcal P}$ costs $\mathcal O(m)$ operations.
At each pseudo-time step, the two affine evaluations and the
final multiplication by $\mathcal D$ each require
$\mathcal O(mr)$ operations, while the componentwise
exponentials and products cost $\mathcal O(m)$.
The overall online optical-pumping evaluation thus costs
$\mathcal O(mr)$, independently of the full-order dimension.

\subsection{Computational complexity}
\label{sec:ROM_complexity}

% The complete offline workflow is summarized in Algorithm~\ref{alg:offline}, with the NQE and DEIM constructions detailed in Algorithms~\ref{alg:offline-nqe} and~\ref{alg:offline-deim}, respectively.
The ROM offers a computational advantage over the FOM by fundamentally decoupling the online evaluation from the high-fidelity dimension.
We analyze each stage individually and provide explicit complexity estimates to quantify computational savings.

\paragraph{Full-order model}
Each pseudo-time iteration of Algorithm~\ref{alg:FOM} consists entirely of sparse matrix-vector products and pointwise operations on vectors of length $N_h$.
Grouping of contributions:
\begin{equation}
    \underbrace{\mathcal{O}(N_h)}_{\text{diffusion}}
    + \underbrace{\mathcal{O}(N_h)}_{\text{SD, FD}}
    + \underbrace{\mathcal{O}(N_h)}_{\text{spin-exchange}}
    + \underbrace{\mathcal{O}(N_xN_yN_z)}_{\text{light propagation and optical pumping}}
    = \mathcal{O}(N_h) \quad \text{per iteration.}
\label{eq:fom-cost}
\end{equation}
All atomic-state operators are either local or nearest-neighbor stencil-based and therefore scale linearly with the number of spatial nodes. 
The light-propagation solver performs an axial sweep over $N_z$ nodes for every transverse coordinate $(x_i,y_j)$. 
Since there are $N_xN_y$ axial rays, one complete light-field update costs
\begin{equation*}
    \mathcal O(N_xN_yN_z)=\mathcal O(N_h).
\end{equation*}
Thus, both the atomic-state and the light-field updates scale linearly with $N_h$.
Hence, the total FOM cost over $N_{\mathrm{iter}}$ pseudo-time steps is
\begin{equation}
    \mathcal{C}_{\mathrm{FOM}} = \mathcal{O} \bigl(N_{\mathrm{iter}}\,N_h\bigr),
    \qquad 
    N_h \sim 10^6.
\end{equation}

\begin{table}%
\centering
\caption{Computational complexity of the full-order, offline, and online stages.
        Here $N_h$ is the number of spatial nodes ($8N_h$ FOM degrees of freedom), $N_\mu$ is the number of parameter samples and steady-state snapshots, $\bar N_{\mathrm{iter}}$ is the average FOM iteration count for snapshot generation, $r$ is the reduced dimension, and $m$ is the number of DEIM interpolation points.}
\label{tab:complexity}
\begin{tabular}{llc}
    \toprule
    Stage & Dominant operation & Cost \\
    \midrule
    FOM per step (Alg.~\ref{alg:FOM}) & Sparse operations on $\mathbb{R}^{8N_h}$ & $\mathcal{O}(N_h)$ per step \\
    Offline (Algs.~\ref{alg:offline-nqe}, \ref{alg:offline-deim} and \ref{alg:offline}) & High-fidelity snapshot generation & $\mathcal{O}(N_\mu\bar N_{\mathrm{iter}}N_h)$ (dominant) \\
    Online - linear terms (Alg.~\ref{alg:online}) & Reduced mat-vec on $\mathbb{R}^r$ & $\mathcal{O}(r^2)$ \\
    Online - spin-exchange (QE) & $\mathcal{A}(\bm{\alpha}\otimes\bm{\alpha})$ & $\mathcal{O}(r^3)$ \\
    Online - optical pumping (DEIM) & Reduced axial maps and DEIM projection & $\mathcal{O}(mr)$ \\
    \bottomrule
\end{tabular}
\end{table}

\paragraph{Offline phase}
The offline stage (Algorithm~\ref{alg:offline}) is executed \emph{once} for a given parameter domain. 
It comprises high-fidelity steady-state snapshot generation, state-snapshot assembly and POD basis construction, optical-pumping right-hand-side snapshot generation and DEIM basis construction, and the construction of the projected and hyperreduced operators.
The parameter samples $\{\bm\mu^{(i)}\}_{i=1}^{N_\mu}$ are first drawn by Latin hypercube sampling, and the FOM is marched to its converged steady state at every sample.
If $N_{\mathrm{iter}}^{(i)}$ denotes the number of pseudo-time iterations required at $\bm\mu^{(i)}$, then $\mathcal C_{\mathrm{FOM}}(\bm\mu^{(i)})=\mathcal O(N_{\mathrm{iter}}^{(i)}N_h)$ and the complete offline cost can be separated as
\begin{equation}
\begin{aligned}
    \mathcal C_{\mathrm{offline}}
    ={}&\underbrace{\sum_{i=1}^{N_\mu}
        \mathcal C_{\mathrm{FOM}}(\bm\mu^{(i)})}_{
        \text{HF snapshot generation}} 
    +\underbrace{\mathcal O(N_\mu N_h)}_{
        \text{snapshot assembly}}
    +\underbrace{\mathcal O(N_\mu^2N_h)}_{
        \operatorname{SVD}(\mathbb S)}
    +\underbrace{\mathcal O(N_\mu N_h)}_{
        \text{optical-pumping RHS snapshot generation}}
    +\underbrace{\mathcal O(N_\mu^2N_h)}_{
        \operatorname{SVD}(\mathbf S_{\mathrm{OP}})} \\
    &+\underbrace{\mathcal O(r^2N_h+r^3N_h)}_{
        \text{linear projections and QE construction}}
    +\underbrace{\mathcal O(m^2N_h+rmN_h+mN_zr)}_{
        \text{DEIM selection and reduced-map construction}}.
\end{aligned}
\label{eq:offline-cost}
\end{equation}
Writing $\bar N_{\mathrm{iter}}$ for the average number of FOM iterations, the first contribution scales as $\mathcal O(N_\mu\bar N_{\mathrm{iter}}N_h)$.
Since every stored state snapshot requires a complete high-fidelity steady-state solve, high-fidelity snapshot generation is the principal offline expense in the present workflow. 
Snapshot generation is therefore distinct from, and precedes, the construction of the POD basis. 
POD basis construction, optical-pumping RHS snapshot generation and DEIM basis construction, QE tensor construction, and DEIM interpolation-operator construction are subsequent one-time preprocessing costs; none of them includes the cost of generating the high-fidelity solution database.
Once the DEIM basis and interpolation points are determined, the online evaluation cost scales only with the reduced dimensions.
Consequently, the full-order dimension $8N_h$ does not enter the leading-order online cost, enabling substantial speedup in multi-query applications.

\paragraph{Online phase}
Once the offline operators are available, each pseudo-time iteration of Algorithm~\ref{alg:online} operates only in subspaces, without any full-grid operation:
\begin{equation}
\begin{aligned}
    \mathcal C_{\mathrm{online}}
    =\underbrace{\mathcal O(r^2)}_{\text{linear reduced operators}}
    +\underbrace{\mathcal O(r^3)}_{\mathcal A(\bm\alpha\otimes\bm\alpha)}
    +\underbrace{\mathcal O(r^2)}_{\mathcal B\bm\alpha+\mathcal C}
    +\underbrace{\mathcal O(mr)}_{\mathcal E\bm\alpha+\bm e}
    +\underbrace{\mathcal O(mr)}_{\mathcal F\bm\alpha+\bm f}
    +\underbrace{\mathcal O(mr)}_{\mathcal D\mathbf f_{\mathrm{OP},P}}
    =\mathcal O(r^3+mr).
\end{aligned}
\label{eq:online-cost}
\end{equation}
giving a total online cost of
\begin{equation}
    \mathcal{C}_{\mathrm{ROM}} =
    \mathcal{O} \bigl(N_{\mathrm{iter}}\,(r^3+mr)\bigr),
    \qquad r,m=10^1\text{--}10^2
    \quad\text{in the computations reported below}.
\end{equation}

\paragraph{Speedup estimate}
Assuming comparable pseudo-time iteration counts, for
$r, m \ll 8N_h$, the ratio of the ROM online cost to the FOM cost
satisfies
% Since $r, m \ll 8N_h$, the ratio of ROM online to FOM cost satisfies
\begin{equation}
    \frac{\mathcal C_{\mathrm{ROM}}}{\mathcal C_{\mathrm{FOM}}}
    =\mathcal O\!\left(\frac{r^3+mr}{N_h}\right)
    \ll1.
\label{eq:speedup}
\end{equation}

For fixed $r$ and $m$, \eqref{eq:speedup} shows that the online cost
is independent of the full-order dimension.
The practical speedup is reported from measured runtimes in Section~\ref{sec:ROM_efficiency}.
The complexity of each stage is summarized in Table~\ref{tab:complexity}.

\subsection{Numerical results}
\label{sec:ROM_results}

Three test cases are used to assess the accuracy and efficiency of the ROM.
% The test cases were constructed by Latin hypercube sampling (LHS) over the prescribed parameter ranges. 
For clarity, the numerical results are organized according to increasing parametric complexity.

\subsubsection{Case 1}
\label{sec:ROM_case_1}

In the first case, $\mathbb{D}$ and $\eta$ are selected as the varying parameters within the ranges summarized in Table~\ref{tab:ROM_case_1}.
The diffusion coefficient $\mathbb D$ enters the linear diffusion operator, whereas $\eta$ affects both the linear relaxation contribution and the quadratic spin-exchange contribution treated by QE.
The dependence on both parameters remains affine after projection; this case therefore provides the basic test of the ROM under diffusion and spin-exchange parameter variation before the nonlocal optical-pumping parameters are varied.
A total of 100 high-fidelity solutions were collected as snapshots.

\begin{table}
\centering
\caption{Parameter setting for ROM Case 1.}
\label{tab:ROM_case_1}
\begin{tabular}{clcc}
    \toprule
    Category & Parameter & Nominal value & Sampling range \\
    \midrule
    \multirow{2}{*}{Linear}
    & $\mathbb{D}$ & 0.01 & (0.008, 0.012) \\
    & $f_D$ & 0.0 & -- \\
    Mixed & $\eta$ & 150.0 & (40.0, 250.0) \\
    \midrule
    \multirow{3}{*}{Nonlinear}
    & $\mathbb{R}_{0}$ & 3.0 & -- \\
    & $\mathrm{OD}$ & 2.0 & -- \\
    & $w$ & 0.7 & -- \\
    \bottomrule
\end{tabular}
\end{table}

\begin{figure}
    \centering
    \begin{subfigure}[b]{0.45\textwidth}
        \includegraphics[width=1.0\linewidth]{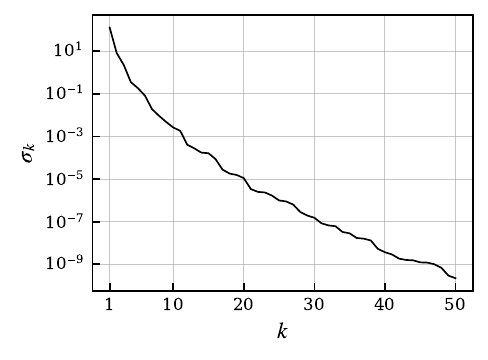}
        \subcaption{POD}
        \label{fig:ROM_POD_01}
    \end{subfigure}
    \hfill
    \begin{subfigure}[b]{0.45\textwidth}
        \includegraphics[width=1.0\linewidth]{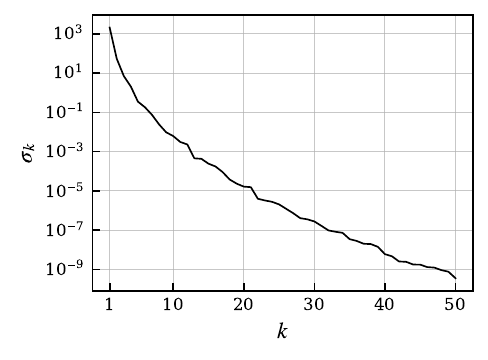}
        \subcaption{DEIM}
        \label{fig:ROM_DEIM_01}
    \end{subfigure}
    \caption{Singular-value decay of the snapshot matrix and the optical-pumping nonlinear snapshots for ROM Case~1.}
    \label{fig:ROM_SVD_01}
\end{figure}

Since the snapshot set is generated by LHS, the nominal parameter values listed in Table~\ref{tab:ROM_case_1} are generally not included in the training samples. 
Accordingly, the nominal parameter set is used here as an out-of-sample test point for ROM validation. 
The same setting is adopted in the remaining ROM cases and will not be repeated hereafter.

Figure~\ref{fig:ROM_SVD_01} shows the singular value decays of the snapshot matrix and the optical-pumping term snapshot matrix in Figs.~\ref{fig:ROM_POD_01} and~\ref{fig:ROM_DEIM_01}, respectively.
In both cases, the singular values decay rapidly, indicating that the high-fidelity solution and the nonlinear term can be represented accurately by low-dimensional bases. 
% This observation suggests that fewer than 10 POD or DEIM modes are already sufficient to achieve high approximation accuracy.
This observation suggests that both POD and DEIM can achieve high approximation accuracy with fewer than 10 modes.

\begin{figure}
    \centering
    \includegraphics[width=0.45\linewidth]{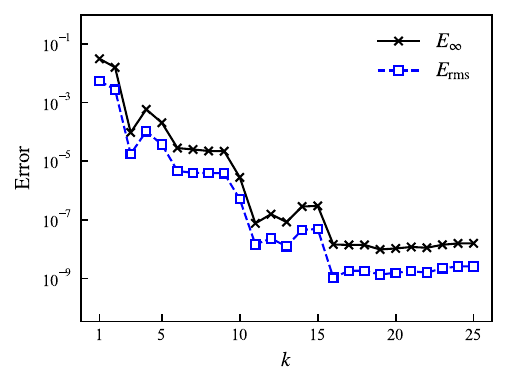}
    \caption{POD basis refinement for ROM Case~1 with $m=50$ DEIM modes.}
    \label{fig:ROM_error_01}
\end{figure}

As additional evidence for the basis selection and validation, Fig.~\ref{fig:ROM_error_01} presents a basis refinement study in which the maximum and average errors between the FOM and ROM solutions are monitored as the POD basis dimension increases. 
In this test, both the FOM and ROM solutions are computed at the nominal parameter values listed in Table~\ref{tab:ROM_case_1}. 
To reduce the influence of nonlinear approximation error, the DEIM basis dimension is fixed at $m = 50$. 
The results show that the ROM error decreases rapidly as the POD basis is enriched, then gradually levels off near the solvers' convergence tolerance, demonstrating the effectiveness of the reduced basis construction.

% The results show that the ROM error decreases rapidly as the POD basis is enriched and then gradually approaches an error level close to the convergence criterion in the solvers, confirming the effectiveness of the reduced basis construction.

\begin{figure}
    \centering
    \begin{subfigure}[b]{0.9\textwidth}
        \includegraphics[width=1.0\linewidth]{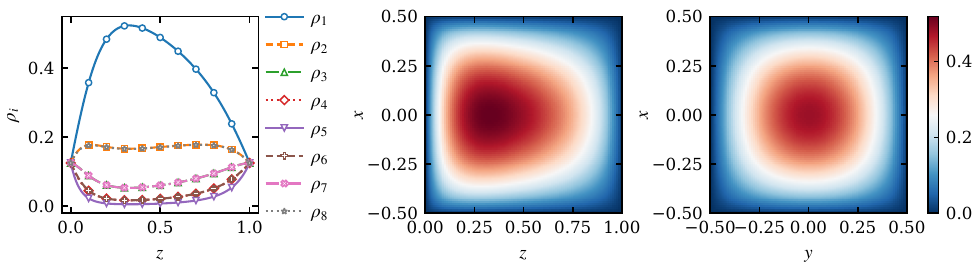}
        \subcaption{FOM}
        \label{fig:ROM_plotfom_01}
    \end{subfigure}
    \hfill
    \begin{subfigure}[b]{0.9\textwidth}
        \includegraphics[width=1.0\linewidth]{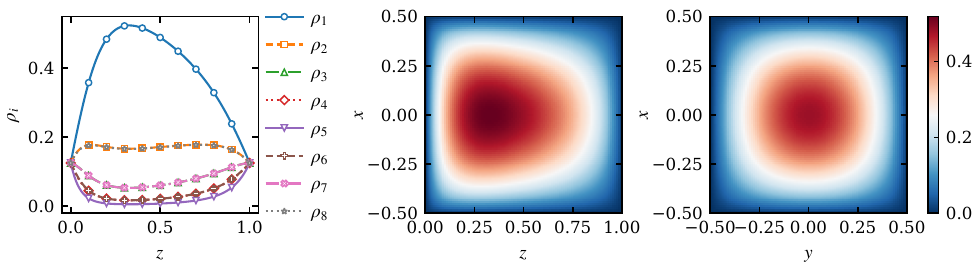}
        \subcaption{ROM}
        \label{fig:ROM_plotrom_01}
    \end{subfigure}
    \caption{Comparison of the FOM and ROM solutions for Case 1 at the nominal parameter values.}
    \label{fig:ROM_plot_01}
\end{figure}

Based on the singular value decay and the basis refinement results, we choose $r=10$ POD modes and $m=10$ DEIM modes for the subsequent ROM prediction and comparison.
Figure~\ref{fig:ROM_plot_01} compares the eight components $\rho_i$ and the polarization $P$ obtained by the FOM and ROM at the nominal parameter values; the FOM and ROM fields are shown in Figs.~\ref{fig:ROM_plotfom_01} and~\ref{fig:ROM_plotrom_01}, respectively.
Each visualization includes the centerline profiles of the eight population components $\rho_i$ along $(x,y)=(0,0)$ and the planar distributions of $P$ on the $x$--$z$ plane at $y=0$ and the $x$--$y$ plane at $z=0.5$.
The close agreement between the FOM and ROM confirms that the reduced model accurately reproduces the main solution features for this affine-parameter test case.
Unless otherwise stated, the same validation protocol, figure layout, and error measures are used for Cases~2 and~3 in Sections~\ref{sec:ROM_case_2} and~\ref{sec:ROM_case_3}, respectively. Repeated descriptions are omitted for brevity.
% Unless otherwise stated, the same validation protocol, figure layout, and error measures are used in Case 2 in Section \ref{sec:ROM_case_2} and Case 3 in Section \ref{sec:ROM_case_3}, and repeated descriptions are omitted for brevity.
The same basis dimensions are also used in the computational-efficiency study reported in Sec.~\ref{sec:ROM_efficiency}.

\subsubsection{Case 2}
\label{sec:ROM_case_2}

Case 2 examines the ROM under variations of parameters controlling the nonlocal optical-pumping nonlinearity.
In contrast to Case 1, $\mathbb D$, $\eta$, and the remaining physical parameters are held fixed, while the incident optical-pumping rate $\mathbb R_0$, optical depth $\mathrm{OD}$ and beam width $w$ are varied over the ranges listed in Table~\ref{tab:ROM_case_2}.
These parameters directly control the magnitude and transverse distribution of the spatially varying optical-pumping contribution; this case therefore assesses the ability of the DEIM  approximation of this nonlocal nonlinear dynamics.
The same validation protocol as in Case 1 is adopted, and only the parameter setting and main observations are reported below.
A total of 200 high-fidelity solutions were collected as snapshots.

\begin{table}
\centering
\caption{Parameter setting for ROM Case 2.}
\label{tab:ROM_case_2}
\setlength{\tabcolsep}{8pt}
\begin{tabular}{clcc}
    \toprule
    Category & Parameter & Nominal value & Sampling range \\
    \midrule
    \multirow{2}{*}{Linear}
    & $\mathbb{D}$ & 0.0013 & -- \\
    & $f_D$ & 0.6 & -- \\
    Mixed & $\eta$ & 1.0 & -- \\
    \midrule
    \multirow{3}{*}{Nonlinear}
    & $\mathbb{R}_{0}$ & 2.2 & $(0.0,\,3.0)$ \\
    & $\mathrm{OD}$ & 0.75 & $(0.5,\, 1.0)$ \\
    & $w$ & 0.6 & $(0.5,\,1.0)$ \\
    \bottomrule
\end{tabular}
\end{table}

\begin{figure}
    \centering
    \begin{subfigure}[b]{0.45\textwidth}
        \includegraphics[width=1.0\linewidth]{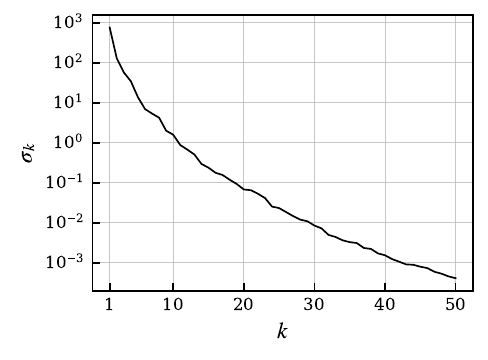}
        \subcaption{POD}
        \label{fig:ROM_POD_02}
    \end{subfigure}
    \hfill
    \begin{subfigure}[b]{0.45\textwidth}
        \includegraphics[width=1.0\linewidth]{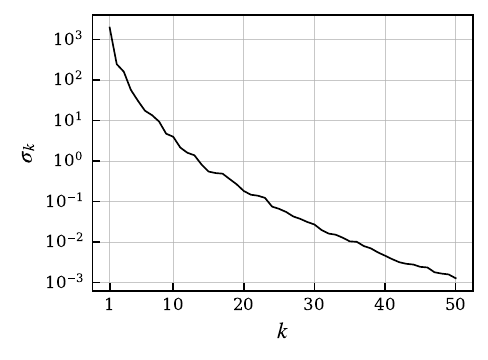}
        \subcaption{DEIM}
        \label{fig:ROM_DEIM_02}
    \end{subfigure}
    \caption{Singular-value decay of the snapshot matrix and the optical-pumping nonlinear snapshots for ROM Case~2.}
    \label{fig:ROM_SVD_02}
\end{figure}

Figure~\ref{fig:ROM_SVD_02} shows that the singular values of both the state (Fig.~\ref{fig:ROM_POD_02}) and optical-pumping term (Fig.~\ref{fig:ROM_DEIM_02}) snapshot matrices decay rapidly, although a slightly larger basis is required to reach the same error level as in Case 1.

\begin{figure}
    \centering
    \includegraphics[width=0.45\linewidth]{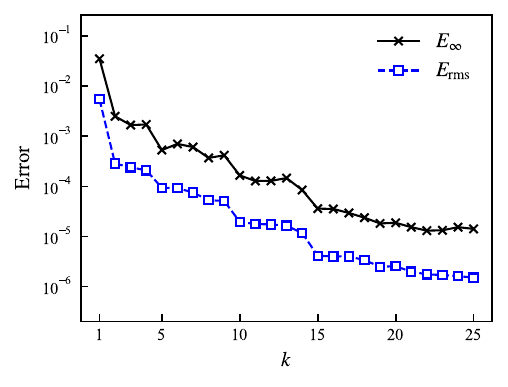}
    \caption{POD basis refinement for ROM Case~2 with $m=50$ DEIM modes.}
    \label{fig:ROM_error_02}
\end{figure}

Figure~\ref{fig:ROM_error_02} reports the corresponding basis-refinement study. To reduce the influence of nonlinear approximation error, the DEIM basis dimension is fixed at $m = 50$.

\begin{figure}
    \centering
    \begin{subfigure}[b]{0.9\textwidth}
        \includegraphics[width=1.0\linewidth]{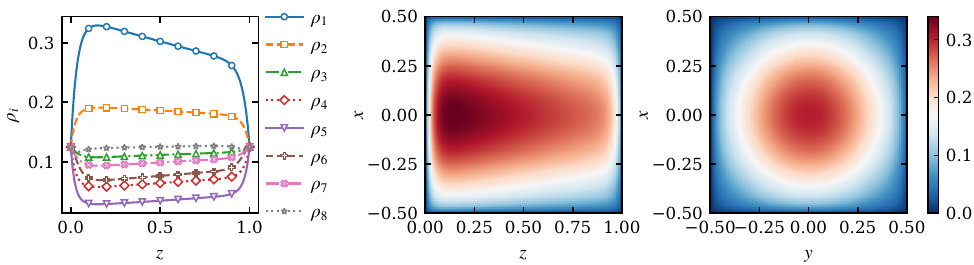}
        \subcaption{FOM}
        \label{fig:ROM_plotfom_02}
    \end{subfigure}
    \hfill
    \begin{subfigure}[b]{0.9\textwidth}
        \includegraphics[width=1.0\linewidth]{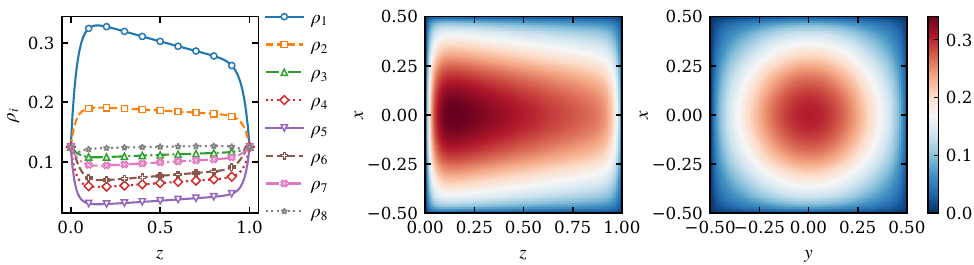}
        \subcaption{ROM}
        \label{fig:ROM_plotrom_02}
    \end{subfigure}
    \caption{Comparison of the FOM and ROM solutions for Case 2 at the nominal parameter values.}
    \label{fig:ROM_plot_02}
\end{figure}

Based on the singular-value decay and the basis refinement results, we choose $r=10$ POD modes and $m=15$ DEIM modes for the subsequent ROM prediction.
Figure~\ref{fig:ROM_plot_02} compares the FOM and ROM solutions in Figs.~\ref{fig:ROM_plotfom_02} and~\ref{fig:ROM_plotrom_02}, respectively, at the nominal parameter values in Table~\ref{tab:ROM_case_2}.
The ROM accurately reproduces the state components and polarization field obtained by the FOM for this three-parameter nonlinear test.

\subsubsection{Case 3}
\label{sec:ROM_case_3}

Case 3 considers the simultaneous variation of the two parameter groups examined separately in Cases 1 and 2.
Specifically, $\mathbb D$ and $\eta$ vary the diffusion and spin-exchange contributions, while $\mathbb R_0$ and $w$ control the nonlocal optical-pumping contribution; the corresponding ranges are listed in Table~\ref{tab:ROM_case_3}.
This mixed setting combines affine variation in the diffusion and spin-exchange parameters with parameter dependence in the optical-pumping nonlinearity and therefore provides a more demanding test of the robustness of the reduced model over the coupled parameter space.
The same validation protocol as in Case 1 is adopted, and only the parameter setting and main observations are reported below.
A total of 300 high-fidelity solutions were collected as snapshots.

\begin{table}
\centering
\caption{Parameter setting for ROM Case 3.}
\label{tab:ROM_case_3}
\setlength{\tabcolsep}{8pt}
\begin{tabular}{clcc}
    \toprule
    Category & Parameter & Nominal value & Sampling range \\
    \midrule
    \multirow{2}{*}{Linear} & $\mathbb{D}$ & 0.01 &  $(0.008, 0.012)$ \\
    & $f_D$ & 0.6 & --- \\
    Mixed & $\eta$ & 120.0 & $(40, 250)$ \\
    \midrule
    \multirow{3}{*}{Nonlinear}
    & $\mathbb{R}_{0}$ & 2.1 & $(0.0, 3.0)$ \\
    & $\mathrm{OD}$ & 2.0 & --- \\
    & $w$ & 0.7 & $(0.2, 0.8)$ \\
    \bottomrule
\end{tabular}
\end{table}

\begin{figure}
    \centering
    \begin{subfigure}[b]{0.45\textwidth}
        \includegraphics[width=1.0\linewidth]{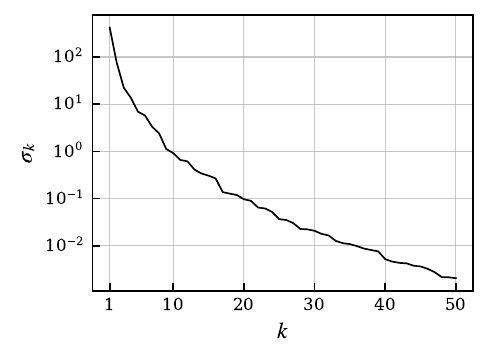}
        \subcaption{POD}
        \label{fig:ROM_POD_03}
    \end{subfigure}
    \hfill
    \begin{subfigure}[b]{0.45\textwidth}
        \includegraphics[width=1.0\linewidth]{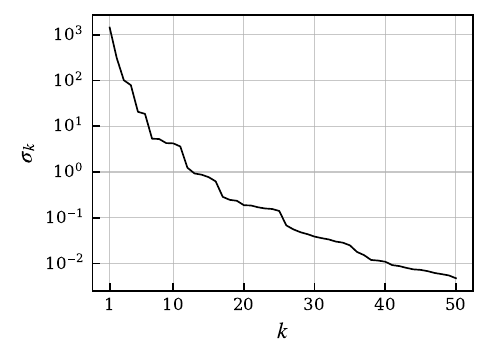}
        \subcaption{DEIM}
        \label{fig:ROM_DEIM_03}
    \end{subfigure}
    \caption{Singular-value decay of the snapshot matrix and the optical-pumping nonlinear snapshots for ROM Case~3.}
    \label{fig:ROM_SVD_03}
\end{figure}

Figure~\ref{fig:ROM_SVD_03} shows that the singular values of both the state (Fig.~\ref{fig:ROM_POD_03}) and optical-pumping term (Fig.~\ref{fig:ROM_DEIM_03}) snapshot matrices decay rapidly, although a larger basis is required to reach the same error level as in Case 1.

\begin{figure}
    \centering
    \includegraphics[width=0.45\linewidth]{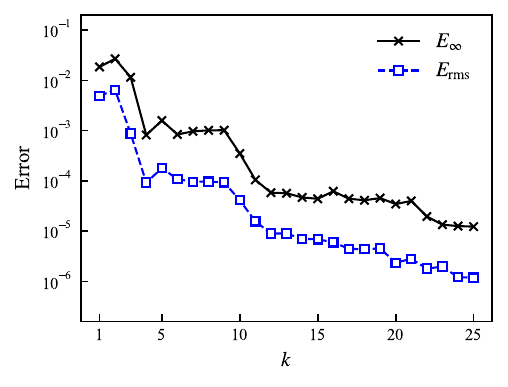}
    \caption{POD basis refinement for ROM Case~3 with $m=50$ DEIM modes.}
    \label{fig:ROM_error_03}
\end{figure}

Figure~\ref{fig:ROM_error_03} reports the corresponding basis-refinement study. To reduce the influence of nonlinear approximation error, the DEIM basis dimension is fixed at $m = 50$.

\begin{figure}
    \centering
    \begin{subfigure}[b]{0.9\textwidth}
        \includegraphics[width=1.0\linewidth]{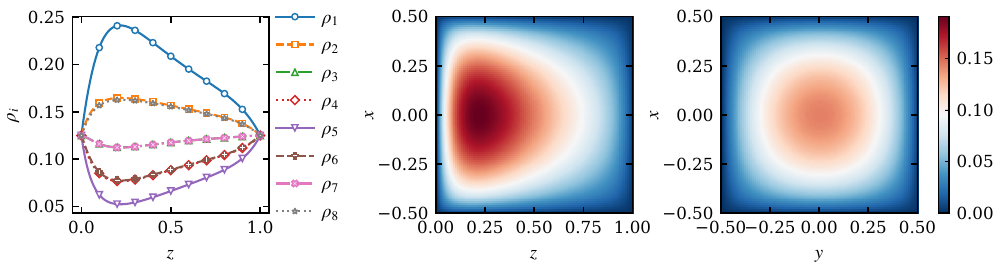}
        \subcaption{FOM}
        \label{fig:ROM_plotfom_03}
    \end{subfigure}
    \hfill
    \begin{subfigure}[b]{0.9\textwidth}
        \includegraphics[width=1.0\linewidth]{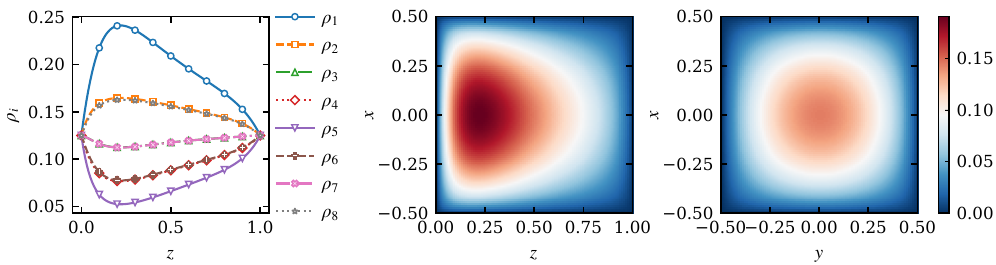}
        \subcaption{ROM}
        \label{fig:ROM_plotrom_03}
    \end{subfigure}
    \caption{Comparison of the FOM and ROM solutions for Case 3 at the nominal parameter values.}
    \label{fig:ROM_plot_03}
\end{figure}

Based on the singular-value decay and the refinement results, we choose $r=10$ POD modes and $m=20$ DEIM modes for the subsequent ROM prediction.
Figure~\ref{fig:ROM_plot_03} compares the FOM and ROM fields shown in Figs.~\ref{fig:ROM_plotfom_03} and~\ref{fig:ROM_plotrom_03}, respectively, for the simultaneous variation of $\mathbb D$, $\eta$, $\mathbb R_0$, and $w$.
The comparison shows that the ROM accurately capture the reported state and polarization fields for this mixed linear--nonlinear parameter case.

\subsection{Computational efficiency}
\label{sec:ROM_efficiency}

Table~\ref{tab:ROM_efficiency} summarizes the computational costs of the FOM and ROM. All simulations were performed on the Tianhe cluster using an NVIDIA A100 GPU or an Intel Xeon Gold 6354 CPU. All CPU runs used a single core. For the low-dimensional ROMs considered here, GPU acceleration is not expected to be beneficial for an individual online solve, as the overheads associated with CUDA initialization, kernel launches, and data transfers can be comparable to or exceed the computational cost of the solve. Accordingly, ROM timings are reported only for CPU execution. Speedup is calculated from the total wall-clock times, with the CPU-based FOM serving as the reference.
% We summarize the computational cost of the FOM and ROM on different computing devices in Table~\ref{tab:ROM_efficiency}. All the simulations in this work are performed on the Tianhe cluster.
% The GPU used in this study is an NVIDIA A100 80GB PCIe, while the CPU is an Intel Xeon Gold 6354 running at 3.00 GHz.
% All CPU tests were performed using a single core.
% For the low-dimensional ROMs considered here, GPU execution of an individual online solve is not expected to be advantageous, as CUDA initialization, kernel-launch, and data-transfer overheads can become comparable to or larger than the online computational cost of the ROM.
% Therefore, only CPU timings are reported for the ROM.
% The speedup is defined based on the total wall-clock time, using the CPU FOM as the reference.

The GPU acceleration and model reduction reported here operate at two distinct levels.
GPU execution provides hardware-level acceleration of the FOM without changing its dimension or mathematical formulation, whereas the ROM provides model-level acceleration by replacing the $8N_h$-dimensional FOM with a low-dimensional ROM.
These mechanisms are therefore complementary rather than mutually exclusive and could, in principle, be combined, for example in batched ROM evaluations or larger-scale online tasks.

The wall-clock times reported in Table~\ref{tab:ROM_efficiency} highlight the substantial differences in computational cost.
Typical FOM simulations take hours on a CPU and minutes on a GPU, while ROM evaluations take seconds.
Relative to the CPU-based FOM, the GPU-accelerated FOM achieves a speedup of approximately $2\times10^2$, and the ROM achieves a speedup exceeding $7\times10^3$.
Hardware acceleration and model reduction offer complementary benefits, suggesting that their combination could further accelerate repeated-query applications.
% The difference in computational scale is evident from the wall-clock times reported in Table~\ref{tab:ROM_efficiency}.
% Typical CPU FOM computations require hours, GPU-accelerated FOM computations require minutes, whereas the ROM completes in seconds.
% The GPU provides a speedup of approximately $2 \times10^2$ relative to the CPU FOM, whereas the ROM achieves over $7\times10^3$.
% Thus, hardware acceleration and model reduction address different sources of computational cost and together provide a potential route toward further acceleration in repeated-query applications.

\begin{table}
    \centering
    \caption{Computational performance comparison of the FOM and ROM.}
    \label{tab:ROM_efficiency}
    % \footnotesize
    \setlength{\tabcolsep}{3pt}
    \begin{tabular}{@{}ccccccccc@{}}
    \toprule
    % Case & Model  & Device& POD modes & DEIM modes & Time (s) & Iterations & Final residual & Speedup \\
    Case & Model  & Device& POD modes & DEIM modes & Time (s) & Iterations & Final increment & Speedup \\
    \midrule
    \multirow{3}{*}{1} & \multirow{2}{*}{FOM} 
        & CPU & -- & -- & 19599.493 & 25800 & 9.123E-9 & -- \\
        && GPU & -- & -- & 118.046 & 25800 & 9.123E-9 & 166.033 \\
        & ROM & CPU & 10 & 10  & 2.728 & 25200 & 9.939E-9 & 7184.565 \\
    \midrule
    \multirow{3}{*}{2} & \multirow{2}{*}{FOM} 
        & CPU & -- & -- & 2260.083 & 2200 & 5.924E-9  & -- \\
        && GPU & -- & -- & 10.018 & 2200 & 5.924E-9 & 225.602 \\
        & ROM & CPU & 10 & 15 & 0.300 & 1800 & 4.482E-9  & 7533.610 \\
    \midrule
    \multirow{3}{*}{3} & \multirow{2}{*}{FOM} 
        & CPU & -- & -- & 11380.310 & 13400 & 9.386E-9 & -- \\
        && GPU & -- & -- & 61.323 & 13400 & 9.386E-9 & 185.580 \\
        & ROM & CPU & 10 & 20 & 1.576 & 14000 & 7.992E-9 & 7221.009 \\
    % \midrule
    % \multirow{3}{*}{4} & \multirow{2}{*}{FOM} 
    %     & CPU & -- & -- & 1225.646 & 2400 & 1.613e-09 & -- \\
    %     && GPU & -- & -- & 9.185 & 2400 & 1.613e-09 & 133.44 \\
    %     & ROM & CPU & 20 & 20 & 0.885 & 9200 & 7.933e-09 & 6232.10 \\
    \bottomrule
    \end{tabular}
\end{table}

% !TeX root = main.tex

\section{Conclusion}
\label{sec:conclusion}

This work presents a computational framework for the coupled
DQME--light system.
The high-fidelity solver combines finite-difference discretization
and explicit pseudo-time marching for the DQME with an exponential
integrator for light propagation.
The resulting scheme is structure-preserving under the derived
restriction on the pseudo-time step and is well suited to tensorized
GPU implementation.
Three complementary verification cases demonstrate agreement with
analytical predictions, asymptotic limits, and an analytical invariant.
The GPU implementation achieves a speedup of more than two orders
of magnitude over a single-core CPU implementation.

To accelerate repeated simulations, we develop an intrusive
POD--Galerkin ROM based on an integro-differential formulation
involving only the atomic populations.
Exact quadratic expansion of the spin-exchange term and DEIM
approximation of the optical-pumping term enable efficient online
evaluation.
Numerical tests show that the ROM closely reproduces the FOM
solutions at parameter points outside the training set while
reducing the online computational cost by more than three orders
of magnitude.

An immediate extension is to the quasi-high-pressure (QHP) regime,
in which the distinct contributions of the resolved ground-state
hyperfine manifolds modify both the optical-pumping operator and
the state-dependent light
absorption~\cite{yan2026opticalpumpingalkalimetalvapor}.
The existing FOM and ROM architectures provide a foundation for
this extension, while the structure-preserving analysis and
hyperreduced optical-pumping formulation will be adapted to the
QHP operators.
Future work will also consider more general external-field
configurations and adaptive model-reduction strategies across
a broader range of operating conditions.

\section*{Code and data availability}
The code used in this work is publicly available at \url{https://github.com/xukeyu-phy/DME_ROM-Collab}. 
Data supporting the findings of this study can be generated using the code and are also available from the corresponding author upon reasonable request.
% All the codes are available at \url{https://github.com/xukeyu-phy/DME_ROM-Collab}.
% Data will be made available on request.

\section*{Acknowledgments}
This work was supported by the National Natural Science Foundation of China (Grant No. 12372284), the Quantum Science and Technology–National Science and Technology Major Project (Grant No. 2025ZD0300802), and the Science Challenge Project (Grant No. TZ2025017).
% This work was supported by the National Natural Science Foundation of China (Grant No. 12372284), Quantum Science and Technology--National Science and Technology Major Project (Grant No. 2025ZD0300802), and Science Challenge Project (Grant No. TZ2025017).
% The authors also acknowledge the National Supercomputer Center in Tianjin for providing GPU computing resources.
% !TeX root = main.tex

% Start of Appendix
\appendix

\section{Explicit superoperators for ${}^{87}\mathrm{Rb}$ and structural properties of the reaction operators}
\label{appendix-sec:evolutionmatrix}

% Under the longitudinal secular approximation adopted in Section~\ref{sec:FOM_govering}, the atomic state is restricted to the eight ground-state populations of ${}^{87}\mathrm{Rb}$ ($I=3/2$). 
Following the population-subspace formulation in Section~\ref{sec:FOM_govering}, the atomic state is restricted to the eight ground-state populations of ${}^{87}\mathrm{Rb}$ ($I=3/2$), spanning the two hyperfine multiplets $F=1$ and $F=2$.
% The population vector is the column vector
% \[
%     \bm\rho=
%     \bigl(\rho_{2,2},\rho_{2,1},\rho_{2,0},\rho_{2,-1},\rho_{2,-2},
%     \rho_{1,-1},\rho_{1,0},\rho_{1,1}\bigr)^\top,
%     \qquad
%     \rho_{F,m_F}=\langle F,m_F|\hat\rho|F,m_F\rangle .
% \]
% The population vector is the column vector
% \[
%     \bm\rho = (\rho_q)_{q=1}^8
%     =\bigl(\rho_{2,2},\rho_{2,1},\rho_{2,0},\rho_{2,-1},\rho_{2,-2},
%     \rho_{1,-1},\rho_{1,0},\rho_{1,1}\bigr)^\top,
%     \qquad
%     \rho_{F,m_F}=\langle F,m_F|\hat\rho|F,m_F\rangle .
% \]
% Here, $q=1,\ldots,8$ indexes the populations in the specified order.
he population vector is ordered as
\[
    \bm\rho = (\rho_q)_{q=1}^8
    =\bigl(
    \rho_{2,2},\rho_{2,1},\rho_{2,0},\rho_{2,-1},
    \rho_{2,-2},\rho_{1,-1},\rho_{1,0},\rho_{1,1}
    \bigr)^\top,
\]
where $\rho_{F,m_F}=\langle F,m_F|\hat\rho|F,m_F\rangle$
and $q=1,\ldots,8$ indexes the populations in the specified order.
Let $S_z$ denote the electron-spin operator (in units of $\hbar$).
The diagonal matrix elements of \(S_z\) define the vector \(\mathbf S_z\in\mathbb R^8\), with \((\mathbf S_z)_q=\langle F,m_F|S_z|F,m_F\rangle\) for the state indexed by \(q\). In the specified population ordering,
\[
\mathbf S_z^{\top} =
\begin{pmatrix}
\frac{1}{2} & \frac{1}{4} & 0 & -\frac{1}{4} &
-\frac{1}{2} & \frac{1}{4} & 0 & -\frac{1}{4}
\end{pmatrix}.
\]
The electron-spin expectation value is then
\[
    \langle S_z\rangle
    =\operatorname{Tr}(\hat\rho S_z)
    =\mathbf S_z^\top\bm\rho.
\]
The components of $\mathbf S_z$ satisfy
$|(\mathbf S_z)_q|\leq 1/2$ for $q=1,\ldots,8$.
This bound is used in Section~\ref{sec:superoperators_structure}
to establish $|P|\leq1$ for physically admissible populations.
% In particular,
% \[
%     -\frac12
%     \leq
%     (\mathbf S_z)_q
%     \leq
%     \frac12,
%     \qquad q=1,\ldots,8,
% \]
% which is used in Section~\ref{sec:superoperators_structure} to establish the physical polarization bound.

The reaction operators below act on $\bm\rho$ from the left and enter the DQME with an overall minus sign. 
% The matrices below act on this vector from the left. 
% In the DQME, the reaction contribution is written with an overall minus sign, so the matrices have nonpositive off-diagonal entries and zero column sums. 
% In the DQME, the reaction contribution is written with an overall minus sign.
% The matrices $A_{\mathrm{SD}}$, $A_{\mathrm{FD}}$, and $A_{\mathrm{OP}}$
% have nonpositive off-diagonal entries, and all four matrices have zero
% column sums.
% This convention is retained here to match the implementation and the structural analysis in Section~\ref{sec:superoperators_structure}. 
We report the explicit matrices required for reproducing the computation; the intermediate angular-momentum reduction is not repeated.

The spin-destruction operator is
\[
A_{\mathrm{SD}} = \frac{1}{16}
% \resizebox{\textwidth}{!}{$
\begin{pmatrix}
8 & -2 & 0 & 0 & 0 & 0 & 0 & -6 \\
-2 & 11 & -3 & 0 & 0 & 0 & -3 & -3 \\
0 & -3 & 12 & -3 & 0 & -1 & -4 & -1 \\
0 & 0 & -3 & 11 & -2 & -3 & -3 & 0 \\
0 & 0 & 0 & -2 & 8 & -6 & 0 & 0 \\
0 & 0 & -1 & -3 & -6 & 11 & -1 & 0 \\
0 & -3 & -4 & -3 & 0 & -1 & 12 & -1 \\
-6 & -3 & -1 & 0 & 0 & 0 & -1 & 11
\end{pmatrix}
% $}
\]

The F-destruction operator is
\[
A_{\mathrm{FD}} =
% \resizebox{\textwidth}{!}{$
\begin{pmatrix}
2 & -2 & 0 & 0 & 0 & 0 & 0 & 0 \\
-2 & 5 & -3 & 0 & 0 & 0 & 0 & 0 \\
0 & -3 & 6 & -3 & 0 & 0 & 0 & 0 \\
0 & 0 & -3 & 5 & -2 & 0 & 0 & 0 \\
0 & 0 & 0 & -2 & 2 & 0 & 0 & 0 \\
0 & 0 & 0 & 0 & 0 & 1 & -1 & 0 \\
0 & 0 & 0 & 0 & 0 & -1 & 2 & -1 \\
0 & 0 & 0 & 0 & 0 & 0 & -1 & 1
\end{pmatrix}
% $}
\]

The optical-pumping operator is
\[
A_{\mathrm{OP}} =
% \resizebox{\textwidth}{!}{$
\begin{pmatrix}
0 & -\frac{1}{4} & 0 & 0 & 0 & 0 & 0 & -\frac{3}{4} \\[0.5em]
0 & \frac{7}{16} & -\frac{3}{8} & 0 & 0 & 0 & -\frac{3}{8} & -\frac{3}{16} \\[0.5em]
0 & 0 & \frac{3}{4} & -\frac{3}{8} & 0 & -\frac{1}{8} & -\frac{1}{4} & 0 \\[0.5em]
0 & 0 & 0 & \frac{15}{16} & -\frac{1}{4} & -\frac{3}{16} & 0 & 0 \\[0.5em]
0 & 0 & 0 & 0 & 1 & 0 & 0 & 0 \\[0.5em]
0 & 0 & 0 & -\frac{3}{16} & -\frac{3}{4} & \frac{7}{16} & 0 & 0 \\[0.5em]
0 & 0 & -\frac{1}{4} & -\frac{3}{8} & 0 & -\frac{1}{8} & \frac{3}{4} & 0 \\[0.5em]
0 & -\frac{3}{16} & -\frac{1}{8} & 0 & 0 & 0 & -\frac{1}{8} & \frac{15}{16}
\end{pmatrix}
\begin{pmatrix}
Q_a & 0 & 0 & 0 & 0 & 0 & 0 & 0 \\
0 & Q_a & 0 & 0 & 0 & 0 & 0 & 0 \\
0 & 0 & Q_a & 0 & 0 & 0 & 0 & 0 \\
0 & 0 & 0 & Q_a & 0 & 0 & 0 & 0 \\
0 & 0 & 0 & 0 & Q_a & 0 & 0 & 0 \\
0 & 0 & 0 & 0 & 0 & Q_b & 0 & 0 \\
0 & 0 & 0 & 0 & 0 & 0 & Q_b & 0 \\
0 & 0 & 0 & 0 & 0 & 0 & 0 & Q_b
\end{pmatrix}
% $}
\]

This fixed matrix is the HP-limit form corresponding to $Q_a=Q_b=1$, as specified in Section~\ref{sec:FOM_govering}; these coefficients are not treated as model parameters in this work.

The longitudinal spin-exchange operator is
\[
A_{\mathrm{SE}} ^{(z)} = \frac{1}{8}
% \resizebox{\textwidth}{!}{$
\begin{pmatrix}
8 & 2 & 0 & 0 & 0 & 0 & 0 & 6 \\
-2 & 4 & 3 & 0 & 0 & 0 & 3 & 0 \\
0 & -3 & 0 & 3 & 0 & 1 & 0 & -1 \\
0 & 0 & -3 & -4 & 2 & 0 & -3 & 0 \\
0 & 0 & 0 & -2 & -8 & -6 & 0 & 0 \\
0 & 0 & -1 & 0 & 6 & 4 & -1 & 0 \\
0 & -3 & 0 & 3 & 0 & 1 & 0 & -1 \\
-6 & 0 & 1 & 0 & 0 & 0 & 1 & -4
\end{pmatrix}
% $}
\]

% In particular,$|(\mathbf S_z)_q|\leq 1/2$ for $q=1,\ldots,8$,
% which is used in Section~\ref{sec:superoperators_structure} to establish the physical polarization bound.

The explicit matrices satisfy the zero-column-sum
identities
\begin{equation}
    \mathbf1^\top A_{\mathrm{SD}}
    =
    \mathbf1^\top A_{\mathrm{FD}}
    =
    \mathbf1^\top A_{\mathrm{OP}}
    =
    \mathbf1^\top A_{\mathrm{SE}}^{(z)}
    =
    \mathbf0^\top.
\label{eq:appendix_column_sums}
\end{equation}
These identities imply the column-conservation property~\eqref{eq:B_column_conservation} of the local reaction matrix.

Furthermore, the off-diagonal entries satisfy
\begin{equation}
    (A_{\mathrm{SD}})_{ij}\leq0,
    \qquad
    (A_{\mathrm{FD}})_{ij}\leq0,
    \qquad
    (A_{\mathrm{OP}})_{ij}\leq0,
    \qquad i\neq j.
\label{eq:appendix_offdiagonal_signs}
\end{equation}
The spin-exchange and spin-destruction operators also satisfy
\begin{equation}
    \left|
        (A_{\mathrm{SE}}^{(z)})_{ij}
    \right|
    \leq
    -2(A_{\mathrm{SD}})_{ij},
    \qquad i\neq j.
\label{eq:appendix_SD_SE_bound}
\end{equation}
These properties are used in Lemma~\ref{lem:frozen_reaction_structure} to establish the sign structure of the frozen
reaction matrix.

\section{Mesh refinement study}
\label{appendix-sec:gridrefinement}

The required spatial resolution is primarily determined by the boundary-layer thickness. 
According to the scaling relation for the governing equation, the boundary-layer thickness is proportional to $\sqrt{\mathbb{D}}$. 
This estimation provides a practical criterion for choosing the grid spacing and, consequently, for assessing whether a given mesh is sufficiently fine for the simulations.

In each spatial direction, the computational domain is discretized into $N$ uniform cells, which correspond to $N+1$ grid nodes and $N-1$ interior nodes. 
Unless otherwise stated, the mesh resolution is specified by the number of cells per direction. 
For example, a mesh with $N=100$ cells in each direction contains 101 grid nodes and 99 interior nodes per direction.

For the production parameter range, a mesh with $N=100$ cells per direction is sufficient for the resolved bulk quantities considered here. It does not fully resolve the boundary-adjacent value in the
extreme case $\mathbb D=10^{-6}$; that case is therefore used only to assess agreement away from the first boundary-adjacent grid point.
To verify this choice, we perform a mesh refinement study.

\begin{figure}[pos=H]
    \centering
    \includegraphics[width=0.9\textwidth]{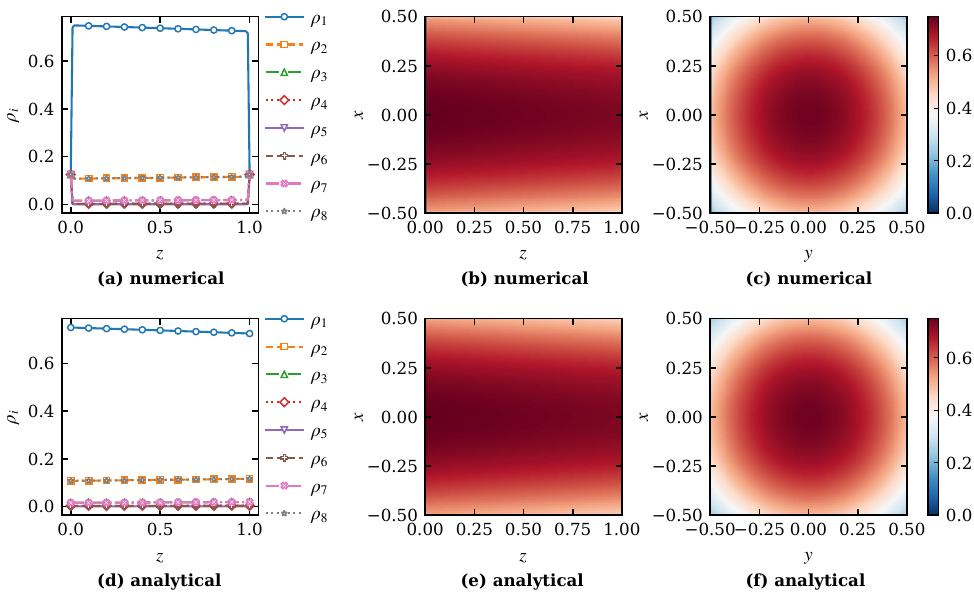}
    \caption{Comparison between the numerical and analytical solutions for $\mathbb{D}=10^{-6}$.}
    \label{fig:Appendix_num_ana_compare}
\end{figure}

Unless otherwise stated, the baseline parameters are fixed as $\eta = 1.0$, $f_D = 0.0$, and $w = 0.5$.
We first consider the most challenging case with an extremely small diffusion coefficient, $\mathbb{D} = 10^{-6}$, for which the boundary layer is particularly thin and the solution exhibits a very steep gradient near the boundary. 
At a grid resolution of $N=100$ cells per direction, the boundary-adjacent value is not fully resolved. 
However, as shown in Fig.~\ref{fig:Appendix_num_ana_compare}, the numerical solution agrees well with the analytical profile away from the grid point closest to the boundary. 
The numerical and analytical $P$ profiles are visually indistinguishable at the scale shown.

\begin{figure}[pos=H]
    \centering
    \includegraphics[width=0.85\textwidth]{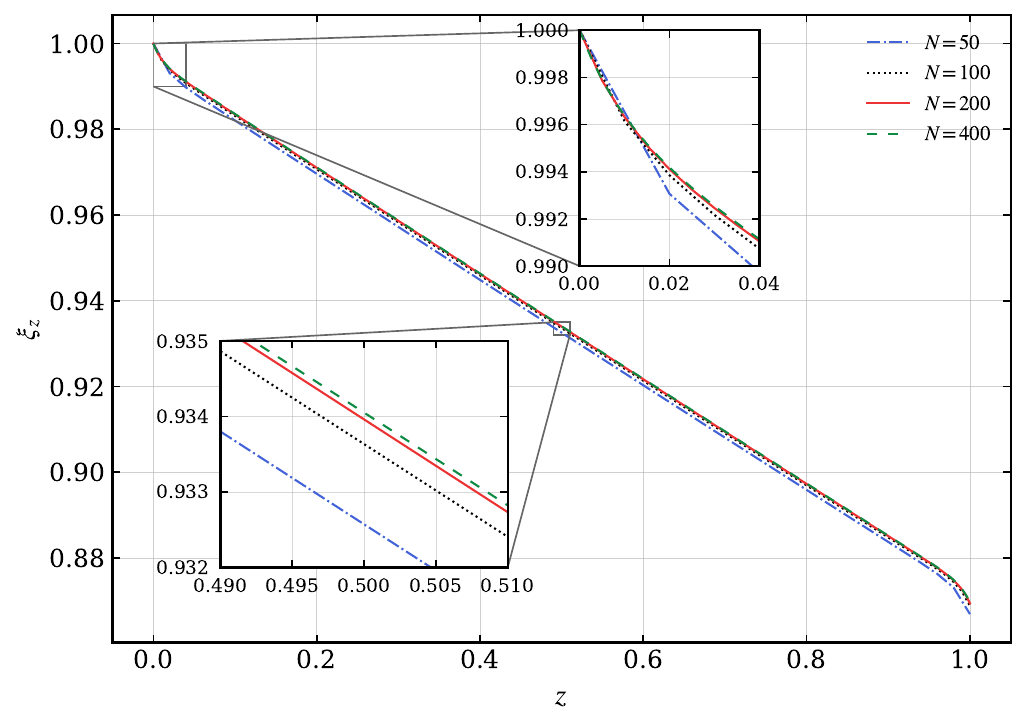}
    \caption{Mesh refinement study for $\mathbb{D}=10^{-4}$. The centerline profiles of $\xi_z$ along $(x,y)=(0,0)$ are shown for different grid resolutions.}
    \label{fig:Appendix_BC_refine}
\end{figure}

We next consider $\mathbb{D}=10^{-4}$, which corresponds to the lower bound of the physically relevant range studied in this work. 
For this case, we use $N=50$, $100$, $200$, and $400$ uniform cells in each spatial direction. 
These meshes correspond to 49, 99, 199, and 399 interior nodes per direction, respectively. 
Figure~\ref{fig:Appendix_BC_refine} shows the centerline profile of $\xi_z$ along $(x,y)=(0,0)$ for these resolutions.
As the mesh is refined, the profiles approach the $N=400$ finest-grid reference. This behavior indicates convergence toward a grid-independent numerical solution.

Using the finest grid solution as the reference, we further quantify the discretization error on the interior nodes. 
Table~\ref{tab:grid_refine_error} reports the $L_\infty$, $L_2$, and relative $L_2$ errors for both $\xi_z$ and $P$. 
The errors decrease consistently as the mesh is refined, providing quantitative evidence of grid convergence for the $\mathbb D=10^{-4}$ test.

\begin{table}
\centering
\caption{Mesh refinement errors for $\xi_z$ and $P$, where the solution on the finest grid with $400^3$ cells ($399^3$ interior nodes) is used as the reference.}
\label{tab:grid_refine_error}
\setlength{\tabcolsep}{3pt}
\renewcommand{\arraystretch}{1.15}
\begin{tabular}{c ccc ccc}
\toprule
\multirow{2}{*}{Cells}
& \multicolumn{3}{c}{$\xi_z$} 
& \multicolumn{3}{c}{$P$} \\
\cmidrule(lr){2-4} \cmidrule(lr){5-7}

& $E_\infty$ & $E_\mathrm{rms}$ & $E_{2,\mathrm{rel}}$
& $E_\infty$ & $E_\mathrm{rms}$ & $E_{2,\mathrm{rel}}$ \\
\midrule
 $50^3$
& $3.809 \times 10^{-3}$ & $1.291 \times 10^{-3}$ & $1.435 \times 10^{-3}$
& $2.345 \times 10^{-2}$ & $4.341 \times 10^{-3}$ & $7.576 \times 10^{-3}$ \\

$100^3$
& $1.289 \times 10^{-3}$ & $3.634 \times 10^{-4}$ & $4.046 \times 10^{-4}$
& $1.045 \times 10^{-2}$ & $1.428 \times 10^{-3}$ & $2.528 \times 10^{-3}$ \\

$200^3$
& $2.858 \times 10^{-4}$ & $7.762 \times 10^{-5}$ & $8.649 \times 10^{-5}$
& $2.702 \times 10^{-3}$ & $3.240 \times 10^{-4}$ & $5.779 \times 10^{-4}$ \\

$400^3$ & \multicolumn{6}{c}{reference solution} \\
\bottomrule
\end{tabular}
\end{table}

\section{Implementation of high-fidelity simulation}
\label{appendix-sec:alg-fom}

\begin{algorithm}[H]
\caption{High-fidelity steady-state solver}
\label{alg:FOM}
\begin{algorithmic}[1]
% \Require $\bm\mu = \{\mathbb{D},\eta,f_D,\mathbb{R}_0,\mathrm{OD},w\},c,\varepsilon,M$
\Function{HighFidelitySolver}{$\bm\mu,\Omega_h,c,\varepsilon,M$}
% \Require Physical parameters $\bm\mu = \{\mathbb{D},\eta,f_D,\mathbb{R}_0,\mathrm{OD},w\}$
% \Require Mesh $\Omega_h$ with $N_h$ nodes, pseudo-time step $\Delta\tau$, convergence tolerance $\varepsilon$, convergence-check interval $M$

\State \textbf{Initialize:} $n \gets 0$,\quad $\mathrm{converged}\gets\mathrm{false}$, \quad
       $\bm{\rho}_h^{0} \gets \tfrac{1}{8}\mathbf{1}_{8N_h}$,\quad
       $\bm{\xi}_z^{0} \gets \Call{LightPropagationSolver}{\bm{\rho}_h^{0}}$
\State Compute the global pseudo-time step: 
$\Delta\tau \gets c
    \left(\frac{6\mathbb D}{h^2}+\frac{3(1+\eta)}{4}+6f_D
    +\mathbb R_0
    +\frac{\eta}{2}\right)^{-1}$
%========================================================
\While{$\mathrm{converged}=\mathrm{false}$}
\Comment{Pseudo-time iteration}

    \State $\mathcal{T}_{\mathrm{dif}} \gets \nabla^2\bm{\rho}^n_h$
           \Comment{Diffusion operator, $\mathcal{O}(N_h)$}
    \State $\mathcal{T}_{\mathrm{SD}}  \gets A_{\mathrm{SD}}\bm{\rho}^n_h$
           \Comment{Spin-destruction, $\mathcal{O}(N_h)$}
    \State $\mathcal{T}_{\mathrm{FD}}  \gets A_{\mathrm{FD}}\bm{\rho}^n_h$
           \Comment{F-destruction, $\mathcal{O}(N_h)$}
    \State $\mathcal{T}_{\mathrm{SE}}  \gets
           \bigl(\mathbf S_z^{\top}\bm{\rho}^n_h\bigr)
           \circ
           \bigl(A_{\mathrm{SE}}^{(z)}\bm{\rho}^n_h\bigr)$
           \Comment{Spin-exchange, $\mathcal{O}(N_h)$}
    \State $\mathcal{T}_{\mathrm{op}}  \gets
           \bm{\xi}_{\perp} \circ \bm{\xi}_z^n\circ\bigl(A_{\mathrm{OP}}\bm{\rho}^n_h\bigr)$
           \Comment{Optical pumping, $\mathcal{O}(N_h)$}
    \State Assemble the right-hand side: \Comment{$\mathcal{O}(N_h)$}
            \[
            \mathcal{R}_{h}^{n} \gets
              \mathbb{D}\mathcal{T}_{\mathrm{dif}}
              -(1+\eta)\mathcal{T}_{\mathrm{SD}}
              -f_D\mathcal{T}_{\mathrm{FD}}
              -\mathbb{R}_0\mathcal{T}_{\mathrm{op}}
              +\eta\mathcal{T}_{\mathrm{SE}}
              \]       
    \State Update the atomic state using the forward Euler method:
            \Comment{$\mathcal{O}(N_h)$}
            \[\bm{\rho}_h^{n+1} \gets \bm{\rho}_h^{n}
              +\Delta\tau\mathcal{R}_{h}^{n}\]

    \State Update the light field using the exponential integrator:
            \Comment{$\mathcal{O}(N_h)$}
    \[\bm{\xi}_z^{n+1} \gets \Call{LightPropagationSolver}{\bm{\rho}_h^{n+1}}\]

    \If{$\mathrm{mod}(n,M)=0$}
    % \Comment{Light update at steps $1,M+1,2M+1,\ldots$}
        \State $\Delta\rho^{n+1} \gets
               \left\| \bm{\rho}_h^{n+1}-\bm{\rho}_h^n \right\|_2$
        \State $\Delta\xi_z^{n+1} \gets
               \left\| \bm{\xi}_z^{n+1} - \bm{\xi}_z^{n} \right\|_2$
    \If{$\Delta\rho^{n+1}<\varepsilon$
        \textbf{and}
        $\Delta\xi_z^{n+1}<\varepsilon$}
        \State $\mathrm{converged} \gets \mathrm{true}$
    \EndIf
    \EndIf
    \State $n\gets n+1$
\EndWhile
\State \Return $\bm{\rho}_h^{n}$, $\bm{\xi}_z^{n}$
\Comment{Converged atomic state and light field}
\EndFunction
\end{algorithmic}
\end{algorithm}

%========================================================
\begin{algorithm}[H]
\caption{Light propagation solver}
\label{alg:LightPropagationSolver}\begin{algorithmic}[1]
\Function{LightPropagationSolver}{$\bm{\rho}_h$}
    
    \For{$i = 1,\ldots,N_x-1$}
    \For{$j = 1,\ldots,N_y-1$}
    \State $\xi_z(i,j,0) \gets 1$ \Comment{Incident B.C.}
    \For{$k = 0,\ldots,N_z-1$}

            \State $\Theta \gets 
                        \exp\Bigl( -\mathrm{OD}\cdot\Delta z_k \bigl[1 - \mathbf S_z^{\top}
                        \bigl(\bm{\rho}_h(i,j,k) + \bm{\rho}_h(i,j,k+1)\bigr)\bigr] \Bigr)$
                       \Comment{Exponential integrator}

            \State $\xi_{z}(i,j,k+1) \gets \Theta \xi_{z}(i,j,k) $
            % \EndCase
        % \EndSwitch
    \EndFor
    \EndFor
    \EndFor
    \State \textbf{return} $\bm{\xi}_z$
    % \State \textbf{return} $\bm{\xi}_z =(\xi_{z}(0),\xi_{z}(1),\ldots,\xi_{z}(N_z))^\top$
\EndFunction
\end{algorithmic}
\end{algorithm}

\section{Implementation of reduced-order model}
\label{appendix-sec:alg-rom}

%========================================================

\begin{algorithm}[H]
\caption{Quadratic expansion of the spin-exchange term}
\label{alg:offline-nqe}
\begin{algorithmic}[1]

\Function{QE}{$\mathbf V,\bar{\bm\rho}$}
\Comment{Total cost: $\mathcal O(r^3N_h)$}

    \For{$i=1,\ldots,r$}
        \For{$j=1,\ldots,r$}
            \State $\bm q_{ij} \gets \mathcal Q(\mathbf v_i,\mathbf v_j)$
            \Comment{$\mathcal O(N_h)$}

            \State $\mathcal A_{:,(i-1)r+j} \gets \mathbf V^\top\bm q_{ij}$
            \Comment{$\mathcal O(rN_h)$}
        \EndFor
    \EndFor
    \Comment{$\mathcal A$: $\mathcal O(r^3N_h)$}

    \For{$i=1,\ldots,r$}
        \State $\bm q_i \gets \mathcal Q(\mathbf v_i,\bar{\bm\rho}) + \mathcal Q(\bar{\bm\rho},\mathbf v_i)$
        \Comment{$\mathcal O(N_h)$}

        \State $\mathcal B_{:,i} \gets \mathbf V^\top\bm q_i$
        \Comment{$\mathcal O(rN_h)$}
    \EndFor
    \Comment{$\mathcal B$: $\mathcal O(r^2N_h)$}

    \State $\bm q_0 \gets \mathcal Q(\bar{\bm\rho},\bar{\bm\rho})$
    \Comment{$\mathcal O(N_h)$}
    
    \State $\mathcal C \gets \mathbf V^\top\bm q_0$
    \Comment{$\mathcal O(rN_h)$}

    \State
    % \Return
    % $\mathcal A\in\mathbb R^{r\times r^2},
    %  \mathcal B\in\mathbb R^{r\times r},
    %  \mathcal C\in\mathbb R^r$
        \Return
    $\mathcal A,\mathcal B,\mathcal C$

\EndFunction

\end{algorithmic}
\end{algorithm}

%========================================================

\begin{algorithm}[H]
\caption{DEIM hyperreduction of the optical-pumping term}
\label{alg:offline-deim}
\begin{algorithmic}[1]
\Function{DEIM}{$\mathbf V,\bar{\bm\rho},\mathcal P_h,
\{\bm\rho_h\left(\bm \mu^{(i)}\right)\}_{i=1}^{N_\mu},\varepsilon_{\mathrm{energy}}$}

    \State Collect snapshots:
    \Comment{$\mathcal{O}(N_\mu N_h)$}
    \begin{equation*}
        \mathbf{S}_{\mathrm{OP}} \gets
    \bigl[\mathbb G_{\mathrm{OP}}\left(\bm\rho_h\left(\bm \mu^{(1)}\right);\bm\mu^{(1)}\right)\bm\rho_h\left(\bm \mu^{(1)}\right) \mid \cdots \mid
    \mathbb G_{\mathrm{OP}}\left(\bm\rho_h\left(\bm \mu^{(N_\mu)}\right);\bm\mu^{(N_\mu)}\right)\bm\rho_h\left(\bm \mu^{(N_\mu)}\right)
    \bigr]
    \end{equation*}
    % $\mathbf{S}_{\mathrm{OP}} \gets
    % \bigl[\mathbb G_{\mathrm{OP}}\left(\bm\rho_h\left(\bm \mu^{(1)}\right);\bm\mu^{(1)}\right)\bm\rho_h\left(\bm \mu^{(1)}\right) \mid \cdots \mid
    % \mathbb G_{\mathrm{OP}}\left(\bm\rho_h\left(\bm \mu^{(N_\mu)}\right);\bm\mu^{(N_\mu)}\right)\bm\rho_h\left(\bm \mu^{(N_\mu)}\right)
    % \bigr].$
    % \Comment{$\mathcal{O}(N_\mu N_h)$}
    \State Extract the POD basis:
    \Comment{$\mathcal O(N_\mu^2N_h)$}
    \begin{align*}
        \mathbf U, \bm\Sigma, \mathbf W &\gets \texttt{SVD}(\mathbb S_{\mathrm{OP}})\\
        m &\gets \min\left\{
        k \in \{1,\ldots,q\}
        \;\middle|\;
        \frac{\sum_{i=k+1}^{q}\sigma_i^2}
             {\sum_{i=1}^{q}\sigma_i^2}
        \leq \varepsilon_{\mathrm{energy}}
    \right\}\\
    \mathbf U_f &\gets\mathbf U(:,1:m)
    \end{align*}

    % \State $\mathbf U\bm\Sigma_f\mathbf W_f^\top
    %     \gets \texttt{Thin SVD}(\mathbf S_{\mathrm{OP}})$
    %     \Comment{$\mathcal O(N_\mu^2N_h)$}
    
    % \State Select the smallest $m$ satisfying
    %     $\displaystyle
    %     \frac{\sum_{i=m+1}^{q}\sigma_{f,i}^2}
    %     {\sum_{i=1}^{q}\sigma_{f,i}^2}
    %     \leq\varepsilon_{\mathrm{energy}}$

    % \State $\mathbf U_f\gets\mathbf U(:,1:m)$

    % \State Determine truncation rank via energy criterion: \Comment{$\mathcal{O}(N_\mu)$}
    %     \[m \gets \min\bigl\{m,\min\{k : 1-E_k \leq \varepsilon_{\mathrm{energy}}\}\bigr\}, \qquad E_k = \frac{\sum_{i=1}^{k}\sigma_i^2} {\sum_{i}\sigma_i^2}\]    
    
    % \State $\mathbf{U}_f \gets [\mathbf{u}_1 \mid \cdots \mid \mathbf{u}_m]$
    % \Comment{$\mathcal{O}(mN_h)$}
    % \State Greedy procedure to select interpolation points:
    \State $p_1 \gets \arg\max_{i}|\mathbf{u}_1|_i$, \quad $\mathbf P_1\gets\mathbf e_{p_1}$
    \Comment{$\mathcal{O}(N_h)$}
    
    \For{$k = 2,\ldots,m$} \Comment{Total cost: $\mathcal{O}(m^2 N_h)$}
        \State $\mathbf{U}_f^{(k-1)} = \left[\mathbf{u}_1 \mid \cdots \mid \mathbf{u}_{k-1}\right]$

    \State Solve
        $\mathbf P_{k-1}^\top\mathbf U_f^{(k-1)}\mathbf c
        =\mathbf P_{k-1}^\top\mathbf u_k$ for $\mathbf c\in\mathbb R^{k-1}$
        \Comment{$\mathcal{O}(k^2)$}

        % \State Solve $\mathbf{U}_f^{(k-1)}\big|_{\mathcal{P}}\mathbf{c}
        %        = \mathbf{u}_k\big|_{\mathcal{P}}$ for $\mathbf{c} \in \mathbb{R}^{k-1}$
        % \Comment{Interpolation on index set $\mathcal{P}$, $\mathcal{O}(k^2)$}
    
        \State $\mathbf{r} \gets \mathbf{u}_k - \mathbf{U}_f^{(k-1)}\mathbf{c}$
        \Comment{DEIM residual, $\mathcal{O}(kN_h)$}
    
        % \State $p_k \gets \arg\max_{i}|\mathbf{r}|_i$, \quad
        %        $\mathcal{P} \gets \mathcal{P}\cup\{p_k\}$
        % \Comment{$\mathcal{O}(N_h)$}

        \State $p_k\gets\arg\max_i|\mathbf r_i|$

        \State $\mathbf P_k\gets[\mathbf P_{k-1},\mathbf e_{p_k}]$
    \EndFor

    \State $\mathbf P\gets\mathbf P_m$
    
    % \State $\mathbf{P} \gets [\mathbf{e}_{p_1},\ldots,\mathbf{e}_{p_m}]\in\{0,1\}^{8N_h\times m}$, \quad $\mathbf{e}_{p_k}$ is the $p_k$-th standard basis vector of $\mathbb{R}^{8N_h}$
    % \Comment{$\mathcal{O}(mN_h)$}

    \State $\mathcal{D} \gets \mathbf{V}^\top\mathbf{U}_f(\mathbf{P}^\top\mathbf{U}_f)^{-1}$
    \Comment{$\mathbf{V}^\top\mathbf{U}_f$: $\mathcal{O}(rmN_h)$,
             $\mathbf{P}^\top\mathbf{U}_f$: $\mathcal{O}(mN_h)$}

    \State Find the matrix indices for the interpolation points:
    \begin{align*}
        \bm \rho_h \left(p_s\right) &\mapsto \left(\bm \rho_h \left(i_s,j_s,k_s\right)\right)_{q_s}\\
        \bm \rho_h \left(\mathbf p_{i_s,j_s,k}\right) &\mapsto \bm \rho_h \left(i_s,j_s,k\right)
    \end{align*}
    
    % \State Map $p_s \mapsto (i_s,j_s,k_s,q_s)$; $\mathbf p_{i,j,k}$ denote the eight components at node $(i,j,k)$

    % \State Form the selected-coordinate vectors
    % $\bm x_m=(x_{i_s})_{s=1}^m$, $\bm y_m=(y_{j_s})_{s=1}^m$, and $\bm z_m=(z_{k_s})_{s=1}^m$

    % \State Compute the matrices and vector for online evaluation:
    
    \State $\bm x_m\gets(x_{i_s})_{s=1}^m,\quad
    \bm y_m\gets(y_{j_s})_{s=1}^m,\quad
    \bm z_m\gets(z_{k_s})_{s=1}^m$

    \For{$s=1,\ldots,m$}
        \State $\displaystyle  \mathcal E_{s,:} \gets \sum_{k=0}^{k_s-1}
        \frac{\Delta z_k}{2} \mathbf S_z^\top
        \left[
            \mathbf V(\mathbf p_{i_s,j_s,k},:) + \mathbf V(\mathbf p_{i_s,j_s,k+1},:)
        \right]$
        \Comment{$\mathcal O(N_zr)$}
    
        \State $\displaystyle e_s \gets \sum_{k=0}^{k_s-1}
        \frac{\Delta z_k}{2} \mathbf S_z^\top
        \left[
            \bar{\bm\rho}(\mathbf p_{i_s,j_s,k}) + \bar{\bm\rho}(\mathbf p_{i_s,j_s,k+1})
        \right]$
        \Comment{$\mathcal O(N_z)$}
    
        \State $\mathcal F_{s,:}\gets
        \mathbf e_{q_s}^\top A_{\mathrm{OP}}
        \mathbf V(\mathbf p_{i_s,j_s,k_s},:)$
        \Comment{$\mathcal O(r)$}
    
        \State $f_s\gets
        \mathbf e_{q_s}^\top A_{\mathrm{OP}}
        \bar{\bm\rho}(\mathbf p_{i_s,j_s,k_s})$
    \EndFor

    % \State $\mathcal E_{s,:} \gets \sum_{k=0}^{k_s-1}\frac{\Delta z_k}{2}\mathbf S_z^\top \left[\mathbf V(i_s,j_s,k)+\mathbf V(i_s,j_s,k+1)\right]$
    % \Comment{$\mathcal O(mN_zr)$}

    % \State $e_s\gets \sum_{k=0}^{k_s-1}\frac{\Delta z_k}{2}\mathbf S_z^\top
    % \left[\bar{\bm\rho}(i_s,j_s,k)+\bar{\bm\rho}(i_s,j_s,k+1)\right], \qquad s=1,\ldots,m$    
    % \Comment{$\mathcal O(mN_z)$}

    % \State $\mathcal F_{sa}\gets
    % \mathbf e_{q_s}^\top A_{\mathrm{OP}}\mathbf v_a(i_s,j_s,k_s)$,
    % $\quad s=1,\ldots,m$, $\ a=1,\ldots,r$
    % \Comment{$\mathcal O(mr)$}

    % \State $f_s\gets
    % \mathbf e_{q_s}^\top A_{\mathrm{OP}}\bar{\bm\rho}(i_s,j_s,k_s)$,
    % $\quad s=1,\ldots,m$
    % \Comment{$\mathcal O(m)$}
    
    \State \Return $\mathcal D,\mathcal E,\bm e,\mathcal F,\bm f,
    \bm x_m,\bm y_m,\bm z_m$
\EndFunction
\end{algorithmic}
\end{algorithm}

\begin{algorithm}[H]
\caption{Offline phase of the reduced-order model}
\label{alg:offline}
\begin{algorithmic}[1]

\Function{Offline}{$\mathcal P_h=\{\bm\mu^{(i)}\}_{i=1}^{N_\mu},\varepsilon_{\mathrm{energy}}$}

\State Generate high-fidelity snapshots:
\Comment{$\mathcal{O}(N_\mu \bar{N}_{\mathrm{iter}} N_h)$}
\ParFor{$i=1,\ldots,N_\mu$}
    \State $\bm\rho_h\left(\bm \mu^{(i)}\right)\gets\Call{HighFidelitySolver}{\bm\mu^{(i)},c,\varepsilon,M}$
    % \Comment{March the FOM to its converged steady state}
\EndFor

\State Compute mean state:
    $\bar{\bm{\rho}} \gets \tfrac{1}{N_\mu}
    \sum_{i=1}^{N_\mu} \bm{\rho}_h\left(\bm \mu^{(i)}\right)$
\Comment{$\mathcal{O}(N_\mu N_h)$}

\State Form snapshot matrix:
    $\mathbb S \gets
    \left[
    \bm\rho_h(\bm\mu^{(1)})-\bar{\bm\rho}
    \mid\cdots\mid
    \bm\rho_h(\bm\mu^{(N_\mu)})-\bar{\bm\rho}
    \right]$
\Comment{$\mathcal{O}(N_\mu N_h)$}

% \State Compute POD basis:
%     $\mathbf{V}\bm{\Sigma}\mathbf{W}^\top \gets \texttt{SVD}(\mathbb S)$
%     \Comment{$\mathcal{O}(N_\mu^2 N_h)$}

\State Extract the POD basis:
\Comment{$\mathcal O(N_\mu^2N_h)$}
\begin{align*}
    \mathbf U, \bm\Sigma, \mathbf W &\gets \texttt{SVD}(\mathbb S)\\
    r &\gets \min\left\{
    k \in \{1,\ldots,q\}
    \;\middle|\;
    \frac{\sum_{i=k+1}^{q}\sigma_i^2}
         {\sum_{i=1}^{q}\sigma_i^2}
    \leq \varepsilon_{\mathrm{energy}}
\right\}\\
\mathbf V & \gets \mathbf U(:,1:r)
\end{align*}
% $\mathbf U, \bm\Sigma, \mathbf W \gets \texttt{SVD}(\mathbb S)$

% \State Determine reduced basis dimension:
% % Select the smallest $r$ satisfying:
%     $r \gets \min\left\{
%     k \in \{1,\ldots,q\}
%     \;\middle|\;
%     \sum_{i=r+1}^{q}\sigma_i^2/
%     \sum_{i=1}^{q}\sigma_i^2
%     \leq \varepsilon_{\mathrm{energy}}
% \right\}.$

% \State $\mathbf V\gets\mathbf U(:,1:r)$

\State Project the linear terms involving the mean state onto the reduced basis:
    \Comment{$\mathcal O(rN_h)$}
    \begin{align*}
        \bar{\mathcal T}_{\mathrm{dif}}^r &\gets
        \mathbf V^\top \left(\Delta_h\otimes\mathbf I_8\right) \bar{\bm\rho}, \\
        \bar{\mathcal T}_{\mathrm{SD}}^r &\gets \mathbf V^\top
        \left(\mathbf I_{N_h}\otimes A_{\mathrm{SD}}\right) \bar{\bm\rho}, \\
        \bar{\mathcal T}_{\mathrm{FD}}^r &\gets
        \mathbf V^\top \left(\mathbf I_{N_h}\otimes A_{\mathrm{FD}}\right) \bar{\bm\rho}.
    \end{align*}

\State Project the linear operators onto the reduced basis:
    \Comment{$\mathcal O(r^2N_h)$}
    \begin{align*}
        \Delta^r &\gets \mathbf V^\top
        \left(\Delta_h\otimes\mathbf I_8\right) \mathbf V, \\
        A_{\mathrm{SD}}^r &\gets
        \mathbf V^\top \left(\mathbf I_{N_h}\otimes A_{\mathrm{SD}}\right) \mathbf V, \\
        A_{\mathrm{FD}}^r &\gets \mathbf V^\top
        \left(\mathbf I_{N_h}\otimes A_{\mathrm{FD}}\right) \mathbf V.
    \end{align*}

\State Hyperreduction of nonlinear terms:
\begin{align*}
    \mathcal{A}, \mathcal{B}, \mathcal{C} &\gets
    \Call{QE}{\mathbf{V}, \bar{\bm{\rho}}}\\
    \mathcal D,\mathcal E,\bm e,\mathcal F,\bm f,
    \bm x_m,\bm y_m,\bm z_m &\gets
    \Call{DEIM}{\mathbf V,\bar{\bm\rho},\mathcal P_h,
    \{\bm\rho_h^{(i)}\}_{i=1}^{N_\mu},\varepsilon_{\mathrm{energy}}}
\end{align*}
           
% \State Quadratic expansion of the spin-exchange term:
% $(\mathcal{A}, \mathcal{B}, \mathcal{C}) \gets$
%     \Call{QE}{$\mathbf{V}, \bar{\bm{\rho}}$}
    
% \State Hyperreduction of the optical-pumping term:
% $(\mathcal D,\mathcal E,\bm e,\mathcal F,\bm f,
%     \bm x_m,\bm y_m,\bm z_m)\gets$
%     \Call{DEIM}{$\mathbf V,\bar{\bm\rho},\mathcal P_h,
%     \{\bm\rho_h^{(i)}\}_{i=1}^{N_\mu},\varepsilon_{\mathrm{energy}}$}

\State \Return $\Theta_{\mathrm{off}} \gets \bigl\{ \mathbf V,\bar{\bm\rho},\mathcal A,\mathcal B,\mathcal C,
    \mathcal D,\mathcal E,\bm e,\mathcal F,\bm f,
    \bm x_m,\bm y_m,\bm z_m,
    \bar{\mathcal T}_{\mathrm{dif}}^r,\bar{\mathcal T}_{\mathrm{SD}}^r,
    \bar{\mathcal T}_{\mathrm{FD}}^r,\Delta^r,A_{\mathrm{SD}}^r,A_{\mathrm{FD}}^r \bigr\}$
\EndFunction
% \item[]
\end{algorithmic}
\end{algorithm}

%========================================================

\begin{algorithm}[H]
\caption{Online phase of the reduced-order model}
\label{alg:online}
\begin{algorithmic}[1]
% \Require $\bm \mu = \{\mathbb{D},\eta,f_D,\mathbb{R}_0,\mathrm{OD},w\},c,\varepsilon,M$
% \Require Precomputed quantities
%         $\mathbf V,\bar{\bm\rho},\mathcal A,\mathcal B,\mathcal C,
%         \mathcal D,\mathcal E,\bm e,\mathcal F,\bm f,
%         \bm x_m,\bm y_m,\bm z_m,
%         \bar{\mathcal T}_{\mathrm{dif}}^r,\bar{\mathcal T}_{\mathrm{SD}}^r,
%         \bar{\mathcal T}_{\mathrm{FD}}^r,\Delta^r,A_{\mathrm{SD}}^r,A_{\mathrm{FD}}^r$
% \Require Pseudo-time step factor $c$, convergence tolerance $\varepsilon$, and convergence-check interval $M$

\Function{Online}{$\bm \mu,\Theta_{\mathrm{off}},\Omega_h,c,\varepsilon,M$}

\State \textbf{Initialize:} $n \gets 0$, \quad
        $\bm{\alpha}^{0} \gets \mathbf{V}^\top
        \bigl(\frac18\mathbf1_{8N_h} - \bar{\bm{\rho}}\bigr)
        \in \mathbb{R}^r$, \quad
        $\mathrm{converged}\gets\mathrm{false}$
% \State Assemble average operator:
%         $\mathcal{AVE}^r 
%         \gets \mathbb{D}\bar{\mathcal{T}}_{\mathrm{dif}}^r
%         -(1+\eta)\bar{\mathcal{T}}_{\mathrm{SD}}^r
%         -f_D\bar{\mathcal{T}}_{\mathrm{FD}}^r$
%         \Comment{$\mathcal{O}(r)$}

\State Compute the global pseudo-time step: 
$\Delta\tau \gets c
    \left(\frac{6\mathbb D}{h^2}+\frac{3(1+\eta)}{4}+6f_D
    +\mathbb R_0
    +\frac{\eta}{2}\right)^{-1}$
    
\State Assemble the parameter-dependent baseline light factors:
\Comment{$\mathcal O(m)$}
    \[
        \bm\xi_{\mathcal P}
        \gets
        \exp\!\left[-\mathrm{OD}\bm z_m
        -\frac{2}{w^2}
        (\bm x_m\circ\bm x_m+\bm y_m\circ\bm y_m)\right]
    \]

\State Assemble the reduced linear operator:
    $\mathbb L_r\gets
    \mathbb D\Delta^r
    -(1+\eta)A_{\mathrm{SD}}^r
    -f_DA_{\mathrm{FD}}^r$
    % \Comment{$\mathcal O(r^2)$}

\State Assemble the reduced constant vector:
    $\mathbb C_r\gets
    \mathbb D\bar{\mathcal T}_{\mathrm{dif}}^r
    -(1+\eta)\bar{\mathcal T}_{\mathrm{SD}}^r
    -f_D\bar{\mathcal T}_{\mathrm{FD}}^r$
    % \Comment{$\mathcal O(r)$}

\While{$\mathrm{converged}=\mathrm{false}$}
\Comment{Pseudo-time iteration}
    % \State $\mathcal{T}_{\mathrm{dif}}^r \gets \Delta^r\bm{\alpha}^n$
    %        \Comment{Diffusion operator, $\mathcal{O}(r^2)$}
    % \State $\mathcal{T}_{\mathrm{SD}}^r  \gets A_{\mathrm{SD}}^r\bm{\alpha}^n$
    %        \Comment{Spin-destruction, $\mathcal{O}(r^2)$}
    % \State $\mathcal{T}_{\mathrm{FD}}^r  \gets A_{\mathrm{FD}}^r\bm{\alpha}^n$
    %        \Comment{F-damping, $\mathcal{O}(r^2)$}
    \State $\mathcal{T}_{\mathrm{SE}}^r  \gets
           \mathcal{A}(\bm{\alpha}^n \otimes \bm{\alpha}^n) + \mathcal{B} \bm{\alpha}^n + \mathcal{C}$
           \Comment{Spin exchange via QE, $\mathcal O(r^3)$}
           
    \State
    \(
        \mathbf f_{\mathrm{OP}}^n
        \gets
        \mathbb R_0 \bm\xi_{\mathcal P}
        \circ\exp\!\left[2\mathrm{OD}(\mathcal E\bm\alpha^n+\bm e)\right]
        \circ(\mathcal F\bm\alpha^n+\bm f)
    \)
        \Comment{$\mathcal O(mr)$}
    
    \State $\mathcal T_{\mathrm{OP}}^r\gets\mathcal D\mathbf f_{\mathrm{OP}}^n$
           \Comment{Optical pumping via DEIM, $\mathcal{O}(mr)$}

    \State Assemble the right-hand side:
        \[\mathcal R_r^n\gets
        \mathbb L_r\bm\alpha^n
        +\mathbb C_r
        +\eta\mathcal T_{\mathrm{SE}}^r
        -\mathcal T_{\mathrm{OP}}^r\]

    % \State Assemble reduced right-hand side:
    %         \[\mathcal{R}_{r}^{n} \gets
    %           \mathbb{D}\mathcal{T}_{\mathrm{dif}}^r
    %           -(1+\eta)\mathcal{T}_{\mathrm{SD}}^r
    %           -f_D\mathcal{T}_{\mathrm{FD}}^r
    %           -\mathcal{T}_{\mathrm{OP}}^r
    %           +\eta\mathcal{T}_{\mathrm{SE}}^r
    %           + \mathcal{AVE}^r\]
    
    \State Forward Euler update:
        \(\bm{\alpha}^{n+1} \gets \bm{\alpha}^n + \Delta\tau\mathcal{R}_{r}^{n}\)

    \If{$\mathrm{mod}(n,M)=0$}
        \State $\Delta\alpha^{n+1} \gets \left\|\bm\alpha^{n+1}-\bm\alpha^n\right\|_2$

        \If{$\Delta\alpha^{n+1}<\varepsilon$}
            \State $\mathrm{converged} \gets \mathrm{true}$
        \EndIf
    \EndIf
    \State $n \gets n + 1$
\EndWhile
\State Reconstruction: $\bm{\rho}_r^n \gets \mathbf{V}\bm{\alpha}^n + \bar{\bm{\rho}}$
\State \Return $\bm{\rho}_r^n$ \Comment{Steady-state ROM solution}
\EndFunction
\end{algorithmic}
\end{algorithm}

% \newpage

% \bibliographystyle{cas-model2-names}
% \bibliographystyle{unsrtnat}
\bibliographystyle{elsarticle-num}
\bibliography{references}

\end{document}